\documentstyle[aps,prb,multicol,epsfig]{revtex}
\newcommand{\revised}[1] {#1}                              % don't mark 

\newcommand{\comment}[1]  {}
\newcommand{\figdep}[1]   {#1}

\newcommand{\sfrac}[2]  { {\textstyle \frac{#1}{#2}} }
\newcommand{\abs}[1]    { {\left| #1 \right|}}
\newcommand{\real}      { {\mathrm Re\,} }
\newcommand{\imag}      { {\mathrm Im\,} }
\newcommand{\sign}      { {\mathrm sign\,} }
\newcommand{\vect}[1]   { {\mathbf #1 } }

\newcommand{\ket}[1]    { \vert   {#1} \rangle }
\newcommand{\mean}[1]   { \left<  \, {#1} \, \right> }
\newcommand{\expect}[3] { \left< {#1} \vert \, {#2} \, \vert {#3} \right> }
\newcommand{\GF}[2]     { \left<\!\left< \, {#1}; \, {#2} \, \right>\!\right> }
\newcommand{\GFome}[2]  { \GF{#1}{#2}_{\omega} }
\newcommand{\comm}[2]   { \left[ {#1}, \, {#2} \right] }
\newcommand{\doco}[3]   { \comm{\comm{#1}{#2}}{#3} }

\newcommand{\vk}        { {\vect{k}} }
\newcommand{\vp}        { {\vect{p}} }
\newcommand{\vq}        { {\vect{q}} }

\newcommand{\Norm}      { {\mathcal{N}} }
\newcommand{\Heff}      { {\mathcal{H}} }
\newcommand{\Kern}      { {\mathcal{K}} }
\newcommand{\phGF}      { {\mathcal{G}} }

\newcommand{\GMF}       { {\mathrm SC}   }
\newcommand{\RPA}       { {\mathrm RPA}  }
\newcommand{\Hgmf}      { \Heff^{\GMF}   }
\newcommand{\Kgmf}      { \Kern^{\GMF}   }
\newcommand{\Krpa}      { \Kern^{\RPA}   }
\newcommand{\corr}      { {\rm c} }

\newcommand{\ChiFr}     { \chi^{\rm free}  }      % free susceptibility
\newcommand{\ChiZe}     { \chi^{\rm 0}  }         % ren. free susceptibility
\newcommand{\ChiCh}     { \chi^{\rm ch} }         % charge susceptibility
\newcommand{\ChiSp}     { \chi^{\rm sp} }         % longitudinal spin susc
\newcommand{\ChiTr}     { \chi^{\rm +-} }         % transverse spin susc

\newlength{\Units}

\begin{document}

% %%%%%%%%%%%%%%%%%%%%%%%%%%%%%%%%%%%%%%%%%%%%%%%%%%%%%%%%%%%%
% title
% %%%%%%%%%%%%%%%%%%%%%%%%%%%%%%%%%%%%%%%%%%%%%%%%%%%%%%%%%%%%
\title{Dyson Equation Approach to Many--Body Greens Functions and
  Self--Consistent RPA \\ First Application to the Hubbard Model} 

\author{
  Steffen Sch\"afer$^{1,2}$\thanks{email: steffen@in2p3.fr}, 
  Peter Schuck$^{1}$} 
\address{
  $^{1}$ 
  Institut des Sciences Nucl\'eaires, 
  Universit\'e Joseph Fourier, CNRS--IN2P3 \\
  53, Avenue des Martyrs, F-38026 Grenoble Cedex, France 
  \\
  $^{2}$
  Physik-Department T32, Technische Universit\"at M\"unchen \\
  James-Franck-Stra\ss{}e 1, D-85747 Garching, Germany}
 
\date{\today }
\maketitle

% %%%%%%%%%%%%%%%%%%%%%%%%%%%%%%%%%%%%%%%%%%%%%%%%%%%%%%%%%%%%
% abstract
% %%%%%%%%%%%%%%%%%%%%%%%%%%%%%%%%%%%%%%%%%%%%%%%%%%%%%%%%%%%%

%\draft command makes pacs numbers print
\draft

\begin{abstract}
\begin{center}
\parbox{14cm}{  
  \revised{An approach for particle--hole correlation functions, based
  on the so-called SCRPA, is developed. }
  This leads to a fully self--consistent RPA-like theory which
  satisfies the $f$--sum rule and several other theorems.
  As a first step, a simpler self--consistent approach, the
  renormalized RPA, is solved numerically in the one--dimensional
  Hubbard model.  
  The charge and the longitudinal spin susceptibility, the momentum
  distribution and several ground state properties are calculated and
  compared with the exact results. 
  Especially at half filling, our approach provides quite promising
  results and matches the exact behaviour apart from a general
  prefactor. 
  The strong coupling limit of our approach can be described
  analytically.   
}
\end{center}
\end{abstract}


\comment{
  71.10.-w Theories and models of many electron systems\\
  75.10.Jm Quantized spin models\\
  72.15.Nj Collective modes (e.g., in one-dimensional conductors) }
% 
% \begin{multicols}{2}
% \narrowtext

% %%%%%%%%%%%%%%%%%%%%%%%%%%%%%%%%%%%%%%%%%%%%%%%%%%%%%%%%%%%%
% Introduction
% %%%%%%%%%%%%%%%%%%%%%%%%%%%%%%%%%%%%%%%%%%%%%%%%%%%%%%%%%%%%
\section{Introduction}
\label{Intro}

The advent of high $T_{c}$ super--conductivity, which remains
unexplained theoretically in its essence, has spurred an enormous quest
for developing many--body approaches capable to describe strongly
correlated Fermion systems. Various formalisms have been applied in
the past, each which its strengths and deficiencies 
(for a review see ref.\cite{Blaizot}). 
\comment{see also article by Vilk/Tremblay}

However, contrary to standard mean--field theory which is a commonly
accepted lowest--order many--body approach, for correlation functions,
such a generic method is missing so far. In this respect, any new and
promising vistas are worthwhile to be pursued and elaborated. Indeed,
in the recent past, a theory for two--body correlation functions has
been developed bearing the characteristics of a generalization of
Hartree--Fock theory to two--body clusters. In its roots this theory
goes back rather far in time and has been promoted independently by
several groups\cite{Rowe,Roth,DukelskySchuck,Roepke}.  In the
literature, it is known under various names such as Self--Consistent
RPA (SCRPA), Cluster Mean Field (CMF) and Equation of Motion Method
(EOM)\cite{Rowe,Roth,DukelskySchuck,Roepke}.  In itself it is an
approximation to the so-called Dyson Equation Approach(DEA) to
correlation functions where one replaces the full mass operator by its
instantaneous part.  However, only recently this method has found the
attention and formal developments it deserves with, indeed, very
promising results. The most outstanding of those is certainly the
exact reproduction of the lowest spin--wave excitation spectrum,
$\omega_{k}=\sfrac{\pi}{2}\abs{\sin k}$, known from the Bethe ansatz,
of the antiferromagnetic Heisenberg chain\cite{Krueger}.  Moreover,
also some simpler models have been treated successfully in this
approach\cite{Dukelsky}.  Encouraged by these results, we thought it
worthwhile to apply the method to the strongly correlated electron
problem within the single--band Hubbard Hamiltonian with on--site
repulsion $U$.

Since the approach, which we hitherto want to call SCRPA, is based on
non--linear equations for non--local two--body correlation functions,
one understands that it is numerically very demanding.   
We therefore choose as a first application the one--dimensional
Hubbard model for several reasons: 
\begin{itemize}
\item 
  The exact solution of the ground state is known from the Bethe
  ansatz\cite{LiebWu}. Therefore, a direct comparison of the SCRPA
  results is possible. 
\item
  The numerical effort in one dimension may be expected to be more
  modest than in higher dimensions.
\item
  \revised{The experience gained in the $1d$ case may help us to
  attack higher dimensions in the future.}
\end{itemize}
The price to pay for this strategy is that one--dimensional problems
are notoriously difficult to describe because of their extreme quantum
character. 
\revised{As our method is not specifically designed for one dimension,
we cannot expect it to reproduce particularities, such as Luttinger
liquid behaviour.}

As we will see, our approach nonetheless permits to obtain interesting
results in one dimension. They should, however, be judged in the light
that in this first application to the Hubbard model we did not solve
the SCRPA equations in full but applied a rather obvious and from the
numerical point of view very simplifying approximation.
Nevertheless, this approximation, known in the literature under the
name of renormalized RPA, keeps the essentials of the
self--consistency aspects.  

We demonstrate in this paper that the formalism allows for the
self--consistent solution of a fully closed system of non--linear 
equations for two--body correlation functions.
Moreover, important formal theorems are respected. Among those, we,
for instance, cite the fulfillment of the $f$--sum rule 
(energy weighted sum rule) and of the Luttinger theorem. 

Other interesting results concern the strong coupling regime of the
half filled chain. For example, the self--consistently calculated
momentum distribution can be found analytically in the large $U$
limit. It obeys $n_{k}\propto \cos k$ with a proportionality factor of
$4/U$ instead of $8\ln 2/U$ of the exact result, known from the large
$U$ expansion of the Bethe ansatz solution, resulting in an error
smaller than $30\%$. This result is the more astonishing as it was
obtained with the renormalized RPA approach. It can be expected that
it will be substantially improved once the full SCRPA solution is
available. 

The paper is organized as follows. 
In section~\ref{GMFtheory}, in order to make our paper
self--contained, we will give a brief overview of the SCRPA theory
starting from the equation of motion for a completely 
general Green's function. 
In particular, the explicit form of the self--consistent and
renormalized RPA particle--hole propagators is derived in terms of a
closed system of non--linear self--consistent equations. 
In section~\ref{HubbardModel}, our approach is applied to charge
and spin correlation functions in the Hubbard model. 
In section~\ref{HubbardChain}, we solve numerically the set of
self--consistent equations for the renormalized RPA response
functions for different fillings and interaction strengths. 
Our results are compared with the Bethe ansatz solution and with Quantum
Monte Carlo calculations. For half filling and large $U$, analytic
expressions are given for the momentum distribution function and the 
susceptibilities of our theory.
In section~\ref{Conclusions}, we draw some conclusions and give an
outlook on some improvements which are planned to be implemented in
our approach.
\revised{In appendix~\ref{Variational}, we outline the connection
between the SCRPA and a variational ansatz which minimizes the energy
weighted sum rule.}
Appendix~\ref{ChiFree} provides the explicit expressions for the free
particle--hole susceptibility in one dimension.
In appendix~\ref{LargeUOcc}, we show how the analytic expressions
which we derived for the strong coupling limit of our theory at half
filling, are a rigorous solution of the renormalized RPA equations.

% %%%%%%%%%%%%%%%%%%%%%%%%%%%%%%%%%%%%%%%%%%%%%%%%%%%%%%%%%%%%
% The Method
% %%%%%%%%%%%%%%%%%%%%%%%%%%%%%%%%%%%%%%%%%%%%%%%%%%%%%%%%%%%%
\section{The Dyson Equation Approach to Many--Body Green's Functions}
\label{GMFtheory}

In this section, we briefly want to review the Dyson Equation Approach
(DEA) to correlation functions\cite{DRS98}. 
The DEA is increasingly used in the many--body community and has recently
produced interesting results in various domains of many--body
physics\cite{DukelskySchuck,Roepke,Krueger,Dukelsky}.  

Let us start with the definition of a general causal (time ordered) or
retarded many--body Green's function at zero temperature and at
equilibrium (the generalization to finite temperature, using the
Matsubara technique, is straightforward),
\comment{Nolting (3.107)}
\begin{eqnarray}
  \label{GFdef}
  G_{ A B }^{c}(t,t') 
  & \equiv & \GF{ A(t) }{ B(t') }^{c} 
  \nonumber \\
  & := & -i \expect{0}{ T_{\epsilon} A(t) B(t')}{0} 
  \\
  G_{ A B }^{\rm ret}(t,t') 
  & \equiv & \GF{ A(t) }{ B(t') }^{\rm ret} 
  \nonumber \\
  & := & -i\,\Theta(t-t')\,\expect{0}{\comm{A(t)}{B(t')}_{-\epsilon}}{0} 
  \; \mbox{,}
\end{eqnarray}
where $ \ket{0} $ is the exact ground--state and $ T_{\epsilon} $ Wick's
time--ordering operator. 

Here, $A(t)$ and $B(t')$ are arbitrary operators built out of any
number of annihilation and/or creation operators of Bosons or Fermions
or mixtures of both. 
Usually $A$ and $B$ will depend on one or several indices, and the
notation $\GF{A}{B}$ has to be considered as a shorthand for the
matrix Green's function $\GF{A_{\alpha}}{B_{\beta}}$ where $\alpha$
and $\beta$ run over the whole set of quantum numbers.
The operators $A$ and $B$ can also be spin operators or even more
general operators such as multi--component operators
$\vect{A}=(A_{1},A_{2},\ldots)$ where the single components are again
chosen according to the problem in question.   
For the derivation of a Dyson equation, however, we will choose
$B=A^{+}$. The case $B \neq A^{+}$ needs further considerations which
may be important for the derivation of integral equations for vertex
functions.

\comment{Nolting (3.108), but without $H-\mu\hat{N}$}
The time dependence of the operators is given in the Heisenberg
picture, $ X(t) = e^{i H t} \, X \, e^{-i H t} $, where the
Hamiltonian $H$ is also completely general. It may describe
relativistic or non--relativistic Fermi, Bose or spin systems or any
system for which a Hamilton operator exists.   

At equilibrium the two time Green's functions (\ref{GFdef}) depend
only on the time difference such that their Fourier transforms are
only functions of one frequency.  
These are the quantities for which we want to derive a Dyson equation. 
As the derivation is the same for either causal or retarded Green's
functions we will from now on omit the upper indices.

Let us establish the equation of motion for $G_{ A B }(t,t')$: 
\comment{Nolting (3.120)}
\begin{equation}
  \label{EqOfMotionTime}
  i \frac{\partial}{\partial t} \GF{A(t)}{B(t')}
  = \delta(t-t') \, \mean{\comm{A}{B}_{-\epsilon}} 
  \;+\; \GF{\comm{A}{H}(t)}{B(t')} 
  \;\mbox{,}
\end{equation}
or, in energy representation,
\comment{Nolting (3.129)}
\begin{equation}
  \label{EqOfMotionFreq}
  \omega \, \GFome{A}{B}
  = \mean{\comm{A}{B}_{-\epsilon}} 
  \;+\; \GFome{\comm{A}{H}}{B}
  \;\mbox{.}
\end{equation}
On the rhs of (\ref{EqOfMotionFreq}) the commutator
\begin{equation}
  \label{Norm}
  \Norm:=\mean{\comm{A}{B}_{-\epsilon}}
\end{equation}
plays the role of a norm matrix, and $\GFome{\comm{A}{H}}{B}$ is a
general Green's function containing the interaction once explicitly.  
For fermion like operators $A$ and $B$ we will use the
anti--commutator Green's function $(\epsilon=-)$ and for boson like
(e.g. a product of an even number of fermion operators) the commutator
Green's  function $(\epsilon=+)$.

Under the assumption that the inverse of $\GFome{A}{B}$ exists, 
an effective ``Hamiltonian'' $\Heff_{A B}(\omega)$ can be defined as  
\begin{equation}
  \label{DefHeff}
  \Heff_{A B}(\omega) = \GFome{\comm{A}{H}}{B} \; \GFome{A}{B}^{-1}
  \; \mbox{.}
\end{equation}
The equation of motion~(\ref{EqOfMotionFreq}) can thus be transformed 
into a Dyson equation,
\begin{equation}
  \label{DysonEq}
  \omega \, \GFome{A}{B} = \Norm \;+\; \Heff_{A B}(\omega) \,  \GFome{A}{B}
  \; \mbox{.} 
\end{equation}
We stress again that the products on the rhs. of eqs.~(\ref{DefHeff})
and (\ref{DysonEq}) are understood to be matrix multiplications.

As we do not yet know how to determine the effective Hamiltonian 
$\Heff_{A B}(\omega)$, the solution of the Dyson eq.~(\ref{DysonEq}),
\begin{equation}
  \label{DysonSolution}
  \GFome{A}{B} = \biggl\{ \omega \,- \, \Heff_{A B}(\omega) \biggr\}^{-1}
  \; \Norm 
  \; \mbox{,} 
\end{equation}
remains for the moment completely formal.
In order to derive a more explicit and useful expression for 
$\Heff_{A B}(\omega)$, we insert the inverse of the formal solution
(\ref{DysonSolution}) in (\ref{DefHeff}): 
\begin{eqnarray}
  \label{Heff12}
  \Heff_{A B}(\omega) & = &
  \GFome{\comm{A}{H}}{B} \; 
  \Norm^{-1} \; \biggl\{ \omega  \,-\, \Heff_{A B}(\omega) \biggr\}
  \nonumber \\
  & \equiv & \qquad
  \Heff_{A B}^{\rm I}(\omega) \qquad - \qquad  \Heff_{A B}^{\rm II}(\omega)
\end{eqnarray}
The first part, 
$\Heff_{A B}^{\rm I}(\omega)\equiv\GFome{\comm{A}{H}}{B}\Norm^{-1}\omega$,
can be obtained from the equation of motion for the higher Green's
function $\GFome{\comm{A}{H}}{B}$ which is set up from ``the right''
this time (i.e. deriving with respect to $t'$ instead of $t$):
\begin{equation}
  \label{EqOfMotion2}
  \GFome{\comm{A}{H}}{B} \, \omega \; = \; 
  \mean{\doco{A}{H}{B}_{-\epsilon}} \;+\; \GFome{\comm{A}{H}}{\comm{H}{B}}
\end{equation}

\comment{see Ring Schuck, Appendix F}

If we adopt $B\equiv A^{+}$, the second part of the effective
Hamiltonian, 
$\Heff_{A B}^{\rm II}(\omega)
\equiv\GFome{\comm{A}{H}}{B}\Norm^{-1}\Heff_{A B}(\omega)$,
contains only $n$-line {\em reducible} contributions\cite{RS}, 
with $n$ being the number of fermion operators in $A$.
Further, it can be shown\cite{RS} that the sole function of 
$\Heff_{A B}^{\rm II}(\omega)$ is to cancel {\em all} reducible
contributions of $\Heff_{A B}^{\rm I}(\omega)$.    
As the double commutator $\mean{\doco{A}{H}{B}_{-\epsilon}}$ has no
reducible contributions, we just have to put an index ``irreducible''
on the Green's function on the rhs. of eq.~(\ref{EqOfMotion2}) to
obtain as the final expression for the effective Hamiltonian:
\begin{eqnarray}
  \label{Heff} 
  \Heff_{A B}(\omega) & = &
  \biggl\{ 
  \mean{\doco{A}{H}{B}_{-\epsilon}} \; + \;
  \GFome{\comm{A}{H}}{\comm{H}{B}}^{\rm irr}
  \biggr\}
  \; \Norm^{-1}
  \nonumber \\
  & \equiv & 
  \qquad \Hgmf_{A B} \qquad \qquad + \qquad \Heff_{A B}^{\rm res}(\omega)
\end{eqnarray} 

We see that the effective Hamiltonian~(\ref{Heff}) splits up in a
natural way in an instantaneous part and in a truly dynamic
(resonant) part. The latter contains scattering processes 
leading to imaginary potentials and corresponding real ones with
a frequency dependence.

To obtain a better understanding of the various terms contributing to
the effective Hamiltonian, let us analyze eq.~(\ref{Heff}) for
the well--known case of the single--particle propagator, that is
$A=a_{1}$ and $B=a^{+}_{1'}$.  
Since later we want to restrict ourselves to a non--relativistic
fermion system let us consider a typical Hamiltonian
\begin{equation}
  \label{genHamiltonian}
  H = \sum \limits_{1 2} t_{12} \, a^{+}_{1} a_{2}
  \;+\; \frac{1}{4} \sum \limits_{1 2 3 4} \bar{v}_{1 2 3 4} \,
  a^{+}_{1} a^{+}_{2} a_{4} a_{3}
\end{equation}
where $a, a^{+}$ are fermion destruction and creation
operators. 
The matrix elements $t_{1 2}$ and $\bar{v}_{1 2 3 4}=v_{1 2 3 4}-v_{1 2 4 3}$  
of the kinetic energy  and the two--particle interaction, respectively,
are expressed in an arbitrary single--particle basis which comprises
for example quantum numbers for momentum, spin, isospin, and so on. 

The norm matrix (\ref{Norm}) is thus given by $\Norm_{1 1'}=\delta_{1 1'}$.
In this case, the effective Hamiltonian is the sum of the
single--particle energy and the full self--energy.  
The static part of the effective Hamiltonian, expressed by the double
commutator $\mean{\doco{a_{1}}{H}{a^{+}_{1'}}_{+}}$,  
yields the Hartree--Fock or mean--field Hamiltonian. We thus have
recovered an important piece of the single--particle Dyson equation. 
Working out the second part of the effective Hamiltonian in
eq.~(\ref{Heff}) yields the following $2p-1h$ Green's function:  
\begin{equation}
  \label{2p1hGF}
  \frac{1}{4} \, \sum \limits_{2 3 4 \atop 2' 3' 4'} 
  \bar{v}_{1 2 3 4} \; 
  \GFome{(a^{+}_{2} \,a_{4}     \,a_{3} )_{t}}%
        {(a^{+}_{3'}\,a^{+}_{4'}\,a_{2'})_{t'}}^{\rm irr}
  \; \bar{v}_{4' 3' 2' 1'} 
  \; \mbox{.}
\end{equation}
As mentioned above, in eq.~(\ref{2p1hGF}), all reducible contributions
to the effective Hamiltonian are removed and we obtain the usual
irreducible self--energy $\Sigma_{1 1'}(\omega)$ of the
single--particle Dyson equation by putting an index ``irreducible'' on
the $2p-1h$ Green's function in eq.~(\ref{EqOfMotion2}).   

The same scenario remains valid if we take for $A$ and $B$ more
complicated operators like e.g. the density operator
$a^{+}_{k}\,a_{k'}$. Again only the $(1p-1h)$ irreducible parts of  
the $2p-2h$ Green's function in eq.~(\ref{EqOfMotion2}) contribute to
the effective Hamiltonian.

%%%%%%%%%%%%%%%%%%%%%%%%%%%%%%%%%%%%%%%%%%%%%%%%%%
\subsection{Self--Consistent Random Phase Approximation}

As discussed above, the effective Hamiltonian splits, also in the
general case, in an instantaneous and an energy dependent part. 
The instantaneous part can be considered as a generalized
Hartree--Fock Hamiltonian (see below). 
Therefore, as a first approximation, one can try to solve this ``HF
problem'', neglecting the resonant part of the effective Hamiltonian.
As we will see later, this allows us to solve e.g.~the two--body 
problem on the level of a Schr\"odinger--like equation for a
single--frequency Green's function, in contrast to the Bethe--Salpeter
case where a three--frequency Green's function has to be determined.  
This means that we can introduce two--particle states \revised{with
shifted energies}.
Therefore, the consideration of the instantaneous part of the effective
Hamiltonian can be understood as a direct generalization of the common 
single--particle HF approximation to the more--body or cluster case.
In the past, this has been called 
{\em Cluster Mean Field} (CMF)\cite{Roepke} or 
{\em Self--Consistent Random Phase Approximation} 
(SCRPA)\cite{DukelskySchuck,DRS98}. In the remainder of this paper, we
will adopt SCRPA as shorthand for our approach, 
\revised{which, for the two-body case, can be connected to a
variational principle (see appendix~\ref{Variational})}.

In analogy to the single--particle Green's function we thus can define
a generalized $n$--body mean--field propagator by substituting the
instantaneous part of the effective Hamiltonian $\Hgmf_{A B}$ back
into the formal solution of the Dyson equation (\ref{DysonSolution}):
\begin{eqnarray}
  \label{docoGF}
  \GFome{A}{B}^{\GMF} 
  & = & \biggl\{ \omega - \Hgmf_{A B} \biggr\}^{-1} \, \Norm
  \nonumber \\
  & = & \biggl\{ \omega - 
  \mean{\doco{A}{H}{B}_{-\epsilon}}\,\Norm^{-1} \biggr\}^{-1} \, \Norm
\end{eqnarray}
Usually it is possible, as we will illustrate in an example below, to
close the system of equations in the following sense: For a full set
of operators $A$ and $B$ and for a two--particle interaction {\em all}
expectation values in $\Hgmf_{A B}$ can be determined
self--consistently from the Green's function (\ref{docoGF}) via the 
spectral theorem.    
For retarded Green's functions, the spectral theorem at temperature 
$\beta=1/(k_{B}T)$ reads\cite{Nolting}:
\comment{Nolting (Chap. 3.2.3: (3.151) and (3.152) with (3.173))}
\begin{eqnarray}
  \label{RetSpectralTheorem}
  \mean{ A\,B }^\corr & \equiv & 
  \mean{ A\,B } \; - \; \mean{A}  \, \mean{B} 
  \nonumber \\ 
  & = & -\frac{1}{\pi} 
  \int \limits_{-\infty}^{\infty} {\rm d} \omega \; 
  \frac{ \imag \GFome{A}{B}^{\rm ret} }{ 1-\epsilon\,{\rm e}^{-\beta\omega} } 
  \nonumber \\ 
  & \stackrel{T \to 0}
  { \unitlength=1ex
    \begin{picture}(5,1)(0,0) \put(0,0.5){\vector(1,0){5}} \end{picture} } & 
  -\frac{1}{\pi} 
  \int \limits_{0}^{\infty} {\rm d} \omega \; \imag \GFome{A}{B}^{\rm ret} 
  \nonumber \\ 
  \mean{ \comm{A}{B}_{-\epsilon} } 
  & = & -\frac{1}{\pi} 
  \int \limits_{-\infty}^{\infty} {\rm d} \omega \;
  \imag \GFome{A}{B}^{\rm ret} 
  \; \mbox{.}
\end{eqnarray}
\revised{The first of the eqs.~(\ref{RetSpectralTheorem}) is the well-known
{\em fluctuation-dissipation theorem}. The superscript ``$\corr$''
indicates that we calculate a correlated expectation value,
i.e. fluctuations of $\mean{A\,B}$ around their classical mean value
$\mean{A}\mean{B}$. This expectation value is known as {\em cumulant}
or, in the sense of Feynman graphs, {\em connected} average\cite{Kubo}.}
As indicated in eq.~(\ref{RetSpectralTheorem}), the Fermi function
in the spectral theorem for correlation functions 
$\mean{ A\,B }^\corr$ reduces to a step function as $T\to 0$.
The commutator expectation values, such as
$\mean{\comm{A}{B}_{-\epsilon}}$, depend only implicitly on
temperature, since the Green's function occurring in the spectral
theorem~(\ref{RetSpectralTheorem}) is temperature dependent.

From now on, we will restrict ourselves to the $T=0$ case, leaving the
finite temperature consideration to forthcoming investigations.

%%%%%%%%%%%%%%%%%%%%%%%%%%%%%%%%%%%%%%%%%%%%%%%%%%
\subsection{Energy Weighted Sum Rule}
\label{EnergySumRule}
\comment{Negele/Orland (Prob. 5.13; (5.139))}

\revised{The well-known energy weighted sum rule or $f$--sum rule
connects the imaginary part of the exact Green's function
$\GFome{A}{B}$ to the expectation value of the double commutator 
$\mean{\doco{A}{H}{B}_{-\epsilon}}$. 
Sometimes it is possible to choose operators $A$ and $B$ such that 
the double commutator $\mean{\doco{A}{H}{B}_{-\epsilon}}$ can be 
evaluated analytically. 
In this case, the sum rule may be used as a rigorous check for any
approximative Green's function.

Let us recall briefly the derivation of the sum rule. 
We can compute} $\mean{\doco{A}{H}{B}_{-\epsilon}}$ using the spectral
theorem (\ref{RetSpectralTheorem}) for the higher Green's function
$\GFome{\comm{A}{H}}{B}$, 
\begin{equation}
  \label{ESumRuleST}
  \mean{\doco{A}{H}{B}_{-\epsilon}} = 
  -\frac{1}{\pi} \int \limits_{-\infty}^{\infty} {\rm d} \omega \; 
  \imag \GFome{\comm{A}{H}}{B}^{\rm ret} 
  \; \mbox{.}
\end{equation}
Inserting the equation of motion (\ref{EqOfMotionFreq}) on the rhs.
and supposing the norm matrix to be real, we find the
well--known energy weighted sum rule\cite{Orland}:
\begin{equation}
  \label{ESumRule}
  \mean{\doco{A}{H}{B}_{-\epsilon}} \;=\; -\frac{1}{\pi} \int
  \limits_{-\infty}^{\infty} {\rm d} \omega \; \omega \; \imag
  \GFome{A}{B}^{\rm ret}
\end{equation}
 
We will now show that the sum rule (\ref{ESumRule}) also holds for
the SCRPA Green's function $\GFome{A}{B}^{\GMF}$.
From eq.~(\ref{docoGF}) we see that $\GFome{A}{B}^{\GMF}$ satisfies
the equation of motion 
\begin{equation}
  \label{GMFEqOfMotion}
  \omega \, \GFome{A}{B}^{\GMF} = \Norm \;+\;
  \mean{\doco{A}{H}{B}_{-\epsilon}} \, \Norm^{-1} \,
  \GFome{A}{B}^{\GMF}
\end{equation}
rather than (\ref{EqOfMotionFreq}).  Again, the norm matrix and the
double commutator on the rhs. are real and $\omega$ independent. 
Inserting (\ref{GMFEqOfMotion}) into (\ref{ESumRule}) and
applying the spectral theorem (\ref{RetSpectralTheorem}) yields the
norm matrix on the rhs.~(which is cancelled by $\Norm^{-1}$).

From the above we see that, because of the double commutator structure
of the effective Hamiltonian $\Hgmf$, the SCRPA Green's function
fulfills the energy weighted sum rule~(\ref{ESumRule}) practically by
\revised{construction}. 

%%%%%%%%%%%%%%%%%%%%%%%%%%%%%%%%%%%%%%%%%%%%%%%%%%
\subsection{Particle--hole propagator}
\label{sec:phGFGen}

As a concrete example, we will derive the SCRPA expression for a
particle--hole Green's function
$\GFome{a^{+}_{p}\,a_{k}}{a^{+}_{k'}\,a_{p'}}^{\rm ret}$ 
in a fermionic system with general two--body interactions as
described by the Hamiltonian (\ref{genHamiltonian}).
Since the operators defining the Green's function are pairs of
fermions, we will use the commutator Green's function
$(\epsilon=+1)$. 

For a homogeneous system, the kinetic energy is diagonal in momentum
space,  
\begin{equation}
  t_{k k'} \; = \; \delta_{k k'} \, \varepsilon_{k} 
  \; \mbox{,}
\end{equation}
with $k$ standing for momentum and other quantum numbers such as
spin. 

The norm matrix is also diagonal,
\begin{eqnarray}
  \label{phNormGen}
  \Norm_{k p k' p'} \; & \equiv & \; 
  \mean{ \comm{a^{+}_{p}\,a_{k}}{a^{+}_{k'}\,a_{p'}}_{-} }
  \nonumber \\
  & = & \; \delta_{k k'} \, \delta_{p p'} \, \left( n_{p}\,-\,n_{k} \right)
  \qquad \mbox{,}
\end{eqnarray}
where 
\begin{equation}
  \label{OccGen}
  n_{k}\,=\,\mean{a^{+}_{k}\,a_{k}}
\end{equation}
stands for the occupation numbers.
The effective SCRPA Hamiltonian, introduced in eq.~(\ref{Heff}), can
be worked out for the Hamiltonian (\ref{genHamiltonian}). 
Using summation convention, this yields
\begin{eqnarray}
  \label{phHgmfGen}
  \Hgmf_{k p k' p'} \; & \equiv &  
  \; \mean{ \doco{a^{+}_{p}\,a_{k}}{H}{a^{+}_{k'}\,a_{p'}}_{-} }
  \,\left( n_{p'}\,-\,n_{k'} \right)^{-1} 
  \nonumber \\
  & = & 
  \; \delta_{k k'} \, \delta_{p p'} \; 
  \left( \epsilon_{k} - \epsilon_{p} \right)
  \;+\; \left( n_{p}\,-\,n_{k} \right) \, \bar{v}_{p' k k' p}
  \nonumber \\
  & & + \biggl[ \;
  \frac{1}{2}  \, \delta_{p p'} \, \bar{v}_{k q_{2} q_{3} q_{4}} 
  \, \mean{ a^{+}_{k'} a^{+}_{q_{2}} a_{q_{3}} a_{q_{4}} }^\corr
  \;
  +\frac{1}{2} \, \delta_{k k'} \, \bar{v}_{q_{1} q_{2} q_{3} p} 
  \, \mean{ a^{+}_{q_{1}} a^{+}_{q_{2}} a_{q_{3}} a_{p'} }^\corr
  \nonumber \\
  & & \qquad
  +\frac{1}{2} \, \bar{v}_{p' k q_{3} q_{4}} 
  \, \mean{ a^{+}_{k'} a^{+}_{p} a_{q_{3}} a_{q_{4}} }^\corr
  \;
  +\frac{1}{2} \, \bar{v}_{q_{1} q_{2} p k'} 
  \, \mean{ a^{+}_{q_{1}} a^{+}_{q_{2}} a_{k} a_{p'} }^\corr
  \nonumber \\
  & & \qquad
  - \bar{v}_{k q_{2} q_{3} k'} 
  \, \mean{ a^{+}_{p} a^{+}_{q_{2}} a_{q_{3}} a_{p'} }^\corr
  \;
  - \bar{v}_{p' q_{2} q_{3} p} 
  \, \mean{ a^{+}_{k'} a^{+}_{q_{2}} a_{q_{3}} a_{k} }^\corr
  \biggr] \, \left( n_{p'}\,-\,n_{k'} \right)^{-1}
  \; \mbox{,}
\end{eqnarray}
where $\epsilon_{k}$ denote the Hartree--Fock corrected
single--particle energies,  
\begin{equation}
  \label{phHFepsGen}
  \epsilon_{k} \; = \; \varepsilon_{k} \,+\, \bar{v}_{k q k q} \, n_{q}  
  \; \mbox{.}
\end{equation}
As we will see in section~\ref{sec:phRPAGen}, the second term in
eq.~(\ref{phHgmfGen}), 
$\left( n_{p}\,-\,n_{k} \right) \, \bar{v}_{p' k k' p}$, will lead us
to a RPA--like theory. 
\revised{We will use the term RPA in a slightly broader sense than
usual and already account for the exchange term of the interaction,
see eq.~(\ref{phHFepsGen}).
The term in brackets, in contrast, contains exclusively correlated
expectation values (i.e. cumulant averages\cite{Kubo}),}
\begin{equation}
  \label{axaxaa_corr}
  \mean{ a^{+}_{q_{1}} a^{+}_{q_{2}} a_{q_{3}} a_{q_{4}} }^\corr
  \equiv   
  \mean{ a^{+}_{q_{1}} a^{+}_{q_{2}} a_{q_{3}} a_{q_{4}} } -
  \left[ 
    \mean{ a^{+}_{q_{1}} a_{q_{4}} } \mean{ a^{+}_{q_{2}} a_{q_{3}} }
    -\mean{ a^{+}_{q_{1}} a_{q_{3}} } \mean{ a^{+}_{q_{2}} a_{q_{4}} }
  \right]
\end{equation}
not taken into account by usual RPA--like theories. 

The SCRPA Green's function 
\begin{equation}
  \label{phGFGen}
  \phGF_{k p k' p'}^{\GMF}(\omega) \, = \, 
  \GFome{a^{+}_{p}\,a_{k}}{a^{+}_{k'}\,a_{p'}}^{\GMF}
\end{equation}
defined in eq.~(\ref{docoGF}), can now be obtained by inverting the 
matrix $\left[\omega-\Hgmf_{k p k' p'}\right]$.
Once it is determined, all elements of the effective
Hamiltonian~(\ref{phHgmfGen}) and the norm matrix~(\ref{phNormGen}) 
can be calculated via the spectral theorem (\ref{RetSpectralTheorem}).
Moreover, it is possible to derive an explicit expression for the
occupation numbers $n_{p}$ by summing the diagonal elements of the
norm matrix, $n_{p}-n_{k}$, over the index $k$. 

With the commutator spectral theorem (\ref{RetSpectralTheorem}) and
the $k$--space volume 
\begin{equation}
  \label{kVol}
  {\mathcal V} = \sum\limits_{k} 1
  \; \mbox{,}
\end{equation}
we get for the occupation numbers
\begin{equation}
  \label{OccSpectralTheoremGen}
  n_{p}  \, = \, \frac{1}{\mathcal V} \sum\limits_{k} n_{k} 
  - \frac{1}{\pi{\mathcal V}}
  \sum\limits_{k} 
  \int \limits_{-\infty}^{\infty} {\rm d} \omega \;
  \imag \GFome{a^{+}_{p}\,a_{_{k}}}{a^{+}_{_{k}}\,a_{p}}^{\rm ret}
  \; \mbox{.}
\end{equation}
In a continuous system, it is necessary to introduce a cutoff in
order to keep the $k$--space volume ${\mathcal V}$ finite. 
In lattice systems, as will be seen in section~\ref{HubbardModel},
${\mathcal V}$ is finite, since the $k$-sum is restricted to the
first Brillouin zone. 

The correlation functions 
$\mean{ a^{+}_{q_{1}} a^{+}_{q_{2}} a_{q_{3}} a_{q_{4}} }^\corr$
in eq.~(\ref{phHgmfGen}) are connected to those, which are
accessible via the spectral theorem (\ref{RetSpectralTheorem}),
\begin{equation}
  \label{axaaxa_corr}
  \mean{ (a^{+}_{q_{1}} a_{q_{4}}) \cdot (a^{+}_{q_{2}} a_{q_{3}}) }^\corr
  \equiv 
  \mean{ a^{+}_{q_{1}} a_{q_{4}} a^{+}_{q_{2}} a_{q_{3}} }
  - \mean{ a^{+}_{q_{1}} a_{q_{4}} } \mean{ a^{+}_{q_{2}} a_{q_{3}} }
  \; \mbox{.}
\end{equation}
This yields
\begin{equation}
  \label{axaxaaSpectralTheoremGen}
  \mean{ a^{+}_{q_{1}} a^{+}_{q_{2}} a_{q_{3}} a_{q_{4}} }^\corr
  = - \frac{1}{\pi} 
  \int \limits_{0}^{\infty} {\rm d} \omega \;
  \imag \GFome{a^{+}_{q_{1}}\,a_{q_{4}}}{a^{+}_{q_{2}}\,a_{q_{3}}}^{\rm ret}
  \,-\,\mean{ a^{+}_{q_{1}} a_{q_{3}} } 
  \left[ \delta_{q_{2} q_{4}} - \mean{ a^{+}_{q_{2}} a_{q_{4}} } \right]
  \; \mbox{,}
\end{equation}
where for a homogeneous system the last term on the rhs.~is related to
the occupation numbers by momentum conservation.

The system of equations is now closed and can be iterated to
self--consistency. We therefore start with an assumption for the
expectation values in $\Hgmf$ and $\Norm$. 
The SCRPA Green's function, obtained by matrix inversion, then allows us
to calculate new values for the elements of the effective Hamiltonian
and the norm matrix by applying the spectral
eqs.~(\ref{OccSpectralTheoremGen}) and (\ref{axaxaaSpectralTheoremGen}). 
The new $\Hgmf$ and $\Norm$ lead us to the next approximation for the
Green's function and so forth.  

%%%%%%%%%%%%%%%%%%%%%%%%%%%%%%%%%%%%%%%%%%%%%%%%%%
\subsection{RPA and Self--Consistent RPA}
\label{sec:phRPAGen}

We now analyze the contributions to the SCRPA Green's function 
by rewriting eq.~(\ref{DysonEq}) as an integral equation:
\comment{linearized Bethe--Salpeter equation, see Ring/Schuck (8.131)} 
\begin{equation}
  \label{phBetheGen}
  \phGF_{k p k' p'}^{\GMF}(\omega)
  \; = \;
  \phGF_{k p k' p'}^{0}(\omega) \,
  + \, \sum\limits_{ k_{1} p_{1} \atop k_{2} p_{2} }
  \phGF_{k p k_{1} p_{1}}^{0}(\omega) \, 
  \Kgmf_{k_{1} p_{1} k_{2} p_{2}} \, 
  \phGF_{k_{2} p_{2} k' p'}^{\GMF}(\omega)
  \; \mbox{,}
\end{equation}
where $\phGF_{k p k' p'}^{0}(\omega)$ has the structure of a free
particle--hole Green's function,  
\begin{equation}
  \label{phG0Gen}
  \phGF_{k p k' p'}^{0}(\omega) 
  \; = \; \delta_{k k'} \, \delta_{p p'} \; 
  \frac{ n_{p}\,-\,n_{k} }
  {\omega \,-\, \left(\epsilon_{k}-\epsilon_{p}\right) \,+i0^{+}} 
  \; \mbox{.}
\end{equation}
The integral kernel $\Kgmf_{k_{1} p_{1} k_{2} p_{2}}$ represents the
interaction occurring in the effective Hamiltonian~(\ref{phHgmfGen}),
\begin{equation}
  \label{phKgmfGen}
  \Kgmf_{k p k' p'}
  \; = \;
  \left( n_{p}\,-\,n_{k} \right)^{-1} \;
  \left[
    \, 
    \Hgmf_{k p k' p'} 
    \,-\, \delta_{k k'} \, \delta_{p p'}
    \left(\epsilon_{k}-\epsilon_{p}\right) 
    \,
  \right]
  \; \mbox{.}
\end{equation}
Since the kernel $\Kgmf$ splits up into a RPA--like part
\begin{equation}
  \label{phKrpaGen}
  \Krpa_{k p k' p'} \, \equiv \,  \bar{v}_{p' k k' p}
\end{equation}
and a remainder, $\Kern^\corr$, which only contains
correlated two--body densities, it is convenient to rewrite
eq.~(\ref{phBetheGen}) in a different way (using matrix notation)
\begin{eqnarray}
  \label{phBethe2Gen}
  \phGF^{\GMF} & = & \phGF^{\RPA} \, + \, 
  \phGF^{\RPA}\,\Kern^\corr\,\phGF^{\GMF} 
  \\
  \label{phBethe1Gen}
  \phGF^{\RPA} & = & \phGF^{0} \, + \, 
  \phGF^{0}\,\Krpa\,\phGF^{\RPA} 
\end{eqnarray}
At this point, we emphasize that the eq.~(\ref{phBethe1Gen}) has
exactly the same structure as the usual RPA equation. 
However, in our theory the occupation numbers can and will, even at
zero temperature, be different from the Hartree-Fock values
\begin{equation}
  \label{OccHFGen}
  n_{k}^{\rm HF} \, = \, \Theta(E_{\rm F}-\epsilon_{k}) 
  \; \mbox{.}
\end{equation}

In the following, we will consider the eq.~(\ref{phBethe1Gen}) as
generic for the RPA whatever the occupation numbers will be. 
We will label it {\em pure RPA} if the occupation numbers are fixed to
their HF values (\ref{OccHFGen}).
In contrast, a theory in which the occupation numbers are determined 
self--consistently from the RPA Green's function or contain
correlations in any other way will be called 
{\em renormalized RPA}\cite{Rowe}.  

In SCRPA, eq.~(\ref{phBethe2Gen}) is coupled to the RPA
eq.~(\ref{phBethe1Gen}), upgrading the (renormalized) RPA to the
Self--Consistent RPA. 
Comparing eq.~(\ref{phBetheGen}) to eq.~(\ref{phBethe1Gen}) clearly
shows  that the RPA structure of the solution is preserved when
passing over to self--consistent RPA:
\begin{eqnarray}
  \label{phSolutionGen}
  \phGF^{\GMF} (\omega)
  & = & 
  \left[ 1 \,-\,\phGF^{0}(\omega)\,\Kern^{\GMF} \right]^{-1}\,
  \phGF^{0}(\omega)
  \nonumber \\
  \phGF^{\RPA} (\omega)
  & = & 
  \left[ 1 \,-\,\phGF^{0}(\omega)\,\Kern^{\RPA} \right]^{-1}\,
  \phGF^{0}(\omega)
\end{eqnarray}

\revised{The improvement contained in the integral kernel
$\Kern^\corr$ can be interpreted by the following connected diagrams:}
\begin{equation}
  \label{fig:phKcorrGen}
  \leavevmode
  \setlength{\Units}{1cm}
  \epsfig{file=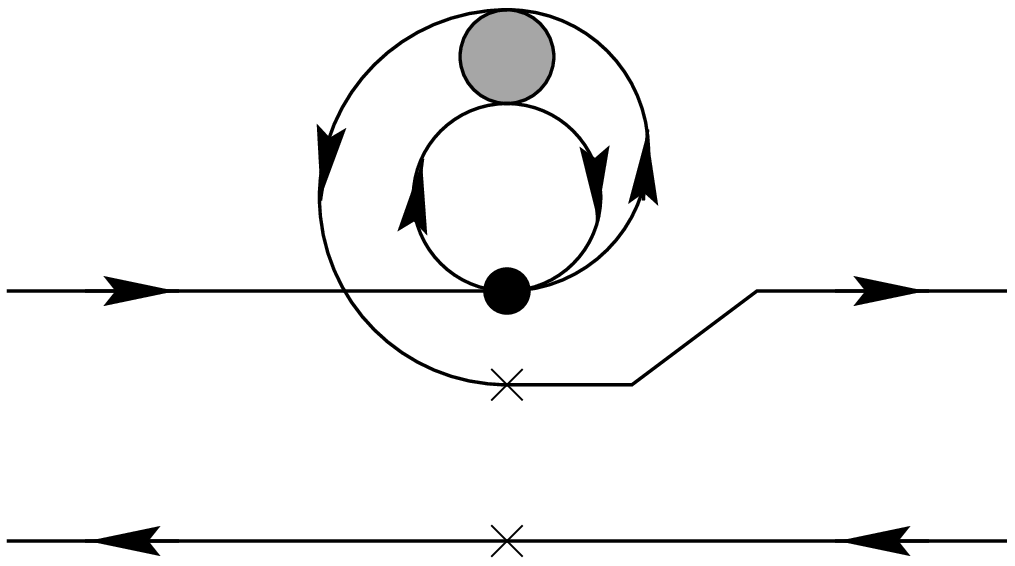,width=4\Units}
  \hspace{2\Units}
  \epsfig{file=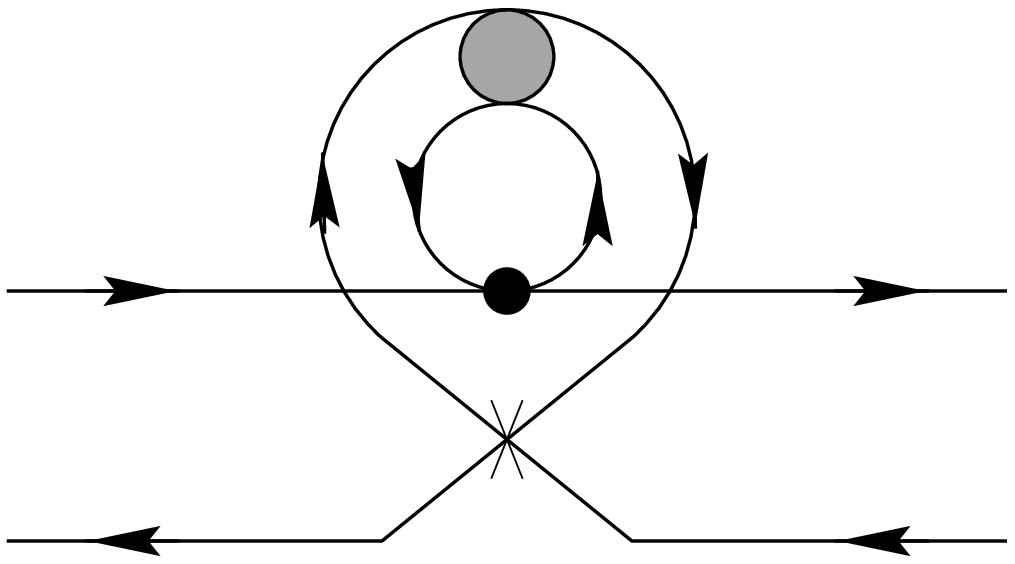,width=4\Units}
\end{equation}
The hatched circle represents a correlated particle--hole propagator
and the dot stands for the antisymmetrized interaction $\bar{v}$. The
crosses represent Kronecker symbols in momentum and other quantum
numbers, and $\delta$--functions in time. 
We see that the first graph in eq.~(\ref{fig:phKcorrGen}) corresponds
to a coupling of the single--particle motion to the density
fluctuations, i.e. a self--energy correction, whereas the second graph
describes an induced (screened) interaction. Of course analogous
graphs exist where the interaction is attached to the hole
line. Again, we want to emphasize that {\em all} terms in
eq.~(\ref{fig:phKcorrGen}) are instantaneous. 
We obtain the second--order contributions in replacing the hatched circle
in eq.~(\ref{fig:phKcorrGen}) by an interaction dot:
\begin{equation}
  \label{fig:phKcorr2Gen}
  \leavevmode
  \setlength{\Units}{1cm}
  \epsfig{file=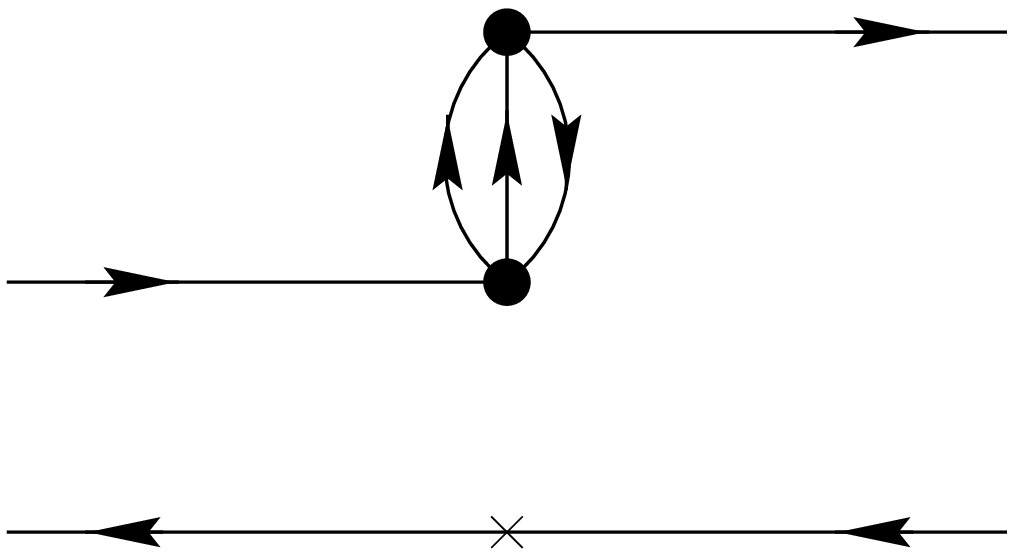,width=4\Units}
  \hspace{2\Units}
  \epsfig{file=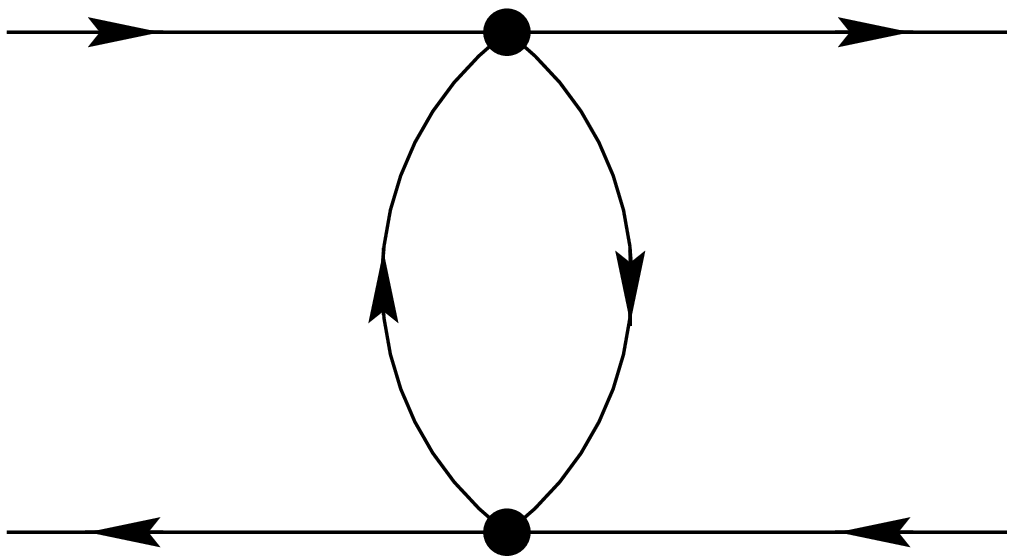,width=4\Units}
\end{equation}
Cutting the graphs in eq.~(\ref{fig:phKcorr2Gen}) between the two
interactions occurring one an infinitesimal time after the other
illustrates the instantaneous coupling of the $1p1h$ and $2p2h$
spaces. 
Solving the eqs.~(\ref{phBethe1Gen}) and (\ref{phBethe2Gen}) 
self--consistently thus constitutes a partial resummation of the 
interaction to a very high order. 

Just as the Hartree--Fock self--energy for a single particle can be
constructed from a two--particle interaction term by attaching an
outgoing to an incoming line, viz.
\begin{equation}
  \label{fig:selfenergy}
  \leavevmode
  \setlength{\Units}{1cm}
  \epsfig{file=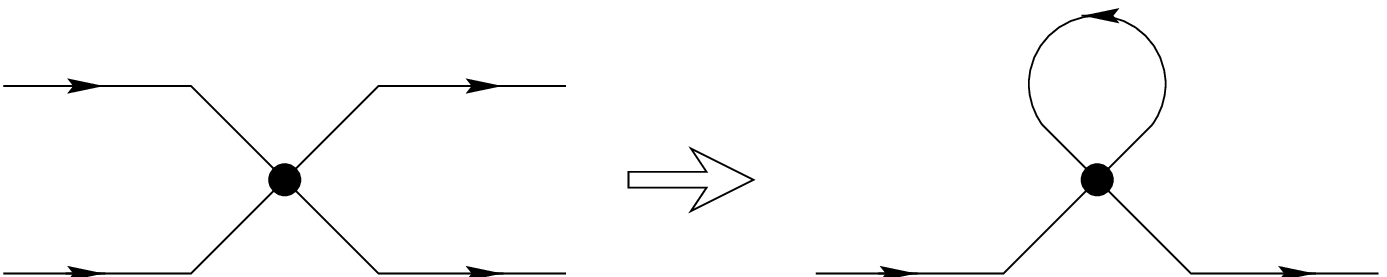,width=11\Units}
  \; \mbox{,}
\end{equation}
we may interpret eq.~(\ref{fig:phKcorrGen}) as the Hartree--Fock
field for density--density fluctuations 
(this point of view has actually been adopted in ref.\cite{Roepke}). 
In analogy, we can reconstruct the loops in eq.~(\ref{fig:phKcorrGen})
by closing two density fluctuation lines in the following first--order
terms for the interaction (which can be obtained from perturbation theory):
\begin{equation}
  \label{fig:2p2hinteraction}
  \leavevmode
  \setlength{\Units}{1cm}
  \epsfig{file=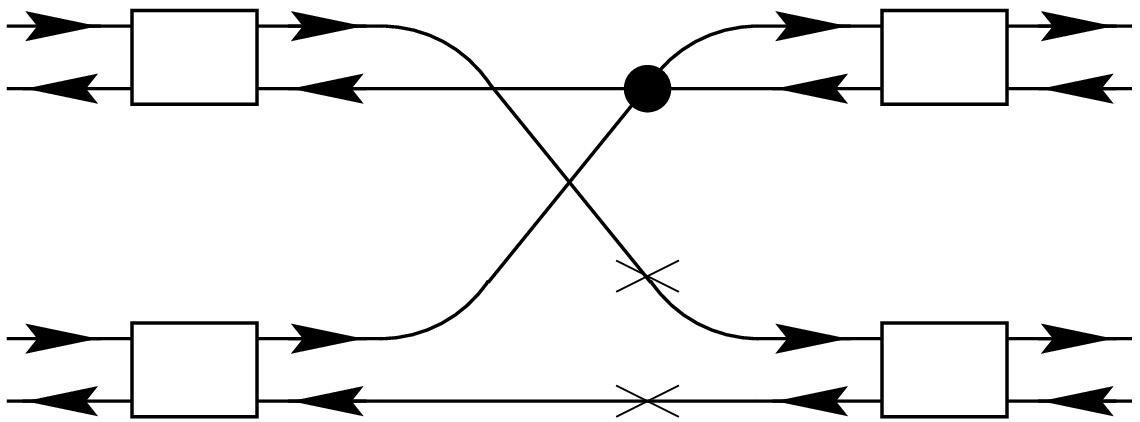,width=4.5\Units}
  \hspace{2\Units}
  \epsfig{file=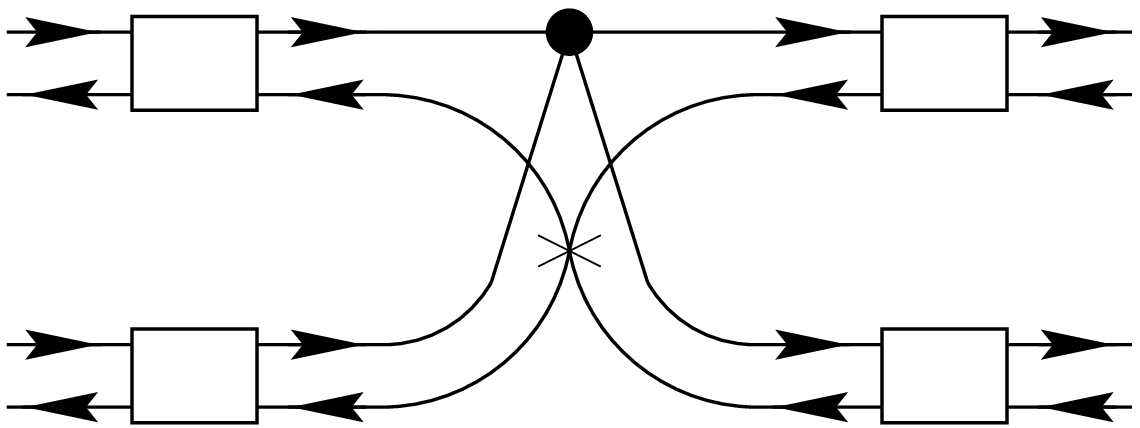,width=4.5\Units}
\end{equation}
Considering all exchange terms it is possible to reconstruct exactly
the effective Hamiltonian~(\ref{phHgmfGen}) which therefore represents
the mean--field Hamiltonian of a gas of quantal fluctuations present in
any many fermion system.
 
% %%%%%%%%%%%%%%%%%%%%%%%%%%%%%%%%%%%%%%%%%%%%%%%%%%%%%%%%%%%%
% Hubbard Model
% %%%%%%%%%%%%%%%%%%%%%%%%%%%%%%%%%%%%%%%%%%%%%%%%%%%%%%%%%%%%
\section{Application to the Hubbard Model}
\label{HubbardModel}

In this section, we will apply the SCRPA, developed in
section~\ref{GMFtheory}, to density--density correlation functions in 
the Hubbard model. 

The single--band Hubbard Hamiltonian describes electrons hopping on a
lattice with an on--site interaction $U$\cite{Hubbard}:  
\comment{Nolting (4.1)}
\begin{equation}
  \label{HubbardCoordSpace}
  H \,=\, \sum \limits_{i j \sigma} t_{i j}\,a^{+}_{i \sigma} a_{j \sigma}
  \,+\, U \sum \limits_{i} \hat{n}_{i \uparrow} \hat{n}_{i \downarrow}
\end{equation}
where $a^{+}_{i \sigma}$ and $a_{i \sigma}$ denote the creation and
destruction operators for an electron with spin $\sigma$ on site $i$,
respectively. 
The occupation number operator on site $i$ is defined as 
$\hat{n}_{i\sigma}=a^{+}_{i\sigma}a_{i\sigma}$. 
In the following, we will restrict ourselves to nearest neighbour hopping 
\begin{equation}
  \label{NearestNeighbour}
  t_{i j} = -t \left( \delta_{j,i+1} + \delta_{j,i-1} \right)
  \; \mbox{,}
\end{equation}
repulsive interactions $U>0$, and zero temperature.
We will work with $\hbar=1$, set the lattice spacing to unity ($a=1$)
and measure energies in units of the hopping integral ($t=1$).

After Fourier transformation, the Hamiltonian reads
\begin{equation} 
  \label{Hubbard}
  H \,=\, 
  \sum \limits_{\vk \sigma} \varepsilon_{\vk}\,
  a^{+}_{\vk \sigma} a_{\vk \sigma}
  \,+\, 
  \frac{U}{N} \sum \limits_{\vk \vp \vq}
  a^{+}_{\vk \uparrow  } a_{\vk+\vq \uparrow}
  a^{+}_{\vp \downarrow} a_{\vp-\vq \downarrow}
  \; \mbox{.}
\end{equation}
Notice that $N$ is the number of sites (ions, not electrons) and all
momentum sums run over the first Brillouin zone unless indicated
differently. 
The single--particle dispersion relation in the hyper--cubic lattice is
given by the Fourier transform of the nearest neighbour hopping matrix
element~(\ref{NearestNeighbour}),
\begin{equation}
  \label{dispersion}
  \varepsilon_{\vk} \,=\, -2 \sum\limits_{i=1}^{d} \cos k_{i}
  \; \mbox{,}
\end{equation}
where $d$ denotes the dimension.

%%%%%%%%%%%%%%%%%%%%%%%%%%%%%%%%%%%%%%%%%%%%%%%%%%
\subsection{Charge-- and longitudinal spin--density correlations}
\label{phGFcharge}

In the following, we will examine the behaviour of charge-- and
spin--density fluctuations in the Hubbard model. 
Therefore, we will introduce \revised{the density operator}
\comment{Nolting (4.128)}
\begin{equation}
  \label{RhoSigma}
  \rho_{\vq\sigma} \, = \, \sum_{\vk} a^{+}_{\vk\sigma} a_{\vk+\vq\sigma}
  \; \mbox{,}
\end{equation}
which is the Fourier transform of the Wannier number operator
$\hat{n}_{i\sigma}$. It will be used to describe charge and
longitudinal spin fluctuations.  
Summing over the spins gives rise to the charge susceptibility
\comment{Nolting chap. 4.2.2}
\begin{equation}
  \label{DefChargeSusc}
  \ChiCh (\vq,\omega) \, = \, \sfrac{1}{N}
  \GFome{ ( \rho_{\vq\uparrow}    +\rho_{\vq\downarrow} ) }
        { ( \rho_{\vq\uparrow}^{+}+\rho_{\vq\downarrow}^{+} ) 
        }^{\rm ret}
  \; \mbox{.}
\end{equation}
As from now on we will use only retarded Green's functions, we will
omit the superscript ``${\rm ret}$'' on correlation functions. 

The $z$--component of the spin on site $i$ can be expressed as the
difference between the number of $\uparrow$--spins and
$\downarrow$--spins on that site, or, after Fourier transformation,
\comment{Nolting (Afg. 4.1.6)}
\begin{equation}
  \label{Sz}
  S^{z}_{\vq} \, = \, 
  \sfrac{1}{2}\,(\rho_{\vq\uparrow}-\rho_{\vq\downarrow} ) 
  \; \mbox{.}
\end{equation}
Correlations between the $z$--components of the spins are described by
the longitudinal spin susceptibility, 
\comment{Nolting (3.69)}
\begin{eqnarray}
  \label{DefLongSpinSusc}
  \ChiSp (\vq,\omega) 
  & = & \sfrac{1}{N} \GFome{ S^{z}_{\vq} }{ {S^{z}_{\vq}}^{+} }
  \nonumber \\ 
  & = & \sfrac{1}{N}
  \GFome{ \sfrac{1}{2} \, ( \rho_{\vq\uparrow}    -\rho_{\vq\downarrow} )     }
        { \sfrac{1}{2} \, ( \rho_{\vq\uparrow}^{+}-\rho_{\vq\downarrow} ^{+}) }
  \; \mbox{.}
\end{eqnarray}
  
\revised{We will now examine the charge and longitudinal spin
susceptibilities, $\ChiCh(\vq,\omega)$ and $\ChiSp(\vq,\omega)$ in SCRPA.}  
Therefore, we introduce the particle--hole Green's function  
\begin{equation}
  \label{phGF}
  \phGF_{\vk\sigma\,\vp\sigma'}(\vq,\omega) \,\equiv\, 
  \GFome{a^{+}_{\vk\sigma}     \,a_{\vk+\vq\sigma}}
  {      a^{+}_{\vp+\vq\sigma'}\,a_{\vp\sigma'}   }
  \; \mbox{.}
\end{equation}
In contrast to the general particle--hole propagator defined in
section~\ref{sec:phGFGen}, we will now account for momentum
conservation right from the beginning.
As the derivation of the SCRPA propagator is completely analogous to
section~\ref{sec:phGFGen}, we will only state the results.

The norm matrix (\ref{phNormGen}) is given by 
\begin{equation}
  \label{phNorm}
  \Norm_{\vk\sigma\,\vp\sigma'}(\vq) \,=\, 
  \delta_{\vk \vp} \delta_{\sigma \sigma'} \,
  \left( n_{\vk\sigma} - n_{\vk+\vq\sigma} \right)
  \; \mbox{,}
\end{equation}
where $n_{\vk\sigma}$ denotes the occupation number for the Bloch
state $\vk$ with spin $\sigma$ defined in analogy to eq.~(\ref{OccGen}). 
The effective Hamiltonian, which was defined for a general two--body
interaction in eq.~(\ref{phHgmfGen}), 
reads for the Hubbard Hamiltonian
\begin{eqnarray}
  \label{phHgmf}
  \Hgmf_{\vk\sigma\,\vp\sigma'}(\vq) & = &
  \delta_{\vk \vp} \delta_{\sigma \sigma'} \, 
  \left[ \varepsilon_{\vk+\vq} - \varepsilon_{\vk} \right]
  \,+\, \delta_{\sigma,-\sigma'} \, 
  \left( n_{\vk\sigma} - n_{\vk+\vq\sigma} \right) \, \frac{U}{N} 
  \nonumber \\
  & & +\,\biggl[\;
  -\delta_{\vk \vp} \delta_{\sigma \sigma'} \, \frac{U}{N} 
  \sum\limits_{\vq'}\mean{ 
    \left(  a^{+}_{\vk+\vq-\vq'\sigma} a_{\vk     \sigma} 
      \,+\, a^{+}_{\vk+\vq     \sigma} a_{\vk+\vq'\sigma} \right)
    \, \rho_{\vq-\vq',-\sigma} }^\corr
  \nonumber \\
  & & \qquad
  +\delta_{\sigma \sigma'} \, \frac{U}{N} 
  \mean{  a^{+}_{\vk    \sigma} a_{\vp    \sigma} \,\rho_{\vk-\vp,-\sigma} 
    \,+\, a^{+}_{\vp+\vq\sigma} a_{\vk+\vq\sigma} \,\rho_{\vp-\vk,-\sigma} 
    }^\corr
  \nonumber \\
  & & \qquad
  +\delta_{\sigma,-\sigma'} \, \frac{U}{N} 
  \sum\limits_{\vq'} 
  \Big\langle                                 % left \mean
    \left(  a^{+}_{\vk     \sigma} a_{\vk+\vq-\vq'\sigma} 
      -     a^{+}_{\vk+\vq'\sigma} a_{\vk+\vq     \sigma} \right) 
    \nonumber \\
    & & \qquad
    \cdot
    \left(  a^{+}_{\vp+\vq-\vq',-\sigma} a_{\vp     ,-\sigma} 
      -     a^{+}_{\vp+\vq     ,-\sigma} a_{\vp+\vq',-\sigma} \right) 
    \Big\rangle^\corr\,                       % right \mean
  \biggr]
%  \nonumber \\
%  & & 
  \cdot
  \left( n_{\vp\sigma'} - n_{\vp+\vq\sigma'} \right)^{-1} 
  \; \mbox{,} 
\end{eqnarray}
with $\rho_{\vq\sigma}$ being the density operator introduced in
eq.~(\ref{RhoSigma}). The Hartree--Fock corrections to the
single--particle energies cancel because of the on--site (and thus
momentum independent) interaction.    

The spectral theorem yields for the occupation numbers 
(see eq.~(\ref{OccSpectralTheoremGen}))
\begin{equation}
  \label{OccSpectralTheorem}
  n_{\vk\sigma} \, = \, \mean{ n_{\sigma} } \,
  - \frac{1}{\pi N} \sum \limits_{\vq}
  \int \limits_{-\infty}^{\infty} {\rm d} \omega \;
  \imag \phGF_{\vk\sigma\,\vk\sigma}(\vq,\omega)
  \; \mbox{,}
\end{equation}
where $\mean{ n_{\sigma} }$ denotes the number of $\sigma$--electrons
per site. 
In the paramagnetic phase, spin--broken expectation values like
$\mean{ a^{+}_{\vk\uparrow} a_{\vk\downarrow} }$ vanish, and we obtain
from equation (\ref{axaxaaSpectralTheoremGen}) for the correlation
functions occurring in the effective Hamiltonian (\ref{phHgmf}):
\begin{equation}
  \label{axaaxaSpectralTheorem}
  \mean{a^{+}_{\vk    \uparrow  } a_{\vk+\vq\uparrow} \,
        a^{+}_{\vp+\vq\downarrow} a_{\vp    \downarrow} }^\corr
  \, = \, -\frac{1}{\pi} 
  \int \limits_{0}^{\infty} {\rm d} \omega \;
  \imag \phGF_{\vk\uparrow\,\vp\downarrow}(\vq,\omega)
\end{equation}

As in section~\ref{sec:phGFGen}, the system of equations is now closed
and can be iterated to self--consistency, 
since, on one hand, we are able to calculate all elements of the
effective Hamiltonian $\Hgmf_{\vk\sigma\,\vp\sigma'}(\vq)$ from the
particle--hole propagator, and, on the other hand, the latter by a
matrix inversion for every $\vq$ and $\omega$: 
\begin{equation}
  \label{phGFsolution}
  \phGF^{\GMF}_{\vk\sigma\,\vp\sigma'}(\vq,\omega) \, = \,
  \left[ \omega-\Hgmf_{\vk\sigma\,\vp\sigma'}(\vq) \right]^{-1}
  \, \left( n_{\vp\sigma'} - n_{\vp+\vq\sigma'} \right)
\end{equation}
We will see in section~\ref{phGFspin} that the corresponding system
of equations for spin--density correlations is not closed onto itself,
but couples back to the charge--density correlations.

As shown in section~\ref{sec:phGFGen} 
(see eqs.~(\ref{phBethe2Gen},\ref{phBethe1Gen})), it is advised to first
calculate the RPA particle--hole propagator before solving the full
SCRPA problem (\ref{phGFsolution}).  
For the Hubbard interaction, the RPA kernel $\Krpa$ defined in
eq.~(\ref{phKrpaGen}) is nothing but the interaction per site 
\begin{equation}
  \label{phKrpa}
  \Krpa_{\vk\sigma\,\vp\sigma'}(\vq) \, = \,  
  \delta_{\sigma, -\sigma'} \frac{U}{N}
  \; \mbox{.}
\end{equation}
Eq.~(\ref{phBethe1Gen}) can therefore be written as an integral
equation coupling $\phGF^{\RPA}_{\vk\sigma\,\vp\sigma}(\vq,\omega)$
and $\phGF^{\RPA}_{\vk -\sigma\,\vp\sigma}(\vq,\omega)$:
\begin{equation}
  \label{phBetherpa}
    \phGF^{\RPA}_{\vk\sigma\,\vp\sigma'}(\vq,\omega)
    \; = \;
    \delta_{\vk \vp} \delta_{\sigma \sigma'}\,\phGF^{0}_{\vk\sigma}(\vq,\omega)
    \; + \;
    \phGF^{0}_{\vk\sigma}(\vq,\omega) \, \frac{U}{N} \, 
    \sum\limits_{\vk'} \phGF^{\RPA}_{\vk'-\sigma\,\vp\sigma'}(\vq,\omega)
  \; \mbox{,}
\end{equation}
where $\phGF^{0}_{\vk\sigma}(\vq,\omega)$ defines the renormalized free
particle--hole propagator 
\begin{equation}
  \label{phGFzero}
  \phGF^{0}_{\vk\sigma}(\vq,\omega) \, = \,
  \frac{\left( n_{\vk\sigma} - n_{\vk+\vq\sigma} \right)}
  {\omega\,-\,\left[\varepsilon_{\vk+\vq}-\varepsilon_{\vk}\right]\,+\,i0^{+}}
  \; \mbox{.}
\end{equation}
In terms of Feynman graphs, this means substituting the SCRPA kernel
$\Kgmf_{\vk\sigma\,\vp\sigma'}(\vq)$ in the integral eq.~(\ref{phBetheGen}) 
by its RPA expression, which is nothing but a spin--flip interaction,
represented by a dot,
%%%%%%%%%%%%%%%%%%%%%%%%%%%%%%%%%%%%%%%%
% RPA Bethe-Salpeter equation
%%%%%%%%%%%%%%%%%%%%%%%%%%%%%%%%%%%%%%%%
\begin{equation}
  \label{phBetherpaFeyn}
  \unitlength=3ex 
  \setlength{\Units}{\the\unitlength} 
  \begin{picture}(23,3)(0,-1.5)
    \put( 0,+1.2){\makebox(0,0)[b]{\scriptsize $\vk,\sigma$}}
    \put( 0,-1.2){\makebox(0,0)[t]{\scriptsize $\vk+\vq,\sigma$}}
    \put( 6,+1.2){\makebox(0,0)[b]{\scriptsize $\vp,\sigma'$}}
    \put( 6,-1.2){\makebox(0,0)[t]{\scriptsize $\vp+\vq,\sigma'$}}
    \put( 0,-1.1){\epsfig{file=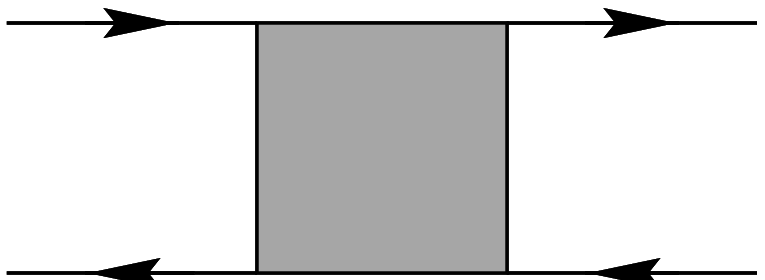, width=6\Units}}
    \put( 7, 0.0){\makebox(0,0)[c]{$=$}}
    \put(10,+1.2){\makebox(0,0)[b]{\scriptsize $\vk,\sigma$}}
    \put(10,-1.2){\makebox(0,0)[t]{\scriptsize $\vk+\vq,\sigma$}}
    \put( 8,-1.1){\epsfig{file=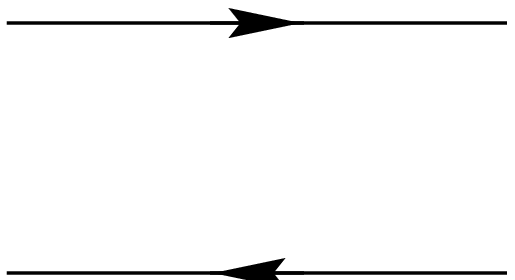,width=4\Units}}
    \put(13, 0.0){\makebox(0,0)[c]{$+$}}
    \put(14,+1.2){\makebox(0,0)[b]{\scriptsize $\vk,\sigma$}}
    \put(14,-1.2){\makebox(0,0)[t]{\scriptsize $\vk+\vq,\sigma$}}
    \put(14,-1.1){\epsfig{file=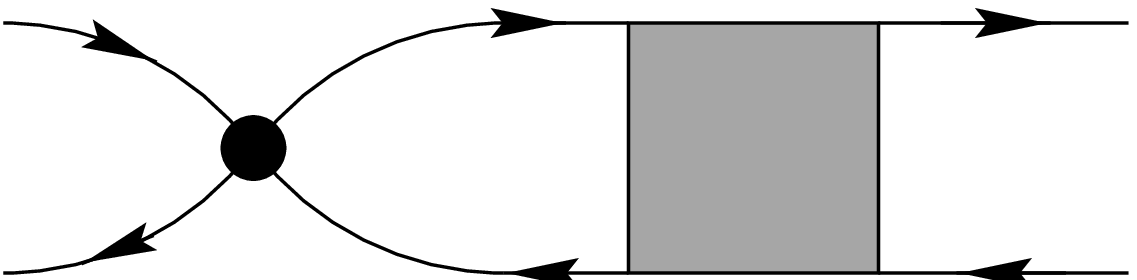,width=9\Units}}
    \put(16,+1.2){\makebox(0,0)[b]{$U$}}
    \put(18,+1.2){\makebox(0,0)[b]{\scriptsize $-\sigma$}}
    \put(18,-1.2){\makebox(0,0)[t]{\scriptsize $-\sigma$}}
    \put(23,+1.2){\makebox(0,0)[b]{\scriptsize $\vp,\sigma'$}}
    \put(23,-1.2){\makebox(0,0)[t]{\scriptsize $\vp,\sigma'$}}
  \end{picture}
  \qquad \mbox{.}
\end{equation}

\begin{figure}[htbp]
  \begin{center}
    \leavevmode
    \epsfig{file=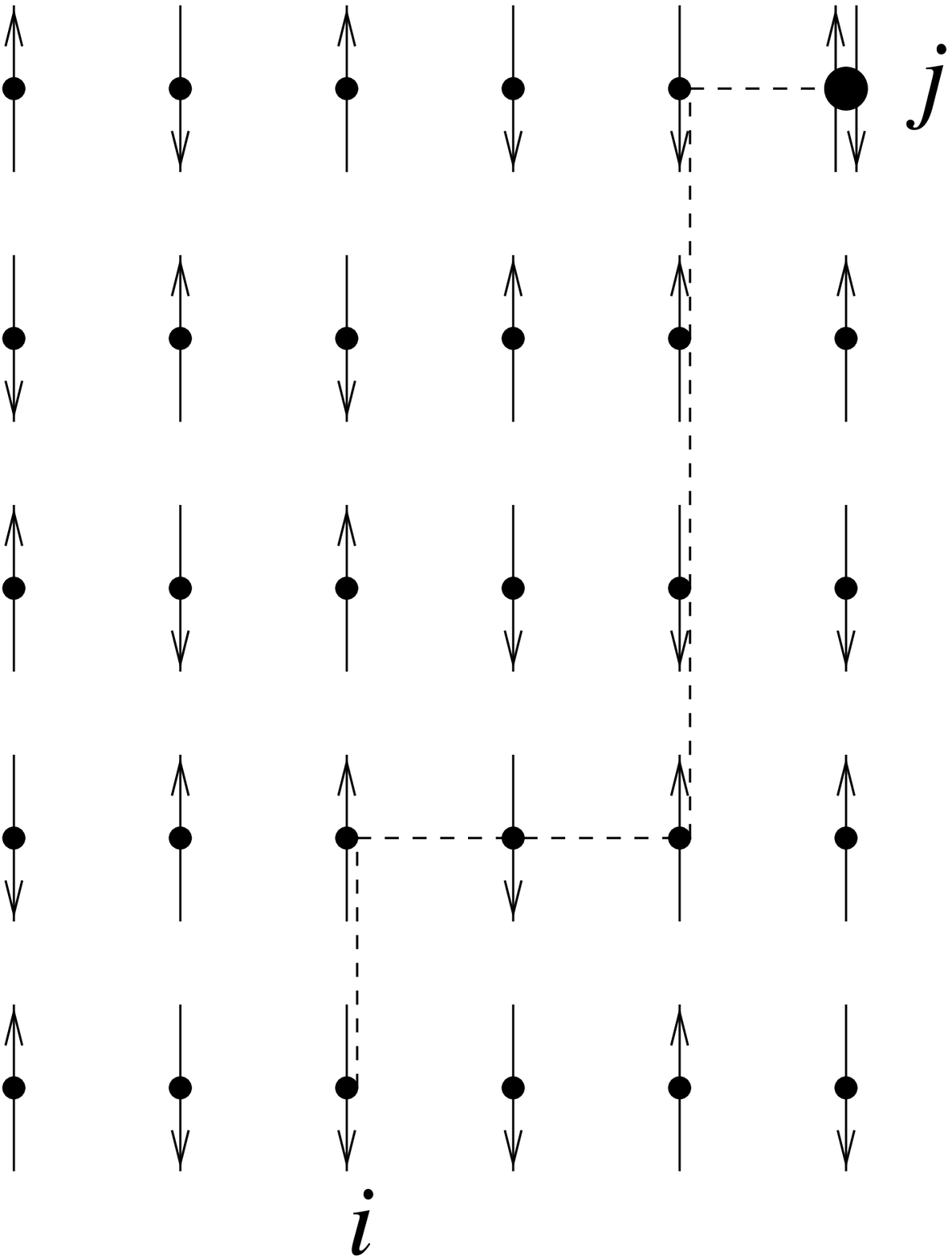,width=5cm}
    \caption{An electron propagating in an antiferromagnetic lattice from
      site $i$ to site $j$} 
    \label{fig:antiferro}
  \end{center}
\end{figure}

The physical interpretation of eq.~(\ref{phBetherpa}) is that in RPA
the Hubbard interaction flips the spin of the electron on each and every
scattering process. 
This is certainly a good approximation for an electron propagating in
an antiferromagnetically ordered state. In this case, we see from
Fig.~\ref{fig:antiferro} that a $\downarrow$--electron added to the
$\uparrow$--electron on site $i$ cannot hop off because all their
neighbours have spin~$\downarrow$, too. 
Therefore it must be the $\uparrow$--electron which hops to a
neighbouring site. 
Arriving there, it can only hop the same way back to its original
site. Otherwise, the $\uparrow$--electron is surrounded by other
$\uparrow$--electrons and thus frozen in.
As this process continues, an extra electron propagating in an
antiferromagnet from site $i$ to site $j$ flips every spin on its
trajectory. 
Thus, neglecting higher--order loop trajectories, the electron's path is
completely retraceable. 
This case was first examined by Brinkman and Rice\cite{Brinkman} who
showed that this ``retraceable path approximation'' is accurate for 
walks up to length twelve for the analogous case of an extra hole
propagating in an antiferromagnetic spin configuration. 
Moreover, they showed that even if the spins are randomly distributed
rather than antiferromagnetically ordered, the dominant contribution
to the hole Green's function comes from retraceable paths.

In this line of reasoning, the dimensionality of the system plays a
crucial role. 
\revised{In one dimension, antiferromagnetic long--range order is
forbidden by the Mermin--Wagner theorem\cite{MerminW}. Nevertheless,
as there are no loop trajectories in one dimension, the retraceable
path approximation becomes exact for any spin configuration.

In dimensions $d\ge3$, antiferromagnetic ordering is possible.
As the number of nearest neighbors increases, loop trajectories become
less probable. Therefore, the retraceable path approximation gets
exact to order ${\mathcal O}(1/d^4)$\cite{Florian}. 
\comment{Florian S.211, ${\mathcal O}(1/d^4)$ private communication} 

Moreover, in higher dimensions the correlations are weaker than in
lower ones. 
As we derived the RPA kernel $\Krpa_{\vk\sigma\,\vp\sigma'}(\vq)$ by
neglecting the correlations present in the SCRPA, we expect it to be
more accurate in higher dimensions than in lower ones.
We will see in section~\ref{HubbardChain} that, even in one dimension,
the renormalized RPA solution shows a Mott--Hubbard transition at a
finite critical interaction strength, which, for half filling is of
the order of the bandwidth. 
This scenario, which can be considered as generic for the RPA, is
certainly wrong for the one dimensional case. 
Nevertheless, in higher dimensions it could be quite realistic. 
There, this viewpoint is also supported by methods like e.g. the
Hubbard-III approximation\cite{HubbardIII,Cyrot}.}

Iterating the integral eq.~(\ref{phBetherpa}), we can decouple the
equations for $\phGF^{\RPA}_{\vk\sigma\,\vp\sigma}(\vq,\omega)$ and   
$\phGF^{\RPA}_{\vk\sigma\,\vp -\sigma}(\vq,\omega)$,
\begin{eqnarray}
  \label{phBetherpa2}
  \phGF^{\RPA}_{\vk\sigma\,\vp\sigma}(\vq,\omega)
  & = &
  \qquad\; \delta_{\vk \vp} \, \phGF^{0}_{\vk\sigma}(\vq,\omega) \qquad\;
  \; + \;
  \phGF^{0}_{\vk\sigma}(\vq,\omega) 
  \, U \, \ChiZe_{-\sigma}(\vq,\omega) \, \frac{U}{N} \, 
  \sum\limits_{\vk'} \phGF^{\RPA}_{\vk'\sigma\,\vp\sigma}(\vq,\omega)
  \nonumber \\
  \phGF^{\RPA}_{\vk\sigma\,\vp-\sigma}(\vq,\omega)
  & = &
  \phGF^{0}_{\vk\sigma}(\vq,\omega) \, \frac{U}{N} \,
  \phGF^{0}_{\vp -\sigma}(\vq,\omega)
  \; + \;
  \phGF^{0}_{\vk\sigma}(\vq,\omega) 
  \, U \, \ChiZe_{-\sigma}(\vq,\omega) \, \frac{U}{N} \, 
  \sum\limits_{\vk'} \phGF^{\RPA}_{\vk'\sigma\,\vp -\sigma}(\vq,\omega)
  \; \mbox{,}
\end{eqnarray}
where we introduced the renormalized non--interacting susceptibility
\begin{equation}
  \label{phChiZero}
  \ChiZe_{\sigma}(\vq,\omega) \, = \, 
  \sfrac{1}{N}\sum\limits_{\vk} \phGF^{0}_{\vk\sigma}(\vq,\omega)
  \; \mbox{.}
\end{equation}
Eq.~ (\ref{phBetherpa2}) can be solved explicitly for the
particle--hole Green's function, yielding
\begin{eqnarray}
  \label{phGFrpa}
  \phGF^{\RPA}_{\vk\sigma\,\vp\sigma}(\vq,\omega)
  & = & \phGF^{0}_{\vk\sigma}(\vq,\omega) 
  \left[
    \, \delta_{\vk \vp}  
    \,+\, \frac{U}{N} \, \phGF^{0}_{\vp\sigma}(\vq,\omega) \,
    \frac{U\ChiZe_{-\sigma}(\vq,\omega)}
    {1-U\ChiZe_{\uparrow}(\vq,\omega)\,U\ChiZe_{\downarrow}(\vq,\omega)}
    \,
  \right]
  \qquad \mbox{and}
  \nonumber \\
  \phGF^{\RPA}_{\vk\sigma\,\vp-\sigma}(\vq,\omega)
  & = & \qquad\qquad
  \phGF^{0}_{\vk\sigma}(\vq,\omega) 
  \frac{U}{N} \, \phGF^{0}_{\vp-\sigma}(\vq,\omega) \,
  \frac{1}
  {1-U\ChiZe_{\uparrow}(\vq,\omega)\,U\ChiZe_{\downarrow}(\vq,\omega)}
  \; \mbox{.}
\end{eqnarray}

Finally, we have to determine the occupation numbers $n_{\vk\sigma}$
self--consistently from the RPA Green's function~(\ref{phGFrpa}).
As will be explained in more detail in section~\ref{NumericalMethod},
this is a somewhat delicate procedure. Indeed, initiating the
iteration cycle, at small $U$, with the Fermi step for
$n_{\vk\sigma}$, one inevitably picks up some spin instabilities when
summing over $\vq$ in eq.~(\ref{OccSpectralTheorem}). These
instabilities correspond to poles in the RPA Green's function at
purely imaginary frequencies. This entails that we may create some
unphysical values when applying the spectral theorem to such a Green's
function. For example, integrating
$\imag\phGF^{\RPA}_{\vk\sigma\,\vp-\sigma}(\vq,\omega)$ over the  
whole $\omega$--axis yields an unphysical finite value if the Green's
function has poles at imaginary frequencies. 
As this integral is connected via the spectral
theorem~(\ref{RetSpectralTheorem}) to a commutator, we know that it  
has to be zero.
When iterating the system of non--linear equations, it turns out that
this pathology is cured. The occupation numbers get rounded thus
weakening the interaction in such a way that finally the imaginary
spin poles disappear. 
At self--consistency, the integral of
$\imag\phGF^{\RPA}_{\vk\sigma\,\vp-\sigma}(\vq,\omega)$ over all
frequencies vanishes for every $\vq$, as it is expected from the
spectral theorem.  

For completeness, we state that {\em fixing} the occupation to their
Hartree--Fock values brings us from 
\revised{{\em renormalized} back to {\em pure} RPA}: 
\begin{equation}
  \label{OccHF}
  n_{\vk\sigma}^{\rm HF} \, = \, \Theta(E_{\rm F}-\varepsilon_{\vk})
\end{equation}

In section~\ref{HubbardEnergySumRule}, an energy weighted sum rule
is shown to be fulfilled in both, renormalized and Self--Consistent RPA.
In pure RPA it is only fulfilled as long as all eigenfrequencies are real.  
As will be pointed out in section~\ref{HubbardESumRule1d}, this is
closely connected to the problem of RPA instabilities. 

If we restrict ourselves to the paramagnetic phase, we have
$n_{\vk\uparrow}=n_{\vk\downarrow}$ 
implying $\ChiZe_{\uparrow}(\vq,\omega)=\ChiZe_{\downarrow}(\vq,\omega)$.
We can thus recover the usual RPA structure for the susceptibilities 
$\ChiCh(\vq,\omega)$ and $\ChiSp(\vq,\omega)$ by combining the two
equations~(\ref{phGFrpa}) and summing over $\vk$ and $\vp$:
\begin{eqnarray}
  \label{cdwRPASusc}
  \ChiCh(\vq,\omega) & = & \frac{1}{N} \sum\limits_{\vk \vp \sigma}
  \left[ 
    \phGF^{\RPA}_{\vk\sigma\,\vp \sigma}(\vq,\omega)  \, + \,
    \phGF^{\RPA}_{\vk\sigma\,\vp-\sigma}(\vq,\omega) 
  \right]
  \nonumber \\
  & = & \frac{2\ChiZe(\vq,\omega)}{1-U\ChiZe(\vq,\omega)} 
  \nonumber \\
  \ChiSp(\vq,\omega) & = & \frac{1}{4 N} \sum\limits_{\vk \vp \sigma}
  \left[ 
    \phGF^{\RPA}_{\vk\sigma\,\vp \sigma}(\vq,\omega)  \, - \,
    \phGF^{\RPA}_{\vk\sigma\,\vp-\sigma}(\vq,\omega) 
  \right]
  \nonumber \\
  & = & \frac{\sfrac{1}{2}\ChiZe(\vq,\omega)}{1+U\ChiZe(\vq,\omega)} 
\end{eqnarray}

%%%%%%%%%%%%%%%%%%%%%%%%%%%%%%%%%%%%%%%%%%%%%%%%%%
\subsection{Energy weighted sum rule} 
\label{HubbardEnergySumRule}

\comment{(see Orland/Negele: Problem 5.13)}

\revised{
The fact that the density operators $\rho_{\vq\sigma}$ commute with
the interaction term of the Hubbard Hamiltonian~(\ref{Hubbard}) gives
us the possibility to evaluate  
$\mean{\doco{\rho_{\vq\sigma}}{H}{\rho^{+}_{\vq\sigma'}}_{-}}$
analytically. In analogy to section~\ref{EnergySumRule}, we thus can
establish an energy weighted sum rule for the exact susceptibilities
$\ChiCh(\vq,\omega)$ and $\ChiSp(\vq,\omega)$.
We will now discuss the fulfillment of the sum rule for the SCRPA, the
renormalized RPA, and the pure RPA Green's functions.

In the first place, we calculate the double commutator
\begin{eqnarray}
  \label{HubbESumRuleDoco}
  \mean{ \doco{\rho_{\vq\sigma}}{H}{\rho^{+}_{\vq\sigma'}}_{-} }
  & = & \delta_{\sigma \sigma'} \, \sum\limits_{\vk}\,
  \left[ \varepsilon_{\vk+\vq} - \varepsilon_{\vk} \right]  \, 
  \left( n_{\vk\sigma} - n_{\vk+\vq\sigma} \right)
  \nonumber \\
  & = & \delta_{\sigma \sigma'} \;
  2 \, \sum\limits_{i=1}^{d}\,\left(\cos q_{i} - 1\right) \,
  \sum\limits_{k_{i}}\,\varepsilon_{\{k_{i}\}}\,n_{\{k_{i}\}\sigma}
\end{eqnarray}
where $k_{i}$ are the components of the vector $\vk$. 
The only contributions to the double commutator come from the kinetic
term of the Hamiltonian~(\ref{Hubbard}), since, as mentioned above,
$\rho_{\vq\sigma}$ commutes with the interaction. 
Remark that for a $\vk^{2}/(2 m)$ dispersion law the rhs. of
(\ref{HubbESumRuleDoco}) yields the well--known result
$\mean{n_{\sigma}} \vq^{2}/m$ (see ref.\cite{Orland}). 

As pointed out in section~\ref{EnergySumRule} for the general case,
the expectation value of the double commutator~(\ref{HubbESumRuleDoco})
is related to the imaginary part of the exact retarded susceptibility
by an energy weighted integral (see eq.~(\ref{ESumRule})):
\begin{equation}
  \label{HubbESumRule}
  \mean{ \doco{\rho_{\vq\sigma}}{H}{\rho^{+}_{\vq\sigma'}} }
  \; = \; -\frac{1}{\pi} 
  \int\limits_{-\infty}^{\infty} {\rm d} \omega \; \omega \; 
  \imag\GFome{\rho_{\vq\sigma}}{\rho^{+}_{\vq\sigma'}}^{\rm ret}
\end{equation}
The SCRPA Green's function was shown to satisfy the sum
rule~(\ref{HubbESumRule}) as well, since the double commutator on 
the lhs. can be expressed by the SCRPA Hamiltonian~(\ref{phHgmf}) and
the norm matrix (\ref{phNorm}):   
\begin{eqnarray}
  \label{HubbDocoByHgmf}
  \mean{ \doco{\rho_{\vq\sigma}}{H}{\rho^{+}_{\vq\sigma'}} }
  & = & 
  \sum\limits_{\vk \vp}\, \sum\limits_{\vp_{1} \sigma_{1}}\,
  \Hgmf_{\vk\sigma\,        \vp_{1}\sigma_{1}}(\vq) \, 
  \Norm_{\vp_{1}\sigma_{1}\,\vp\sigma'       }(\vq)
  \nonumber \\
  & = &
  \sum\limits_{\vk \vp}\,\Hgmf_{\vk\sigma\,\vp\sigma'}(\vq) 
  \, \left( n_{\vp\sigma'} - n_{\vp+\vq\sigma'} \right)
\end{eqnarray}
In the view of our formalism this may seem evident. However, theories
which generalize the RPA approach do not necessarily fulfill the
$f$-sum rule.
Above all, we notice that in eq.~(\ref{HubbDocoByHgmf}) {\em all}
terms containing correlations from the effective Hamiltonian
$\Hgmf_{\vk\sigma\,\vp\sigma'}(\vq)$ cancel when summing over $\vk$
and $\vp$.  
In {\em renormalized} RPA, on the other hand, we neglect these
correlations right from the beginning 
(see eqs.~(\ref{phKrpa},\ref{phBetherpa})). 
Consequently, the renormalized RPA susceptibilities fulfill the energy
weighted sum rule~(\ref{HubbESumRule}), too.
Moreover, it is well-known\cite{RS,Thouless} that the {\em pure} RPA
susceptibilities satisfy the sum rule~(\ref{HubbESumRule}) if the
expectation value on the lhs. is evaluated with the Hartree--Fock
ground state wave function. 
However, this statement only holds true as long as all RPA frequencies
are real.  
\comment{Ring/Schuck Chap. 8.7.2}

Finally, we find the energy weighted sum rules for
$\ChiCh(\vq,\omega)$ and $\ChiSp(\vq,\omega)$ by combining
eq.~(\ref{HubbESumRule}) for the different spin configurations
according to the definitions of the charge
susceptibility~(\ref{DefChargeSusc}) and longitudinal spin
susceptibility~(\ref{DefLongSpinSusc}), respectively: 
\begin{eqnarray}
  \label{HubbSuscRule} 
  2\,\sum\limits_{i=1}^{d}\,
  \left(\cos q_{i} - 1\right) \, \mean{ \hat{t^{i}} }
  & = &  -\frac{1}{\pi} \int
  \limits_{-\infty}^{\infty} {\rm d}\omega \; 
  \omega \; \imag \ChiCh(\vq,\omega)
  \nonumber \\
  \sfrac{1}{2}\,\sum\limits_{i=1}^{d}\,
  \left(\cos q_{i} - 1\right) \, \mean{ \hat{t^{i}} }
  & = &  -\frac{1}{\pi} \int
  \limits_{-\infty}^{\infty} {\rm d}\omega \; 
  \omega \; \imag \ChiSp(\vq,\omega)
\end{eqnarray}
On the lhs., we introduced the shorthand
\begin{equation}
  \mean{\hat{t^{i}}} 
  \; = \; 
  \sfrac{1}{N}\,\sum\limits_{k_{i}\sigma}\,
  \varepsilon_{\{k_{i}\}}\,n_{\{k_{i}\}\sigma}
  \; \mbox{,}
\end{equation}
standing for the contribution to the mean kinetic energy per site
provided by the electron motion in $i$ direction.
It should be stressed that $\mean{\hat{t^{i}}}$ depends on the
occupation numbers $n_{\vk\sigma}$ and thus implicitly on the Green's
function.  In contrast, for the usual $\vk^{2}/(2 m)$ dispersion law,
the double commutator in eq.~(\ref{HubbESumRuleDoco}) depends on the
mean number of electrons per site (i.e. the filling) instead of
$\mean{\hat{t^{i}}}$, and is therefore a model independent quantity.  
In our case of a cosine dispersion law, however, the $\vq$--dependence
of the sum rule provides a check which does not depend on the Green's
function or any other assumption. 
}

%%%%%%%%%%%%%%%%%%%%%%%%%%%%%%%%%%%%%%%%%%%%%%%%%%
\subsection{Transverse spin--density correlations}
\label{phGFspin}

\revised{In this section, for completeness, we will shortly discuss
the transverse spin response. The spin--flip operators $S^{\pm}_{i}$
may be substituted in the usual way by combination of an annihilation
operator of a $\sigma$ electron and a creation operator of a $-\sigma$
electron. After Fourier transformation to momentum space, we obtain}
\comment{Nolting (Afg. 4.1.6)}
\begin{equation}
  \label{Splusminus} 
  S^{+}_{\vq} \, = \, \sum_{\vk} a^{+}_{\vk\uparrow} a_{\vk+\vq\downarrow}
  \qquad \mbox{and} \qquad
  S^{-}_{\vq} \, = \, \sum_{\vk} a^{+}_{\vk+\vq\downarrow} a_{\vq\uparrow}
  \; \mbox{.}
\end{equation}
\revised{Correlations between the spin--flip operators} give rise to the
transverse spin susceptibility  
\comment{Nolting chapt. 3.1.3, (3.72)}
\begin{equation}
  \label{DefTransSpinSusc}
  \ChiTr (\vq,\omega) 
  \, = \, \sfrac{1}{N} \GFome{ S^{+}_{\vq} }{ S^{-}_{\vq} }
  \; \mbox{.}
\end{equation}
\revised{In order to examine the transverse spin susceptibility in
SCRPA, we introduce the following correlation functions:}
\begin{equation}
  \label{phSpinGF}
  \phGF^{\mathrm trans}_{\vk\sigma\,\vp\sigma'}(\vq,\omega) \, \equiv \,
  \GFome{a^{+}_{\vk      \sigma} \,a_{\vk+\vq,-\sigma }}
  {      a^{+}_{\vp+\vq,-\sigma'}\,a_{\vp      \sigma'}}
\end{equation}

In complete analogy to the charge--density case
(see section~\ref{phGFcharge}), we calculate the norm matrix by
evaluating the commutator of the two operators defining the Green's
function (see eq.~(\ref{phNormGen})):
\begin{equation}
  \label{phSpinNorm}
  \Norm^{\mathrm trans}_{\vk\sigma\,\vp\sigma'}(\vq) \,=\, 
  \delta_{\vk \vp} \delta_{\sigma \sigma'} \,
  \left( n_{\vk\sigma} - n_{\vk+\vq,-\sigma} \right)
\end{equation}
The effective SCRPA Hamiltonian reads in analogy to
eq.~(\ref{phHgmfGen})  
\begin{eqnarray}
  \label{phSpinHgmf}
  \Heff^{\mathrm trans}_{\vk\sigma\,\vp\sigma'}(\vq) & = &
  \delta_{\vk \vp} \delta_{\sigma \sigma'} \, 
  \left[ \epsilon_{\vk+\vq,-\sigma} - \epsilon_{\vk\sigma} \right]
  \,-\, \delta_{\sigma \sigma'} \,
  \left( n_{\vk\sigma} - n_{\vk+\vq,-\sigma} \right) \, \frac{U}{N}
  \nonumber \\
  & & +\,\biggl[ \;
  -\delta_{\vk \vp} \delta_{\sigma \sigma'} \, \frac{U}{N} 
  {\displaystyle \sum\limits_{\vq'}}\mean{ 
    a^{+}_{\vk+\vq,-\sigma} a_{\vk+\vq',-\sigma} \, \rho_{\vq-\vq'\sigma} 
    \,+\, 
    a^{+}_{\vk+\vq-\vq'\sigma} a_{\vk\sigma}  \, \rho_{\vq-\vq',-\sigma} 
    }^\corr
  \nonumber \\
  & & \qquad
  -\delta_{\sigma \sigma'} \, \frac{U}{N} 
  \sum\limits_{\vq'}
  \mean{
    \left(  a^{+}_{\vk       \sigma} a_{\vk+\vq-\vq' \sigma} 
      - a^{+}_{\vk+\vq',-\sigma} a_{\vk+\vq,    -\sigma} \right) 
    \cdot
    \left(  a^{+}_{\vp+\vq-\vq'  \sigma} a_{\vp       \sigma} 
      - a^{+}_{\vp+\vq     ,-\sigma} a_{\vp+\vq',-\sigma} \right) 
    }^\corr
  \nonumber \\
  & & \qquad
  -\delta_{\sigma,-\sigma'} \, \frac{U}{N} 
  \mean{  a^{+}_{\vk  \sigma}   a_{\vp    ,-\sigma} \,S^{\sigma}_{\vk-\vp} 
    \,+\, a^{+}_{\vp+\vq\sigma} a_{\vk+\vq,-\sigma} \,S^{\sigma}_{\vp-\vk} 
    }^\corr
  \biggr]
  \; \cdot \left( n_{\vp\sigma'} - n_{\vp+\vq,-\sigma'} \right)^{-1} 
  \; \mbox{,} 
\end{eqnarray}
with $\rho_{\vq\sigma}$ denoting the density operator introduced in
eq.~(\ref{RhoSigma}) and $S^{\sigma}_{\vq}$ being a spin--flip operator
defined in analogy to $S^{+}_{\vq}$  
(see eq.~(\ref{Splusminus})): 
\begin{equation}
  \label{SpinFlip}
  S^{\sigma}_{\vq} \, = \, \sum_{\vk} a^{+}_{\vk\sigma} a_{\vk+\vq,-\sigma}
\end{equation}
The Hartree--Fock corrected single--particle energies defined in 
eq.~(\ref{phHFepsGen}) are given by
\begin{equation}
  \label{phHFeps}
  \epsilon_{\vk\sigma} \; = \; \varepsilon_{\vk} \,+\, U\,\mean{ n_{-\sigma} }
  \; \mbox{.}
\end{equation}

In contrast to the charge--density case, not {\em all} elements of the
effective Hamiltonian~(\ref{phSpinHgmf}) can be determined
self--consistently from the Green's functions~(\ref{phSpinGF}).
The calculation of terms like
$ \mean{ a^{+}_{\vk\uparrow} a_{\vk+\vq-\vq'\uparrow} \, 
         a^{+}_{\vp+\vq-\vq'\uparrow} a_{\vp\uparrow} }^\corr$,
for example, cannot be performed with the spectral theorem for
spin--flip Green's functions, since they contain always the same
number of $\uparrow$ and $\downarrow$ spins. 
These terms can, however, be determined from the charge--density
Green's function introduced section~\ref{phGFcharge}. By this means,
the transverse spin susceptibility is coupled to the charge--density
susceptibilities $\ChiCh$ and $\ChiSp$. 
In this work, however, we will not further investigate the transverse
spin excitations.

% %%%%%%%%%%%%%%%%%%%%%%%%%%%%%%%%%%%%%%%%%%%%%%%%%%%%%%%%%%%%
% 1d Hubbard Model
% %%%%%%%%%%%%%%%%%%%%%%%%%%%%%%%%%%%%%%%%%%%%%%%%%%%%%%%%%%%%
\section{Results for the charge and the longitudinal spin response 
  in the Hubbard Chain}  
\label{HubbardChain}

As a first application of our general formalism, we will calculate the
charge and longitudinal spin correlation functions
$\ChiCh(q,\omega)$ and $\ChiSp(q,\omega)$ in the one--dimensional
Hubbard model. \revised{This will also serve as a test of whether our
formalism is well behaved in a numerical sense.}

It was explained in section~\ref{phGFcharge} that the first step will
consist in calculating the Green's function
$\phGF_{k\sigma\,p\sigma'}(q,\omega)$, introduced in eq.~(\ref{phGF}),
on the level of the {\em renormalized} RPA.  
In this paper, we will not go beyond this approximation.
Indeed, the numerical solution of the full SCRPA problem turns out to
be quite enormous and certainly needs a major numerical effort which is
intended to be invested in future work. Nevertheless, the main
characteristics of the self--consistency cycle are already present on
the level of the renormalized RPA.

We thus will determine self--consistently the RPA Green's function
(\ref{phGFrpa}) together with the occupation numbers $n_{k\sigma}$. 
Due to the self--consistency, the occupation numbers are ``renormalized''
which modifies the Green's function $\phGF^{0}_{k\sigma}(q,\omega)$ and
the susceptibility $\ChiZe(q,\omega)$ occurring in the RPA propagator
(\ref{phGFrpa}).  

In contrast, in pure RPA, i.e. if we fix the occupation numbers to
their Hartree--Fock values given by eq~(\ref{OccHF}),  
$\phGF^{0}_{k\sigma}(q,\omega)$ and $\ChiZe(q,\omega)$
are identical with the free particle--hole propagator and
susceptibility, respectively. 

After a brief overview of the numerical method we will discuss the
results for the renormalized RPA in the infinite Hubbard chain in the
paramagnetic phase, i.e. $n_{k\uparrow}=n_{k\downarrow}$.

%%%%%%%%%%%%%%%%%%%%%%%%%%%%%%%%%%%%%%%%%%%%%%%%%%
\subsection{Numerical Method}
\label{NumericalMethod}

In order to determine the renormalized RPA Green's function, we have to
solve the RPA eqs.~(\ref{phGFzero},\ref{phChiZero},\ref{phGFrpa})
consistently with the equation for the occupation numbers,
(\ref{OccSpectralTheorem}).

Therefore, we start with an assumption for the initial occupation
numbers and set up the following iteration cycle:
\begin{itemize}
\item 
  Set up the renormalized free Green's function~(\ref{phGFzero})
  from the current set of occupation numbers.
\item 
  Calculate the renormalized free susceptibility~(\ref{phChiZero}) by
  integrating $\phGF^{0}_{k\sigma}(q,\omega)$ over $k$. Note that
  for the computation of $\imag \ChiZe(q,\omega)$ it is convenient to
  introduce a small but finite imaginary part ``$i0^{+}$'' in the
  denominator of $\phGF^{0}_{k\sigma}(q,\omega)$.
\item
  Set up the RPA Green's function~(\ref{phGFrpa}) from
  $\phGF^{0}_{k\sigma}(q,\omega)$ and $\ChiZe(q,\omega)$.
\item 
  Compute a new set of occupation numbers by performing the
  spectral integral in eq.~(\ref{OccSpectralTheorem}). 
  This task is also simplified by assuming a finite ``$i0^{+}$''.
\item
  Repeat the iteration cycle from the beginning until
  self--consistency is  achieved.
\end{itemize}

Usually, we will start with a small interaction, e.g. $U=1$, and a
Fermi step for the occupation numbers. 
For this $U$, the RPA equations are then iterated to self--consistency.  
The result is used to initialize the occupation numbers of an
iteration cycle with a slightly higher interaction.
We thus increase the interaction in small steps, iterating each time
to self--consistency, until the desired value for $U$ is reached.
This procedure has the advantage that the spin response, in contrast
to pure RPA, remains stable throughout the whole calculation.

The momentum integrations are performed by summing over a grid of
uniformly distributed points in the \revised{first} Brillouin
zone. The number of points is typically 256.

Energy integrals are computed using a grid of points obeying a
Lorentzian distribution peaked at $\omega=0$. The number of energy
points is typically 2048 and the half width half maximum of the
distribution is about twice the bandwidth, i.e. 8.

As mentioned above, it is convenient for computational purposes to
introduce a small but finite imaginary part ``$i0^{+}$'' in the
denominator of the renormalized free Green's function
$\phGF^{0}_{k\sigma}(q,\omega)$. 
Typically we use values of the order of magnitude of $i/16$ which
corresponds to $1/64$ the bandwidth. It can be shown that within our
resolution the results are not affected by slight variations of this
value. 

%%%%%%%%%%%%%%%%%%%%%%%%%%%%%%%%%%%%%%%%%%%%%%%%%%
\subsection{Occupation numbers}
\label{OccupationNumbers}

\begin{figure}[htbp]
  \begin{center}
    \leavevmode
    \epsfig{file=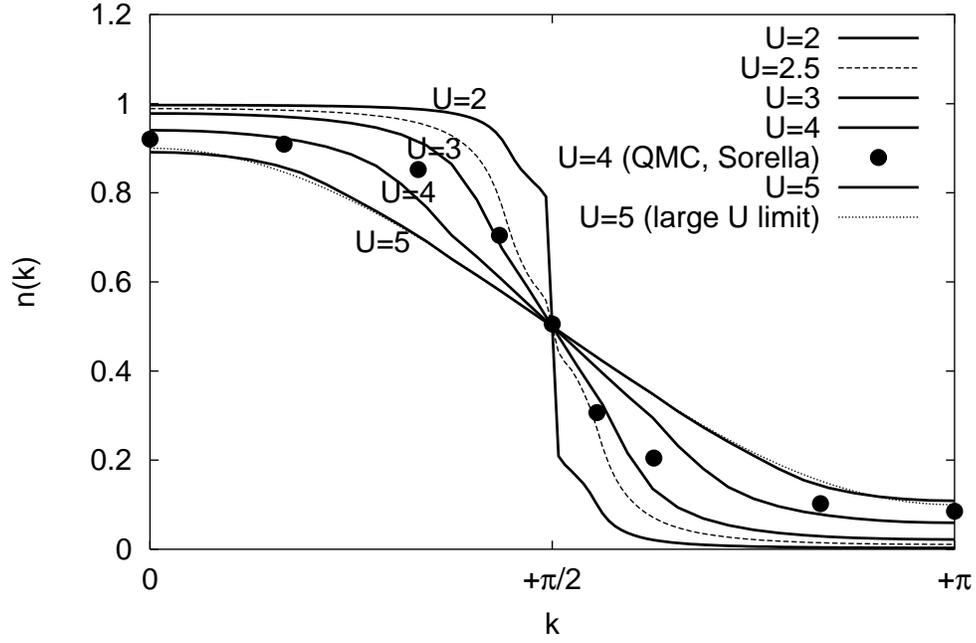,width=13cm}
    \caption{
      Momentum distribution function $n_{k}$ for the
      half--filled Hubbard chain. 
      Quantum Monte--Carlo data from Sorella\cite{Sorella88}. 
      The dotted line shows the large $U$ limit of the renormalized
      RPA, given by eq.~(\ref{nk_UinfRPA}). }
    \label{fig:nk_half}
  \end{center}
\end{figure}

\begin{figure}[htbp]
  \begin{center}
    \leavevmode
    \epsfig{file=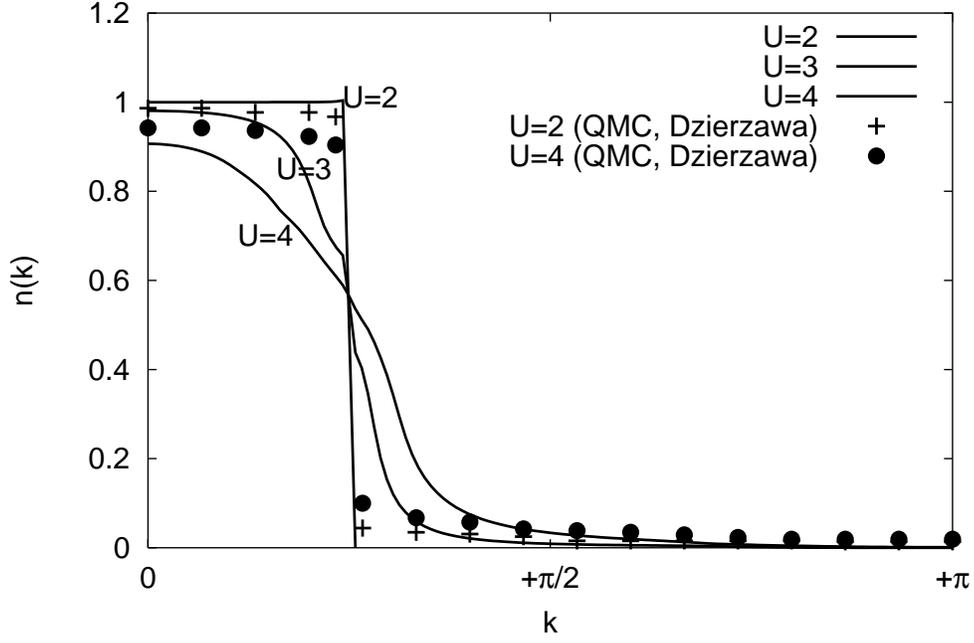,width=13cm}
    \caption{
      Momentum distribution function $n_{k}$ for the
      quarter--filled Hubbard chain. 
      Quantum Monte--Carlo data from Dzierzawa\cite{Dzierzawa}. }
    \label{fig:nk_quarter}
  \end{center}
\end{figure}

\revised{
Fig.\ref{fig:nk_half} shows the momentum distribution function   
\begin{equation}
  \label{nk}
  n_{k} \, = \, \sfrac{1}{2} \, 
  \left(  n_{k\uparrow} +  n_{k\downarrow} \right)
\end{equation}
of the half--filled Hubbard chain. The renormalized RPA results are
compared with Quantum Monte--Carlo calculations.
We see that, for small $U$, the renormalized RPA momentum distribution
has a discontinuity at the Fermi edge. This is typical for Fermi
liquids, and thus indicates a metallic behaviour. 
For large $U$, the renormalized RPA momentum distribution is
continuous, and our theory predicts an insulating ground
state. Increasing $U$ in small steps ($1/4$), we deduce that within  
our numerical momentum resolution the discontinuity at $k_{F}$
vanishes at $U\approx 3$. 
We thus find a Mott metal-insulator transition at an interaction
strength which is slightly smaller than the band width (i.e. $4$). 
This is in good agreement with approximations designed for the Hubbard
model in higher dimensions, like e.g. the Hubbard-III
solutions\cite{HubbardIII,Cyrot}. 
In {\em one dimension}, however, the exact solution\cite{LiebWu}
predicts an insulating ground state for any finite $U$. 
Consequently, the exact results for $n_{k}$, known from
Quantum Monte--Carlo (QMC)\cite{Sorella90} for finite $U$ and from Bethe
ansatz in the limit of large $U$\cite{CarmeloBaeriswyl,BaeriswylLinden},
show a smooth behaviour over the whole $k$ range for all interactions.

This disagreement should be judged in the light that the effective
Hamiltonian of the renormalized RPA neglects {\em all} correlation
functions. 
As discussed in section~\ref{phGFcharge}, this approximation is
expected to be far better in higher dimensions. 
Nevertheless, in the strong coupling limit, our theory reproduces the
cosine--behaviour for $n_{k}$, known from the Bethe ansatz expansion,
apart from the prefactor (see section~\ref{LargeU}). 

Away from half filling, the renormalized RPA predicts a similar
scenario: For small $U$, the momentum distribution function shows a
discontinuity at $k_{F}$, which now persists up to higher interaction
strengths as in the half filled case. For the quarter filled chain,
e.g., this jump lasts up to $U\approx 4$, as can be seen from
Fig.~\ref{fig:nk_quarter}. 
The discontinuity occurs precisely at the same $k=k_{F}$  as the step
in the momentum distribution of the free Fermi gas. Therefore, the
renormalized RPA satisfies the Luttinger theorem in one
dimension\cite{Luttinger}.  
  
Again, the Mott-Hubbard transition predicted by our theory is in
disagreement with the exact behaviour away from half filling, which is
known to be a Luttinger liquid for all interaction strengths. 
The latter is characterized by a power law singularity in the momentum
distribution at the Fermi points, which is expected in the strong
coupling limit from both,  
Bethe ansatz expansions\cite{OgataShiba89,OgataShiba90,Penc} 
and QMC calculations\cite{Sorella90}.  
For finite interaction strengths, it was also detected by QMC studies
of infinite Hubbard chains\cite{Dzierzawa}.

Calculations of finite chains indicate a discontinuity at $k_{F}$
decreasing only very slowly with increasing chain length
$N$\cite{Sorella88,Voit}.
Therefore, they have great difficulties to detect the Luttinger liquid
behaviour. There are certain similarities between finite chains and
calculations using a finite number of points in the Brillouin zone as
e.g.~the numerical solution of the renormalized RPA equations. 
Nevertheless, our calculations indicate a finite slope of the momentum
distribution on both sides of the Fermi points, and we do not find any
signature of Luttinger liquid behaviour.}

%%%%%%%%%%%%%%%%%%%%%%%%%%%%%%%%%%%%%%%%%%%%%%%%%%
\subsection{Renormalized free susceptibility $\ChiZe(q,\omega)$}
\label{RenChiFree}

\subsubsection{Imaginary part}
\label{ImRenChiFree}

Before discussing the RPA response functions $\ChiCh(q,\omega)$ and
$\ChiSp(q,\omega)$ let us analyze the renormalization effects in the
free susceptibility $\ChiZe(q,\omega)$.

In the $(q,\omega)$--plane, the imaginary part of $\ChiZe(q,\omega)$
is restricted to the region, where particle--hole excitations exist,
i.e. there must be a $k$ for which
$\omega=\varepsilon_{k+q}-\varepsilon_{k}$ is satisfied, or,   
\begin{equation}
  \label{phExciteB}
  \abs{\omega} \, \le \, \abs{4\sin\frac{q}{2}}
  \qquad \mbox{.}
\end{equation}

If the occupation numbers are step functions like in the free Fermi
gas, a second boundary condition is provided by the fact, that in the
particle--hole continuum described by eq.~(\ref{phExciteB}) there must
be at least one $k$--vector for which the numerator of the free
Green's function~(\ref{phGFzero}),
$\left(n_{k\sigma}-n_{k+q\sigma}\right)$, does not vanish.   
This yields 
\begin{equation}
  \label{StepB}
  \abs{\omega} \, \ge \, 2\,\abs{\cos(k_{F})-\cos(k_{F}-\abs{q})}
  \qquad\mbox{.}
\end{equation}

\begin{figure}[htbp]
  \begin{center}
    \leavevmode
    \epsfig{file=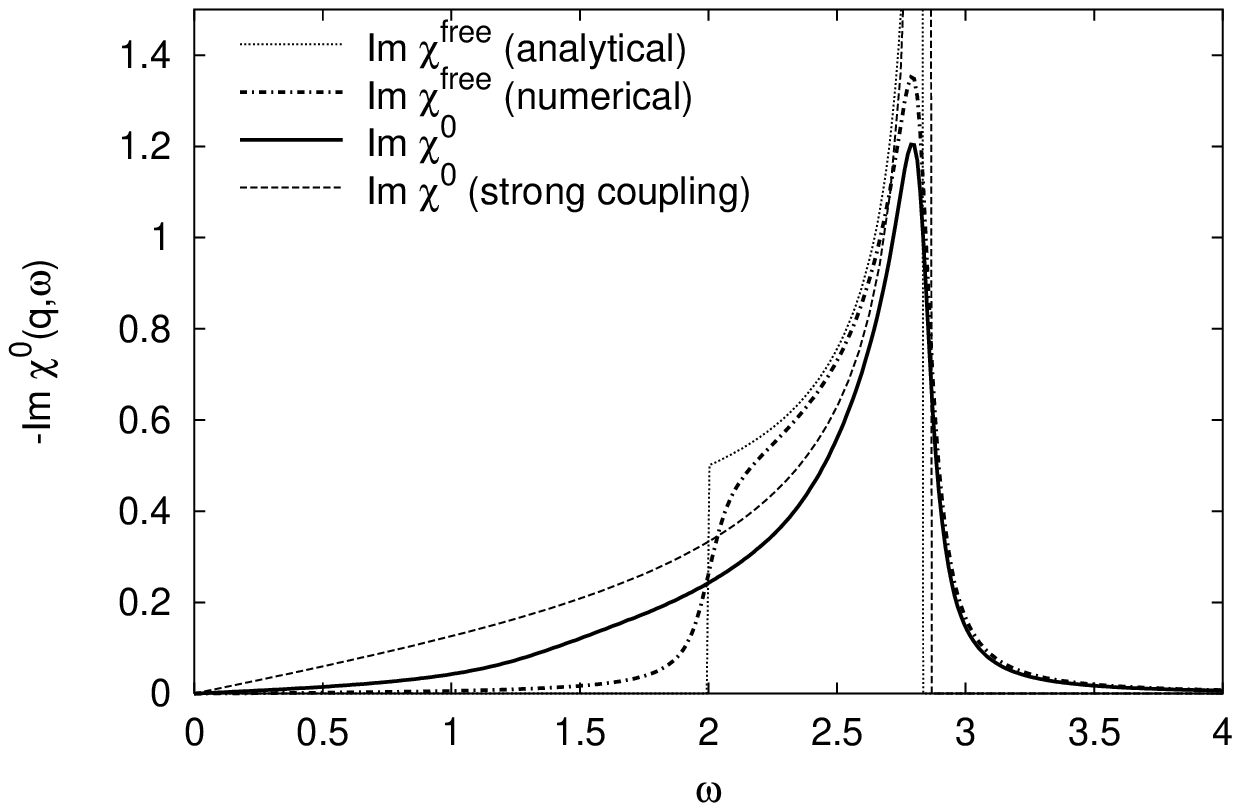,width=13cm}
    \caption{
      Imaginary part of the free and renormalized free susceptibility
      $\ChiZe(q=\pi/2,\omega)$ for the half filled Hubbard chain at $U=3$.}   
    \label{fig:ImChi0_U3}
  \end{center}
\end{figure}

We have seen in section~\ref{phGFcharge} that the Hartree--Fock
corrections to the single--particle energies cancel. Hence, the
denominator of the free Green's function~(\ref{phGFzero}) does not
change throughout the renormalization process, and
eq.~(\ref{phExciteB}) represents a rigid boundary for the 
imaginary part of the free susceptibility. 
On the other hand, any rounding of the occupation numbers will directly
affect the second boundary condition, (\ref{StepB}). 
This behaviour of the imaginary part is illustrated in
Fig.~\ref{fig:ImChi0_U3} for \figdep{half filling and $q=\pi/2$}. 
We see that $\imag \ChiFr(q,\omega)$, represented by the
\figdep{dotted line}, is only nonzero in the domain in between the two
boundaries given by eq.~(\ref{phExciteB}) and eq.~(\ref{StepB}). 

In section~\ref{NumericalMethod}, we explained that for technical
reasons we have to set ``$i0^{+}$'' in the denominator of
$\phGF^{0}_{k\sigma}(q,\omega)$ to a small but finite value in order
to perform the integral in eq.~(\ref{phChiZero}). 
This smoothens the numerically calculated free susceptibility
\figdep{(dot--dashed line)} in comparison to the analytical expression
given in appendix~\ref{ChiFree} \figdep{(dotted line)}.

Let us now study the imaginary part of the renormalized free
susceptibility, $\imag \ChiZe(q,\omega)$, represented by the
\figdep{continuous line} in Fig.~\ref{fig:ImChi0_U3} for
\figdep{half filling, $U=3$ and $q=\pi/2$}. 
On the outer boundary~(\ref{phExciteB}) it behaves essentially in the
same way as $\imag \ChiFr(q,\omega)$, whereas the inner
boundary~(\ref{StepB}) is completely washed out due to the
renormalization of the occupation numbers $n_{k\sigma}$. 

As will be discussed in section~\ref{LargeU}, the strong coupling
limit of our theory can be calculated analytically for half filling. 
Scaling our strong coupling result for $\imag \ChiZe(q,\omega)$ down
to \figdep{$U=3$} leads to the \figdep{dashed line} in
Fig.~\ref{fig:ImChi0_U3}.  
For \figdep{$U=3$}, which is lower than the bandwidth and
has thus to be considered as an intermediate interaction, 
$\imag\ChiZe$ of the complete renormalized RPA calculation
(\figdep{continuous line}) qualitatively already resembles the
properly scaled strong coupling result (\figdep{dashed line}).  

\begin{figure}[htbp]
  \begin{center}
    \leavevmode
    \epsfig{file=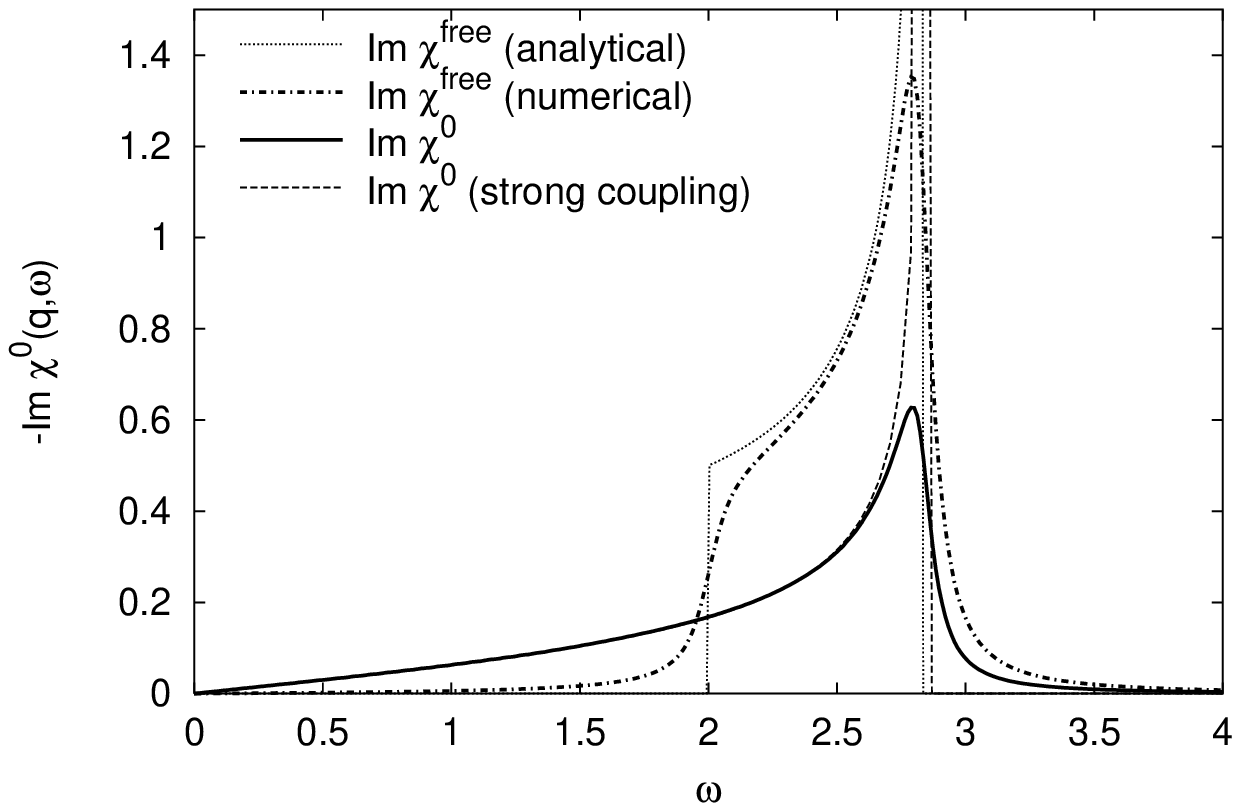,width=13cm}
    \caption{
      Imaginary part of the free and renormalized free susceptibility
      $\ChiZe(q=\pi/2,\omega)$ for the half filled Hubbard chain at $U=6$.}  
    \label{fig:ImChi0_U6}
  \end{center}
\end{figure}

Fig.~\ref{fig:ImChi0_U6} shows $\imag \ChiZe(q,\omega)$ for
\figdep{$U=6$ and half filling}. Comparing the \figdep{continuous} with
the \figdep{dashed line}, we notice that the renormalized RPA result
for $\imag\ChiZe$ now agrees also quantitatively very well with the
corresponding strong coupling result. 
The remaining difference comes from the finite imaginary part
``$i0^{+}$'' used for the numerical computation of
$\imag\ChiZe(q,\omega)$.   

%%%%%%%%%%%%%%%%%%%%%%%%%%%%%%%%%%%%%%%%%%%%%%%%%%
\subsubsection{Real part}
\label{ReRenChiFree}

Let us now examine the real part of the free susceptibility
$\ChiFr(q,\omega)$. As can be seen from the analytic expressions given
in appendix~\ref{ChiFree}, $\real \ChiFr(q,\omega)$ diverges at the two
boundaries of $\imag \ChiFr(q,\omega)$ which demark the particle--hole
continuum. 

The divergence on the upper boundary, given by eq.~(\ref{phExciteB}), makes
the denominator of the pure RPA charge susceptibility 
(see eq.~(\ref{cdwRPASusc})), $1-U\ChiFr(q,\omega)$, 
vanish at an energy above the continuum limit where an undamped plasmon
is created. 

\begin{figure}[htbp]
  \begin{center}
    \leavevmode
    \epsfig{file=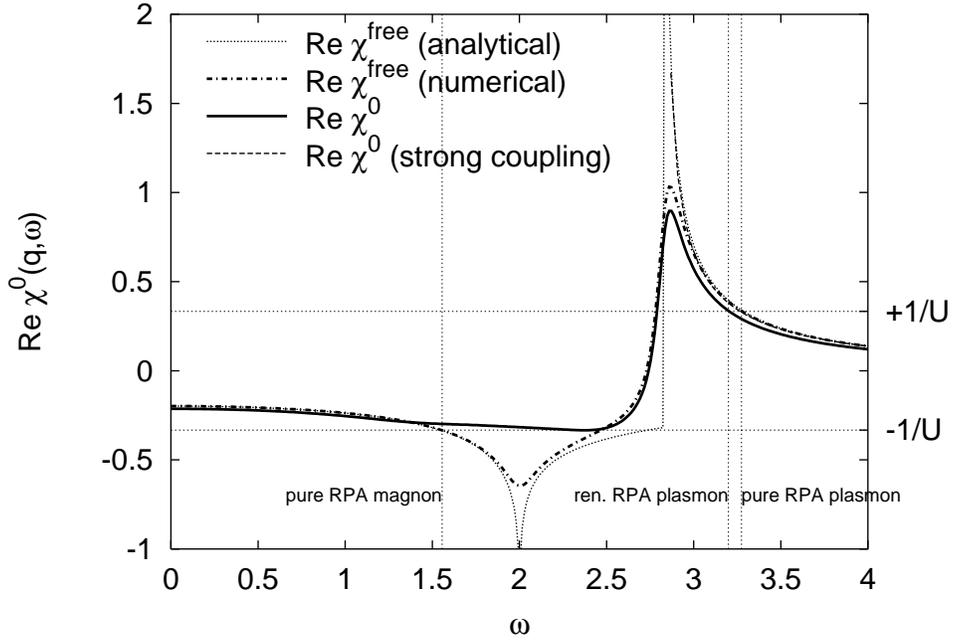,width=13cm}
    \caption{
      Real part of the free and renormalized free susceptibility
      $\ChiZe(q=\pi/2,\omega)$ for the half filled Hubbard chain at $U=3$.
      The intersection with the horizontal lines at $\pm 1/U$ indicate
      where the collective excitations occur (see text). }  
    \label{fig:ReChi0_U3}
  \end{center}
\end{figure}

This is illustrated in Fig.~\ref{fig:ReChi0_U3} by the intersection
of the horizontal line at $+1/U$ with the real part of
$\ChiFr(q,\omega)$ at \figdep{$\omega\approx 3.3$}.
\figdep{The dotted and dot--dashed lines} represent the analytical or
numerical expressions for $\real\ChiFr(q,\omega)$, respectively.  
Hence, as $U$ is increased, the horizontal line at $1/U$ is lowered
and the pure RPA plasmon is shifted towards higher energies.

A similar scenario can be established for the longitudinal spin
response in the pure RPA. Its denominator, $1+U\ChiFr(q,\omega)$,
vanishes at an energy below the lower continuum limit, given by
eq.~(\ref{StepB}), and an undamped magnon is created. Again, this is
shown in Fig.~\ref{fig:ReChi0_U3} by the intersection of the
horizontal line at $-1/U$ with $\real\ChiFr$ 
at \figdep{$\omega\approx 1.6$}. 
As the interaction is increased, the magnon is shifted towards lower
frequencies, and as it reaches $\omega=0$, the system becomes
unstable. This will be discussed in more detail below.
Note that slightly below the upper continuum edge, the
\figdep{dotted line} representing $\real\ChiFr$ also meets the
horizontal line at $-1/U$. However, as $\imag\ChiFr$ is large in this
region near the square--root singularity at the upper continuum
boundary, it will not contribute to the pure RPA spin response. 

\begin{figure}[htbp]
  \begin{center}
    \leavevmode
    \epsfig{file=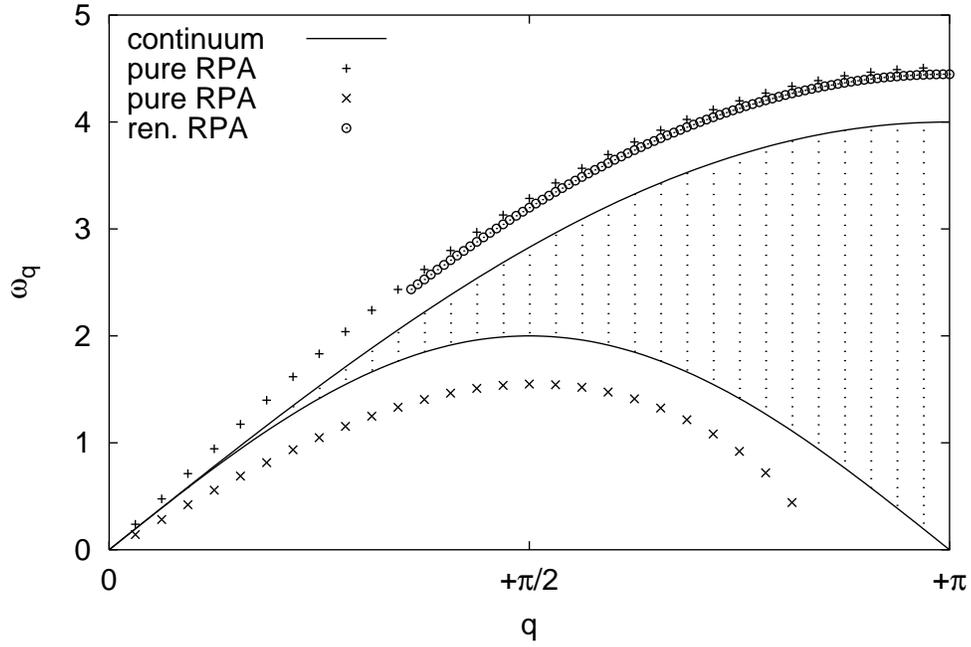,width=13cm}
    \caption{
      Plasmon and magnon dispersion for the half filled Hubbard chain
      at $U=3$. 
      The renormalized RPA plasmons are plotted with circles. 
      ``+''-- and ``x''--symbols stand for the pure RPA plasmons and
      magnons, respectively.
      The dotted area illustrates the non--interacting particle--hole
      continuum. Its boundaries are given by the continuous lines.} 
    \label{fig:plasmon_half}
  \end{center}
\end{figure}

\begin{figure}[htbp]
  \begin{center}
    \leavevmode
    \epsfig{file=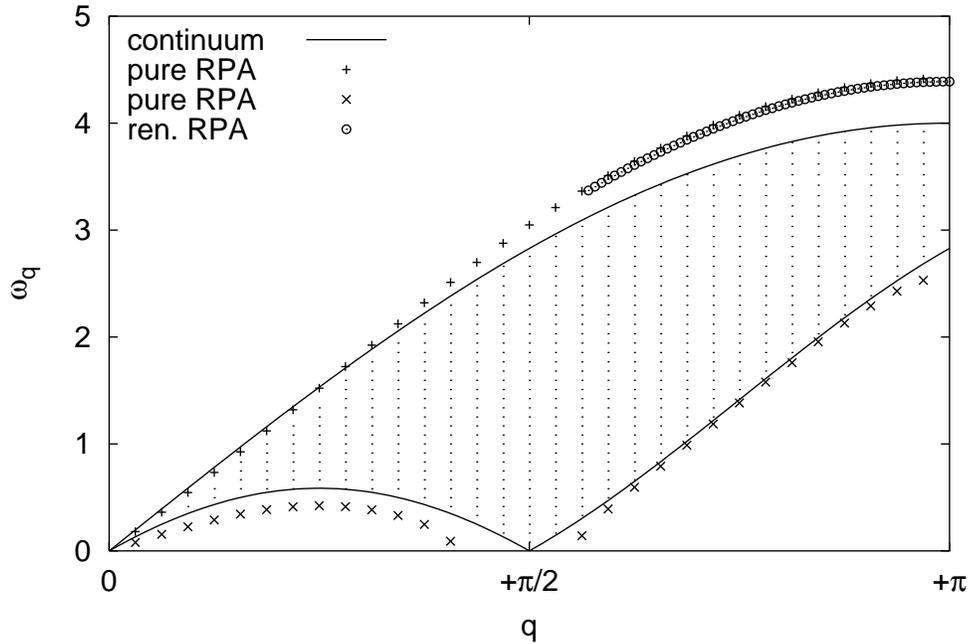,width=13cm}
    \caption{
      Plasmon and magnon dispersion for the quarter filled Hubbard chain
      at $U=3$. Symbols as in Fig.~\ref{fig:plasmon_half}. }
    \label{fig:plasmon_quarter}
  \end{center}
\end{figure}

Fig.~\ref{fig:plasmon_half} and Fig.~\ref{fig:plasmon_quarter} display
the position of the plasmons and the magnons in the
$(q,\omega)$--plane for 
\figdep{$U=3$ and for half and quarter filling, respectively}.  
Once more, we notice that the pure RPA plasmons lie above the
particle--hole continuum, represented by the \figdep{dotted area}. The
pure RPA magnons occur below the lower continuum boundary. As the
lower boundary goes to $\omega=0$ for $q\to 2 k_{F}$, the spin pole
will meet the momentum axis near $2 k_{F}$ producing the well--known
Peierls instability.  
This instability occurs in the pure RPA for infinitesimal $U$ exactly
at $q=2 k_{F}$. 
When increasing the interaction, the region of instability is
enlarged covering a growing interval around $q=2 k_{F}$.

\revised{
It extends to the whole $q$--axis when $U$ exceeds some critical
interaction $U^{\rm Stoner}$, provided by Stoner's mean--field
theory\cite{Stoner}. 
In mean field, the paramagnetic state,
$m=\mean{n_{\uparrow}}-\mean{n_{\downarrow}}=0$, is always a local
extremum in the energy surface as a function of the magnetization $m$.
This extremum is a minimum for
\begin{equation}
  \label{UcMF}
  U\,<\,U^{\rm Stoner}\,=\,2\pi\sin\frac{\pi n}{2},
\end{equation}
with $n$ denoting the number of electrons per site.  
For $U>U^{\rm Stoner}$ it is a maximum. Nevertheless, it should be
remembered that there is a range of interactions 
$U^{\rm MF}<U<U^{\rm Stoner}$ for which the paramagnetic state is a
local but not the global minimum, and that the global minimum is
reached in the fully ferromagnetic state $n=m$.}

We thus conclude that the pure RPA only produces valid results in the
region where no instabilities occur. Calculating quantities having a
contribution from the unstable region around $q=2 k_{F}$, like
e.g. the occupation numbers, does not make sense even for small
interactions. Above $U^{\rm Stoner}$, the pure RPA is unstable for all
$q$ and $\omega$. 
As we will see below, it is the virtue of the renormalized RPA (and
also the SCRPA) to cure these instabilities, rendering possible the
computation of the occupation numbers from eq.~(\ref{OccSpectralTheorem}).  
These arguments will be underlined in section~\ref{HubbardESumRule1d}
by considering the energy weighted sum rule.

Above the outer continuum limit~(\ref{phExciteB}), the real part of
the renormalized free susceptibility, $\real\ChiZe(q,\omega)$, behaves
qualitatively like $\real\ChiFr(q,\omega)$. 
In the same manner as above, we find an undamped plasmon at the energy
where the denominator of the renormalized charge susceptibility in
eq.~(\ref{cdwRPASusc}), $1-U\ChiZe(q,\omega)$, vanishes.  
For small $U$, the plasmon is only slightly shifted to lower energies
with respect to the pure RPA plasmon. 
For larger $U$, the renormalization effects are stronger.
We will see in section~\ref{LargeU} that the frequencies of the
renormalized plasmons remain finite as $U$ goes to infinity. 
The pure RPA plasmons, in contrast, occur at infinite frequencies in
the limit of large $U$. 

Below the outer continuum limit~(\ref{phExciteB}), the
renormalization effects in $\real\ChiZe(q,\omega)$ are more drastic. 
This is shown by the \figdep{continuous line} in
Fig.~\ref{fig:ReChi0_U3} for \figdep{$U=3$ and half filling}. 
The singularity of $\real\ChiFr$ at the lower continuum boundary 
(\figdep{$\omega\approx 2$}) is completely damped and $\real\ChiZe$ is
an almost structure--less function within the renormalized continuum,
i.e. from $\omega=0$ up to almost the continuum edge~(\ref{phExciteB})
at \figdep{$\omega\approx 2.8$}. 

This gives, even for small interactions, rise to qualitative changes
in the longitudinal spin response. 
At the energy where the denominator of the longitudinal spin
susceptibility, $1+U\ChiZe(q,\omega)$, becomes resonant, the imaginary
part of $\ChiZe(q,\omega)$ is large.  
Therefore, the renormalized spin response will show a broad maximum
instead of the undamped magnon found in the pure RPA spin response. 

Apart from the qualitative differences expected between the pure and
the renormalized RPA spin responses, there are important consequences
for the regime where the pure RPA is unstable.
We recall that in this $q$--range, the pure RPA magnon frequency
becomes imaginary and the magnon pole in the pure RPA spin response
disappears.  
As one consequence, we will outline in section~\ref{HubbardESumRule1d}
that the energy weighted sum rule for the pure RPA spin response will
be violated.

In contrast, the broad maximum in the renormalized RPA spin response
will persist even for the $q$--vectors where the pure RPA is unstable.
Therefore, the renormalized RPA spin response fulfills the energy
weighted sum rule for every $q$, even for interactions $U>U^{\rm Stoner}$. 
This will be explained in more detail in section~\ref{HubbardESumRule1d}.
By this means, the Peierls instability occurring in the pure RPA is
cured by the renormalization process.

In the limit of large interactions, the real part of $\ChiZe$ becomes
completely flat within the renormalized particle--hole continuum, and 
$1+U\real\ChiZe(q,\omega)$ vanishes within the {\em whole} continuum. 
As $\imag\ChiZe(q,\omega)$ is smallest within the continuum for
$\omega\to 0$, the broad maximum in the longitudinal spin response is
shifted towards zero frequency. The consequences are discussed in
chapter~\ref{RPALongSpinResponse}. 

We finally remark that even for small $U$, the renormalization
effects are strong enough to lock $\real\ChiZe(q,\omega)$ to values
greater or equal $-1/U$ (see Fig.~\ref{fig:ReChi0_U3}). 
Consequently, the scenario given above is valid, and the renormalized
spin response shows already for weak interaction strengths a broad
continuum peak rather than a sharp magnon pole.

%%%%%%%%%%%%%%%%%%%%%%%%%%%%%%%%%%%%%%%%%%%%%%%%%%
\subsection{Charge response}
\label{RPAChargeResponse}

Due to the rather small renormalization effects of $\ChiZe(q,\omega)$
outside the particle--hole continuum, we expect the charge response in
the pure and in the renormalized RPA to behave similarly.

For intermediate interactions, the pure RPA charge response is given
by a rather small continuum which is limited by the two boundary
conditions~(\ref{phExciteB}) and (\ref{StepB}), and an undamped
plasmon lying above the continuum. 
This is illustrated for \figdep{$U=3$ and half filling} by the
\figdep{dotted line} in Fig.~\ref{fig:ImChiC_U3}, where we used the 
analytical expressions for $\ChiFr$, given in appendix~\ref{ChiFree},
to evaluate the pure RPA charge response.
Evaluating the pure RPA charge response numerically, i.e. with a
finite value of ``$i0^{+}$'' as discussed in
section~\ref{NumericalMethod}, leads to the \figdep{dot--dashed line}
in Fig.~\ref{fig:ImChiC_U3}. 
We observe that, due to the finite ``$i0^{+}$'', the pure RPA plasmon
at \figdep{$\omega\approx 3.3$} is broadened with respect to the
analytical curve, and that the sharp cutoff at the continuum
boundaries is smoothened. 

Comparing the \figdep{continuous line with the dot--dashed line} in
Fig.~\ref{fig:ImChiC_U3} shows that the renormalized charge response
agrees essentially with the pure RPA predictions. 
As explained in section~\ref{ReRenChiFree}, the main differences are
that the tail of the renormalized charge continuum now goes down to
$\omega=0$ and that the plasmon is slightly shifted towards
lower energies. The precise position of the renormalized plasmon is
indicated in Fig.~\ref{fig:ImChiC_U3} by a 
\figdep{continuous vertical line} in the center of the numerically 
computed plasmon pole which, for numerical reasons, has a finite
width. 

Qualitatively, the charge response for $U=3$ is already very similar
to the properly scaled strong coupling limit of our theory. This can
be seen by comparing \figdep{the dashed and the continuous lines}
in Fig.~\ref{fig:ImChiC_U3}. 
Moreover, the plasmon position of the pure RPA and the strong coupling
limit of the renormalized RPA agree very well, such that
the \figdep{dotted and the dashed vertical lines} cannot be resolved
from another. Note, however, that this agreement is purely accidental.

The renormalization effects become more drastic as the interaction
strength is increased. For \figdep{half filling and $U=6$}
Fig.~\ref{fig:ImChiC_U6} illustrates that the renormalized charge
continuum now not only goes down to $\omega=0$ 
(\figdep{continuous line}) but also is much stronger than the pure RPA
charge continuum (\figdep{dot--dashed line}). 
Further, the energy shift of the plasmon obtained from renormalization
is larger than in the weak coupling limit.   

For \figdep{$U=6$}, the charge response already agrees quantitatively
very well with the strong coupling limit of our theory. 
Comparing the \figdep{continuous and the dashed line} shows that both,
the continuum contributions and the positions of the plasmon peaks are
in good agreement.
In section~\ref{LargeU} we will discuss in detail that the
renormalization prevents the plasmon pole in the large $U$ limit from
being shifted to infinite frequencies as this happens for the pure RPA
plasmon. 

Similar results are obtained away from half filling.

%%%%%%%%%%%%%%%%%%%%%%%%%%%%%%%%%%%%%%%%%%%%%%%%%%
\subsection{Longitudinal spin response}
\label{RPALongSpinResponse}

According to the discussion of the real parts of $\ChiFr(q,\omega)$
and $\ChiZe(q,\omega)$ in section~\ref{ReRenChiFree}, we expect the
spin response to change qualitatively when passing from pure to
renormalized RPA, even in the regime of weak and intermediate
coupling. 

\begin{figure}[htbp]
  \begin{center}
    \leavevmode
    \epsfig{file=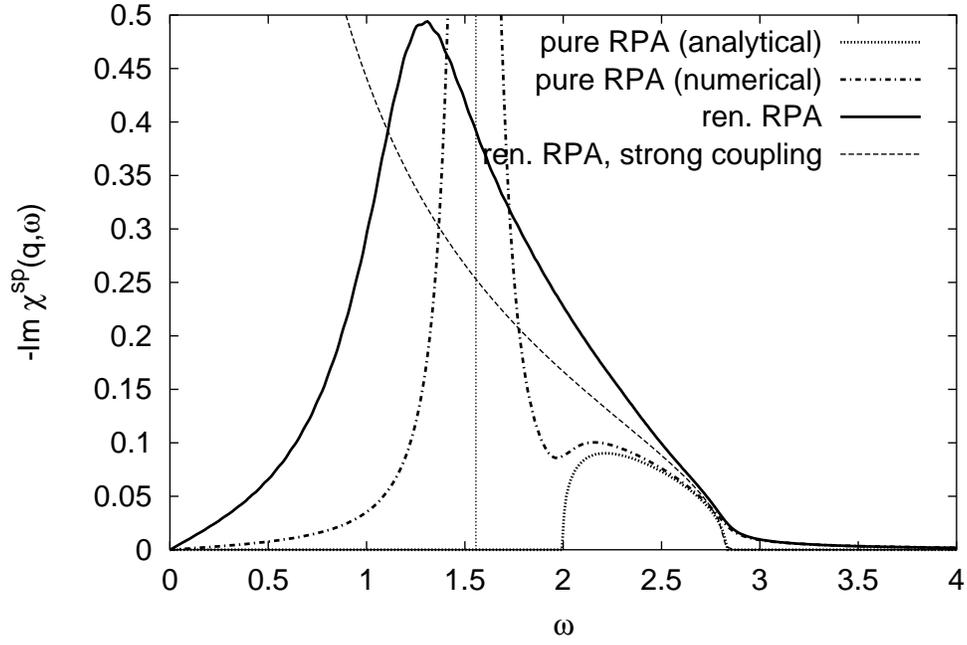,width=13cm}
    \caption{
      Spin response $\ChiSp(q=\pi/2,\omega)$ in pure and
      renormalized RPA for the half filled Hubbard chain at $U=3$.
      The vertical line indicates the pure RPA magnon peak.} 
    \label{fig:ImChiS_U3}
  \end{center}
\end{figure}

This is illustrated in Fig.~\ref{fig:ImChiS_U3} 
for \figdep{$q=\pi/2$, $U=3$ and half filling}. 
For these values there are already significant changes, although the
pure RPA is still stable. 
The pure RPA spin response (\figdep{dot--dashed line}) consists of a
rather small continuum being restricted to the area in between the two
boundaries given by eqs.~(\ref{phExciteB}) and (\ref{StepB}), and an
undamped magnon occurring below this continuum.

As explained before, the broadening of the magnon is of numerical
origin. The corresponding analytic expression can be obtained using
the representation of $\ChiFr$ from appendix~\ref{ChiFree}. 
The latter leads to the \figdep{dotted line} in Fig.~\ref{fig:ImChiS_U3}.

It was argued in section~\ref{ReRenChiFree} that the magnon pole
disappears during the renormalization procedure.
Hence, the renormalized RPA spin response is fully described by a
continuum exhibiting a broad peak (\figdep{continuous line}).
Like the renormalized charge continuum, this spin continuum starts at
$\omega=0$ and goes up to the upper continuum boundary~(\ref{phExciteB}).
The \figdep{dashed line} in Fig.~\ref{fig:ImChiS_U3} shows the spin
response in the strong coupling limit of our theory scaled to
\figdep{$U=3$}. 
A comparison with the renormalized RPA result
(\figdep{continuous line}) shows that especially for low frequencies
there are still important differences. Thus, at \figdep{$U=3$} the
strong coupling limit is not yet reached. 

\begin{figure}[htbp]
  \begin{center}
    \leavevmode
    \epsfig{file=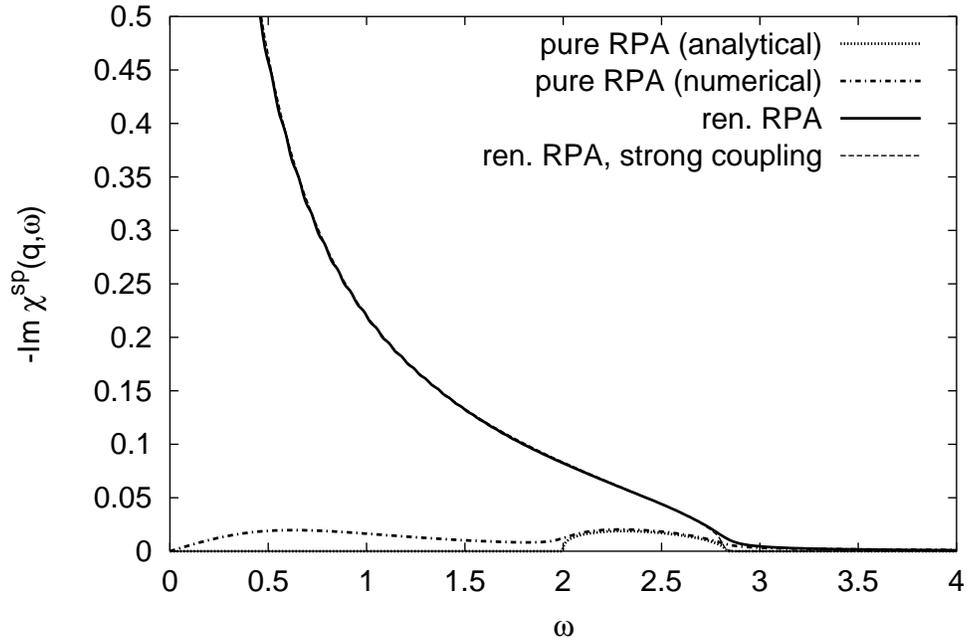,width=13cm}
    \caption{
      Spin response $\ChiSp(q=\pi/2,\omega)$ in pure and
      renormalized RPA for the half filled Hubbard chain at $U=6$.
      The magnon peak, present in Fig.~\ref{fig:ImChiS_U3},
      has vanished due to the instability of the pure RPA (see text). }
    \label{fig:ImChiS_U6}
  \end{center}
\end{figure}

If $U$ is increased or if the $q$--vector is chosen in the domain
around $2 k_{F}$ where the pure RPA is unstable, the pure RPA magnon
frequency will become purely imaginary.  
Fig.~\ref{fig:ImChiS_U6} shows this case for 
\figdep{$U=6$, $q=\pi/2$ and half filling}. 
The pure RPA spin response is represented by 
the \figdep{dot--dashed and the underlying dotted line}, depending on
whether the numerical or the analytical expression is monitored.
We notice that the magnon peak in the pure RPA spin response 
vanishes completely. This corresponds to an unphysical situation,
as will be underlined in section~\ref{HubbardESumRule1d} by sum rule
arguments.

In renormalized RPA, the broad maximum in the longitudinal spin
response is shifted towards zero frequency when the interaction is
increased. 
This can be seen by comparing the \figdep{continuous lines} in
Fig.~\ref{fig:ImChiS_U3} and Fig.~\ref{fig:ImChiS_U6}, which represent
the renormalized spin response for \figdep{$U=3$} and \figdep{$U=6$},
respectively. 
Whereas in Fig.~\ref{fig:ImChiS_U3}, the renormalized spin response
reaches its maximum at a finite frequency, we see from 
Fig.~\ref{fig:ImChiS_U6} that for \figdep{$U=6$} it is strongly peaked
at $\omega=0$. 
This behaviour is characteristic for the strong coupling limit of our
theory, given by the \figdep{dashed line which cannot be resolved from
  the continuous line in the latter graph}.
In renormalized RPA, the energy weighted sum rule is fulfilled for
all $U$, as will be demonstrated in the next subsection. Nevertheless,
we will point out in section~\ref{LargeU} that the singularity of the
spin response at $\omega=0$ may lead to divergences in correlation
functions.  

%%%%%%%%%%%%%%%%%%%%%%%%%%%%%%%%%%%%%%%%%%%%%%%%%%
\subsection{Energy weighted sum rule}
\label{HubbardESumRule1d}

In section~\ref{HubbardEnergySumRule} a sum rule is derived for the
energy weighted spin and charge response. 
In the one--dimensional Hubbard model, eq.~(\ref{HubbSuscRule})
connects an energy weighted integral over the response functions to
the mean kinetic energy per site, $\mean{\hat{t}}$, times a form
factor:
\begin{eqnarray}
  \label{HubbSuscRule1d} 
  S^{\mathrm ch}_{1}(q) : \;
  -2\,\mean{ \hat{t} }\,\left( 1-\cos q \right)
  & = &  -\frac{1}{\pi} \int
  \limits_{-\infty}^{\infty} {\rm d}\omega \; 
  \omega \; \imag \ChiCh(q,\omega)
  \nonumber \\
  S^{\mathrm sp}_{1}(q) : \; 
  -\sfrac{1}{2}\,\mean{ \hat{t} }\,\left( 1-\cos q \right)
  & = &  -\frac{1}{\pi} \int
  \limits_{-\infty}^{\infty} {\rm d}\omega \; 
  \omega \; \imag \ChiSp(q,\omega)
  \; \mbox{.}
\end{eqnarray}

In section~\ref{HubbardEnergySumRule} it is argued that it not only
holds true for the exact response functions $\ChiCh(q,\omega)$ and
$\ChiSp(q,\omega)$ but as well for the response functions in
Self--Consistent and in renormalized RPA if the mean kinetic energy
per site, $\mean{\hat{t}}$, is calculated self--consistently from the
corresponding Green's function. 

Moreover, it is well known that the sum rule for the pure RPA
response functions is fulfilled if $\mean{\hat{t}}$ is calculated with
the Hartree--Fock ground--state 
{\em as long as no instability occurs}\cite{RS,Thouless}.

\begin{figure}[htbp]
  \begin{center}
    \leavevmode
    \epsfig{file=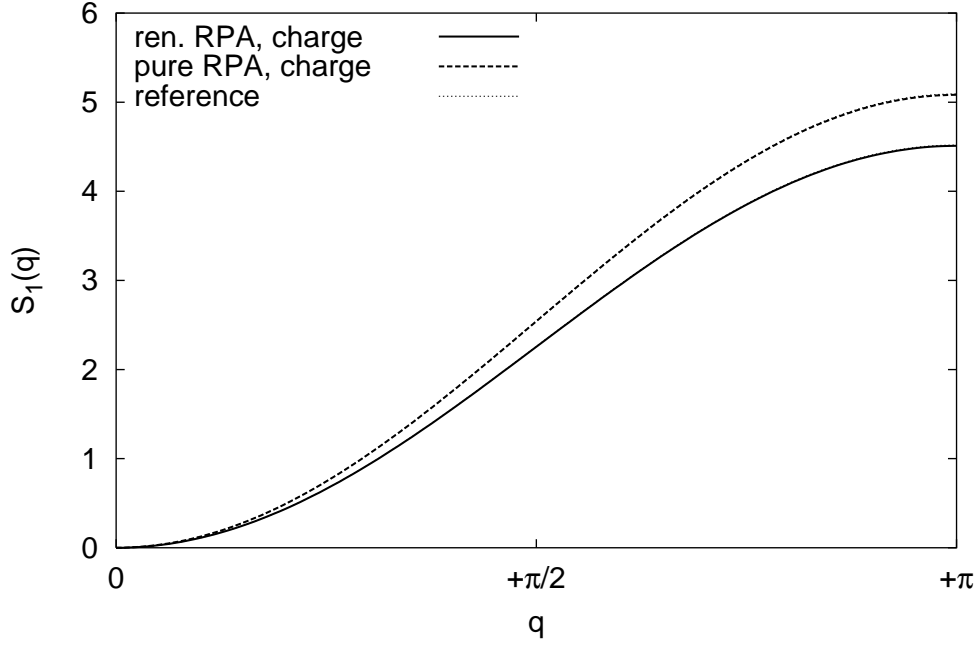,width=13cm}
    \caption{\revised{
      Energy weighted sum rule for the renormalized and pure RPA
      charge susceptibility for the half filled Hubbard chain at
      $U=3$. 
      The dashed and continuous line correspond to the rhs. of
      the charge sum rule $S^{\mathrm ch}_{1}(q)$,
      eq.~(\ref{HubbSuscRule1d}), computed with the pure and
      renormalized RPA charge susceptibility, respectively. 
      The left hand side of the sum rule is plotted with the thin
      dotted ``reference'' lines. 
      As the sum rule is fulfilled in both cases, these reference
      lines cannot be resolved from the corresponding continuous or
      dashed lines.}}
    \label{fig:ESumRuleCh_half}
  \end{center}
\end{figure}

\begin{figure}[htbp]
  \begin{center}
    \leavevmode
    \epsfig{file=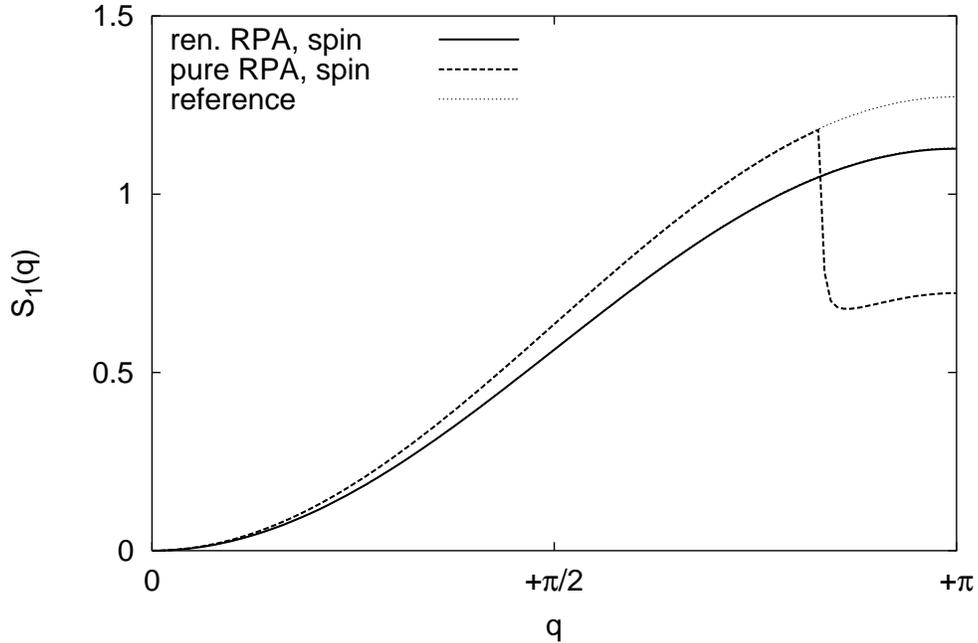,width=13cm}
    \caption{\revised{
      Energy weighted sum rule for the longitudinal spin
      susceptibility in renormalized and in pure RPA for the half
      filled Hubbard chain at $U=3$. 
      The dashed and continuous line correspond to the rhs. of
      the spin sum rule $S^{\mathrm sp}_{1}(q)$,
      eq.~(\ref{HubbSuscRule1d}), computed with the pure and
      renormalized RPA spin susceptibility, respectively. 
      The left hand side of the sum rule is plotted with the thin
      dotted ``reference'' lines. 
      As the sum rule for the renormalized RPA is fulfilled, the
      reference line cannot be resolved from the continuous line.
      For the pure RPA, there is a region around $2 k_{F}$ where the
      sum rule breaks down.}}
    \label{fig:ESumRuleSp_half}
  \end{center}
\end{figure}

\begin{figure}[htbp]
  \begin{center}
    \leavevmode
    \epsfig{file=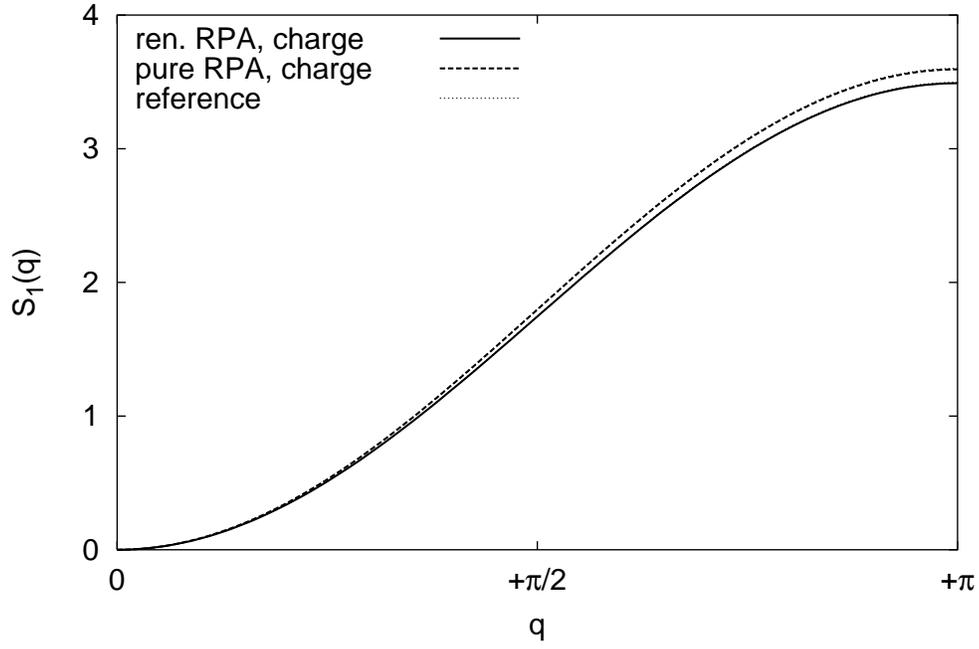,width=13cm}
    \caption{\revised{
      Energy weighted sum rule for the renormalized and pure RPA
      charge susceptibility for the quarter filled Hubbard chain at
      $U=3$. 
      Description as in Fig.~\ref{fig:ESumRuleCh_half}. }}
    \label{fig:ESumRuleCh_quarter}
  \end{center}
\end{figure}

\begin{figure}[htbp]
  \begin{center}
    \leavevmode
    \epsfig{file=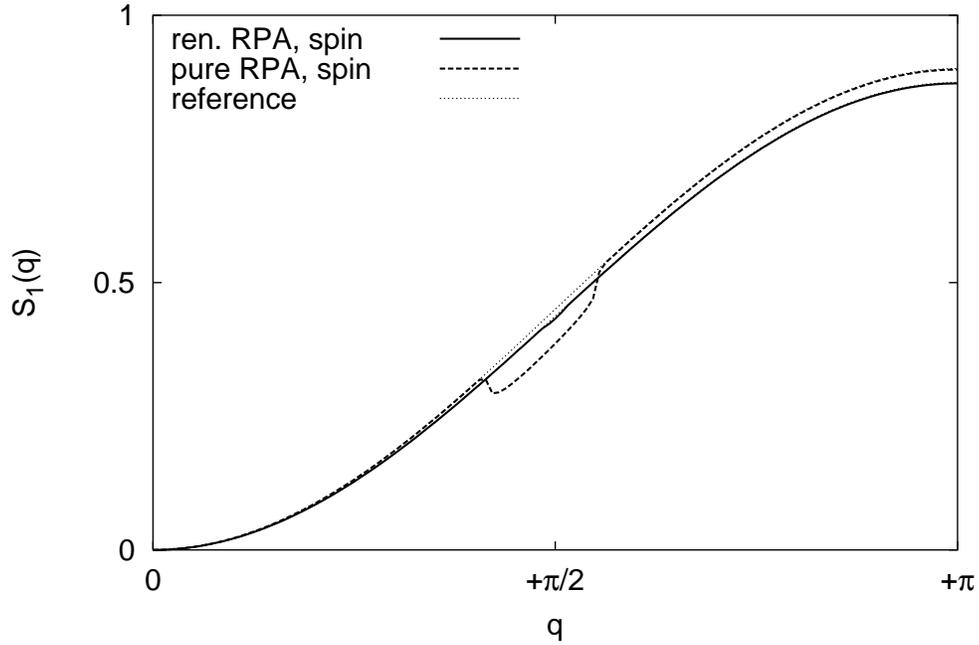,width=13cm}
    \caption{\revised{
      Energy weighted sum rule for the longitudinal spin
      susceptibility in renormalized and in pure RPA for the quarter 
      filled Hubbard chain at $U=3$.
      Description as in Fig.~\ref{fig:ESumRuleSp_half}. }}
    \label{fig:ESumRuleSp_quarter}
  \end{center}
\end{figure}

\revised{
In Fig.~\ref{fig:ESumRuleCh_half}, \ref{fig:ESumRuleSp_half},
\ref{fig:ESumRuleCh_quarter} and \ref{fig:ESumRuleSp_quarter}
we show the result of the sum rule checks for \figdep{$U=3$}, and for
\figdep{half and quarter filling, respectively}.
In all figures, the left hand sides of the eqs.~(\ref{HubbSuscRule1d})
are plotted with the \figdep{dotted lines}. 
As their $(1-\cos q)$--behaviour is independent of the approximations
made in the Green's function, we will consider them as ``reference lines''.  
Nevertheless, they are scaled with a prefactor, $\mean{\hat{t}}$, which
depends on the Green's function. 
In renormalized RPA, the mean kinetic energy per site is less negative
than in Hartree--Fock, since the momentum distribution is smoothened.
Thus, the reference lines for the renormalized RPA sum rules lie
always below the pure RPA lines.

For the {\em renormalized} RPA susceptibilities, the right hand sides of the
eqs.~(\ref{HubbSuscRule1d}) are represented by the \figdep{continuous lines}. 
They cannot be resolved from the corresponding 
\figdep{dotted reference lines}. 
Hence, in renormalized RPA the sum rule is fulfilled for all
$q$--vectors and for both, the charge and the longitudinal spin response. 

Calculating the rhs. of the eqs.~(\ref{HubbSuscRule1d}) with the 
{\em pure} RPA susceptibilities yields the \figdep{dashed lines}.
The sum rule for the charge response is monitored in the figures
\ref{fig:ESumRuleCh_half} and \ref{fig:ESumRuleCh_quarter}. 
There, the \figdep{dashed lines} cannot be resolved from their
\figdep{dotted reference lines}. This means, that the sum rule for the
pure RPA charge response holds true for all $q$--vectors. 

In the figures \ref{fig:ESumRuleSp_half} and
\ref{fig:ESumRuleSp_quarter}, we show the sum rule check for the pure
RPA spin response. The \figdep{dashed lines}, which represent the
energy weighted pure RPA spin response, agree with their reference
lines, apart from a range of $q$--vectors around the Peierls vector 
$2 k_{F}$. This $q$--range coincides with the $q$--range where the
corresponding magnon dispersion in Fig.~\ref{fig:plasmon_half} and
Fig.~\ref{fig:plasmon_quarter}, respectively, goes to $\omega=0$.
Therefore, the energy weighted sum rule for the pure RPA spin
susceptibility is only fulfilled for the range of momenta where the
pure RPA is stable. }

Again, we want to emphasize the quality of the renormalized and the
Self--Consistent RPA to restore the $f$--sum rule for {\em all} 
momenta and interactions.
  
%%%%%%%%%%%%%%%%%%%%%%%%%%%%%%%%%%%%%%%%%%%%%%%%%%
\subsection{The large $U$ limit}
\label{LargeU}

At half filling and for large $U$, the occupation numbers predicted by
renormalized RPA can be fitted very accurately by
\begin{equation}
  \label{nk_UinfRPA}
  n_{k\sigma} \, = \, \frac{1}{2}\,
  \left( 1 \, + \, \frac{4}{U} \cos{k} \right)
  \; \mbox{,}
\end{equation}
whereas the large $U$ expansion of the Bethe ansatz solution
yields\cite{CarmeloBaeriswyl,BaeriswylLinden}
\begin{equation}
  \label{nk_UinfBethe}
  n_{k\sigma} \, = \, \frac{1}{2}\,
  \left( 1 \, + \, \frac{8\ln 2}{U} \cos{k} \right)
  \; \mbox{.}
\end{equation}
Comparing the \figdep{dotted line and the lowest continuous line} in
Fig.~\ref{fig:nk_half} shows that for \figdep{$U=5$}, the
expression~(\ref{nk_UinfRPA}) agrees already very well with the
numerical results for $n_{k}$. We will show in the following that
eq.~(\ref{nk_UinfRPA}) indeed is the fully self--consistent solution
of the renormalized RPA equations in the strong coupling limit.

Based on eq.~(\ref{nk_UinfRPA}), it is possible to give an explicit
expression for the renormalized free susceptibility. As the continuum
boundary condition for step--like occupation numbers,
eq.~(\ref{StepB}), becomes meaningless in the limit of strong
interactions, the only characteristic energy is provided by the upper
continuum limit~(\ref{phExciteB}).
Hence, the explicit expressions for $\ChiZe(q,\omega)$ do not depend
on $q$ and $\omega$ as independent variables anymore, but can be
denoted as a function of one single variable $\xi$, which is the
energy in units of the upper continuum limit~(\ref{phExciteB}):
\begin{equation}
  \label{xi}
  \xi \, = \, \frac{\omega}{\abs{4\sin\frac{q}{2}}}
\end{equation}

Performing the integration in eq.~(\ref{phChiZero}) with occupation
numbers as given by eq.~(\ref{nk_UinfRPA}) yields the renormalized
free susceptibility 
\begin{equation}
  \label{ChiZeroInf}
  \ChiZe(\xi) \, = \, -\frac{1}{U}
  \cases{ 
    1 + i\frac{\xi     }{\sqrt{1-\xi^2}} & \mbox{for $\abs{\xi}<1$} \cr
    1 - \frac{\abs{\xi}}{\sqrt{\xi^2-1}} & \mbox{for $\abs{\xi}>1$} }
  \mbox{.}
\end{equation}
Like for the free susceptibility $\ChiFr$, the imaginary part of
$\ChiZe$ still has a square root singularity at the continuum limit
$\abs{\xi}=1$, whereas the real part of $\ChiZe$ is completely flat
within the particle--hole continuum. 
By scaling the renormalized free susceptibility of the strong coupling
limit, eq.~(\ref{ChiZeroInf}), to \figdep{$U=3$ or $U=6$} we obtain
the \figdep{dashed lines} in Figs.~\ref{fig:ImChi0_U3},
\ref{fig:ImChi0_U6} and \ref{fig:ReChi0_U3}.

Substituting eq.~(\ref{ChiZeroInf}) in eq.~(\ref{cdwRPASusc}) yields
the renormalized RPA response functions. 
The large $U$ charge response is characterized by a small continuum
within $\abs{\xi}<1$,
\begin{equation}
  \label{ImChiChInf}
  \imag \ChiCh(\xi)  \, = \, -\frac{2}{U} \;
  \frac{\xi\,\sqrt{1-\xi^2}}{4\,-3\xi^2}
  \; \mbox{.}
\end{equation}
Outside the particle--hole continuum, the charge response has a
collective pole at $\xi=2/\sqrt{3}$, describing a undamped plasmon
obeying the dispersion relation 
\begin{equation}
  \label{PlasmonDispInf}
  \Omega_{p} \,=\, \frac{8}{\sqrt{3}} \, \abs{\sin\frac{q}{2}}
  \; \mbox{.}
\end{equation}
Notice that the dispersion relation of the renormalized plasmon
becomes independent of $U$ in the large $U$ limit.
The pure RPA, in contrast, predicts in this limit a plasmon dispersion
proportional to $U$. 
The strong coupling charge response scaled to the appropriate $U$ is
represented in Fig.~\ref{fig:ImChiC_U3} and Fig.~\ref{fig:ImChiC_U6}
by the \figdep{dashed lines}.

From the second of the eqs.~(\ref{cdwRPASusc}), in combination with the
renormalized free susceptibility~(\ref{ChiZeroInf}), we obtain the
large $U$ spin response in renormalized RPA.  
It is fully described by a continuum for $\abs{\xi}<1$,
\begin{equation}
  \label{ImChiSpInf}
  \imag \ChiSp(\xi)  \, = \, -\frac{1}{2 U} \;
  \frac{\sqrt{1-\xi^2}}{\xi}
  \; \mbox{,} 
\end{equation}
and vanishes elsewhere. This is plotted with dashed lines in
Fig.~\ref{fig:ImChiS_U3} for \figdep{$U=3$} and
Fig.~\ref{fig:ImChiS_U6} for \figdep{$U=6$}.
No collective excitations occur, since the denominator of the
renormalized RPA spin response, $1+U\ChiZe$, is always finite. 
Nevertheless, we see from eq.~(\ref{ImChiSpInf}) that the spin
continuum diverges at $\omega=0$ with consequences that we will
discuss below. 

It can be shown that the real and imaginary part of $\ChiZe$ from
eq.~(\ref{ChiZeroInf}) fulfill the Kramers--Kronig relations.
Moreover, we will show in appendix~\ref{LargeUOcc} that the occupation
numbers calculated from the large $U$ expression of the renormalized
RPA Green's function via the spectral theorem (\ref{OccSpectralTheorem})
are consistent with eq.~(\ref{nk_UinfRPA}).
This means that the occupation numbers~(\ref{nk_UinfRPA}) fulfill
together with the strong coupling susceptibilities~(\ref{ImChiChInf})
and (\ref{ImChiSpInf}) the self--consistency condition of the
renormalized RPA. 

The number of double occupancies per site are given by
\begin{equation}
  \label{DoubleOcc}
  \sfrac{1}{N} \sum \limits_{i} n_{i\uparrow} n_{i\downarrow}
  \, = \, \mean{ n_{\uparrow} } \mean{ n_{\downarrow} } 
  \, + \, \frac{w^\corr}{U}
  \; \mbox{,}
\end{equation}
where $w^\corr$ stands for the correlated potential energy per
site, 
\begin{equation}
  \label{wcorr}
  w^\corr \, = \, \frac{U}{N^2} \sum \limits_{k p q} 
  \mean{a^{+}_{k    \uparrow  } a_{k+q\uparrow} \,
        a^{+}_{p+q\downarrow} a_{p    \downarrow} }^\corr
  \; \mbox{.}
\end{equation}
$w^\corr$ is independent of $U$, since the large $U$ Green's
function scales with $1/U$.

Nevertheless, as the expectation value in eq.~(\ref{wcorr}) can be
obtained from the spectral theorem~(\ref{axaaxaSpectralTheorem})
by integrating the renormalized RPA Green's function over positive
frequencies, the $1/\omega$ divergence of the spin
response~(\ref{ImChiSpInf}) makes the expectation value 
$w^\corr$ diverge logarithmically. 
This means that both, the total potential energy per site 
and the number of doubly occupied sites are going to $-\infty$ as
$U\to\infty$.  

One way to correct these deficient results is to calculate the
ground--state energy $E(U)$ from the kinetic energy $T(U)$, using the
Hellmann--Feynman--Theorem\cite{Kittel},  
\begin{equation}
  \label{HellmannFeynmanTheorem}
  E(U) \, = \, U\,\int\limits_{U}^{\infty} {\rm d}y \, \frac{T(y)}{y^2}
  \; \mbox{.}
\end{equation}

The behaviour of the kinetic energy in the renormalized RPA can be
obtained from the occupation numbers~(\ref{nk_UinfRPA}). 
For large $U$, this yields
\begin{equation}
  \label{EkinUinfRPA}
  T(U) \, = \, -\frac{4}{U}
  \; \mbox{.}
\end{equation}
Again, only the prefactor differs from the exact result,
\begin{equation}
  \label{EkinUinfBethe}
  T^{\rm exact}(U) \, = \, -\frac{8\ln 2}{U}
  \; \mbox{,}
\end{equation}
known from the large $U$ expansion of the Bethe ansatz
solutions\cite{CarmeloBaeriswyl,BaeriswylCarmeloLuther}.  

\begin{figure}[htbp]
  \begin{center}
    \leavevmode
    \epsfig{file=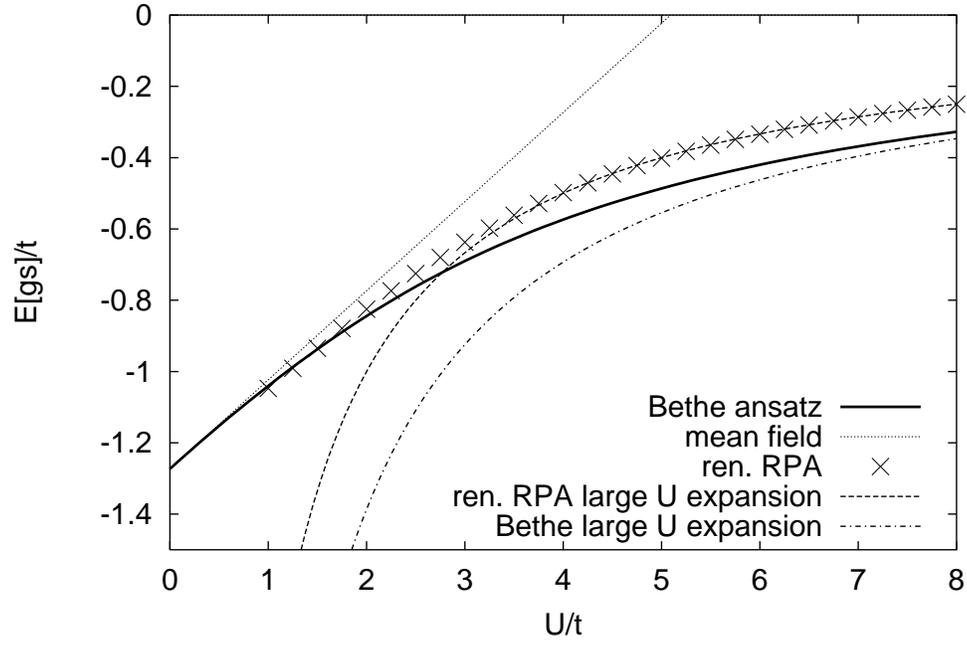,width=13cm}
    \caption{
      Ground state energy per site for the half filled Hubbard chain.
      Bethe ansatz expansion from Baeriswyl et
      al.\cite{BaeriswylCarmeloLuther,CarmeloBaeriswyl}.
      Large $U$ limit of the renormalized RPA from
      eq.~(\ref{EgsUinfRPA}). }
    \label{fig:GSEnergy_half}
  \end{center}
\end{figure}

Calculating the ground--state by eq.~(\ref{HellmannFeynmanTheorem}),
our theory predicts 
\begin{equation}
  \label{EgsUinfRPA}
  E(U) \, = \, -\frac{2}{U} 
\end{equation}
as $U\to\infty$. The large $U$ expansion of the renormalized RPA
ground--state energy of the half filled Hubbard chain is shown in
Fig.~\ref{fig:GSEnergy_half} by the \figdep{dashed line}. 
For arbitrary interactions, the renormalized RPA ground--state energy
calculated with the Hellmann--Feynman theorem is illustrated by the
\figdep{``x''--symbols}. This compares reasonably well with the exact
result, known from Bethe ansatz (\figdep{continuous line}) and its
large $U$ expansion (\figdep{dot--dashed line}). 

\begin{figure}[htbp]
  \begin{center}
    \leavevmode
    \epsfig{file=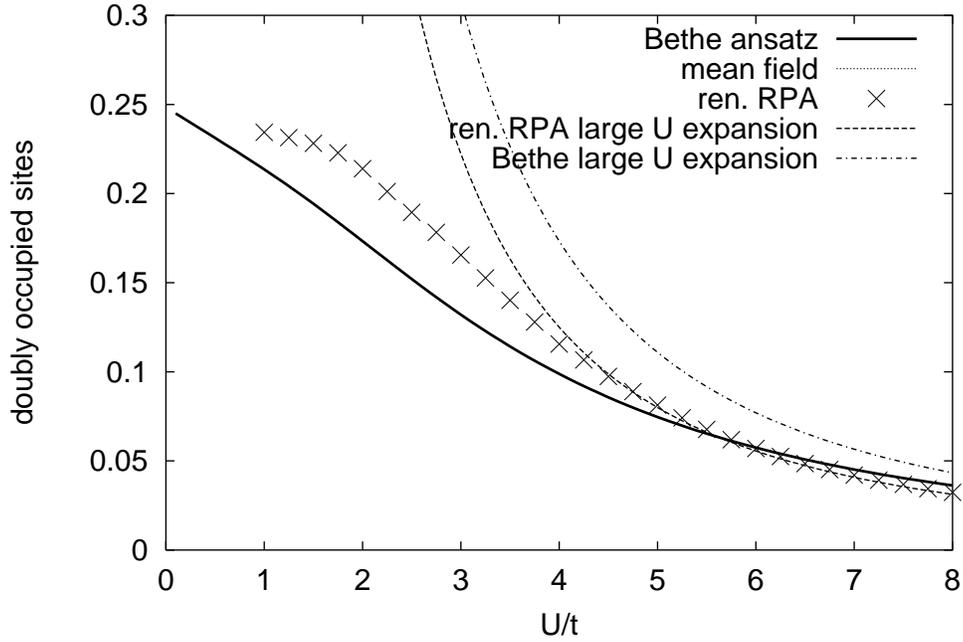,width=13cm}
    \caption{
      Number of double occupancies per site for the half filled
      Hubbard chain.}  
    \label{fig:DoubleOcc_half}
  \end{center}
\end{figure}

We now have access to the potential energy as the difference between
ground--state energy and kinetic energy. It thus behaves as $2/U$ for
large $U$. This implies that the number of double occupancies vanishes
as $2/U^{2}$ for strong interactions. 
In Fig.~\ref{fig:DoubleOcc_half}, the number of double occupancies at
half filling is monitored as a function of $U$. We see that for small
interactions, the renormalized RPA (\figdep{``x''--symbols})
reproduces the exact result (\figdep{continuous line}). 
For large interaction strengths, our theory matches the Bethe ansatz
expansion, $-(4\ln 2)/U^{2}$ (\figdep{dot--dashed line}), apart from
the prefactor. 

% %%%%%%%%%%%%%%%%%%%%%%%%%%%%%%%%%%%%%%%%%%%%%%%%%%%%%%%%%%%%
% Conclusions
% %%%%%%%%%%%%%%%%%%%%%%%%%%%%%%%%%%%%%%%%%%%%%%%%%%%%%%%%%%%%
\section{Discussion, conclusions and outlook}
\label{Conclusions}

In this work, we performed a first application of the Self--Consistent
RPA (SCRPA) theory to the Hubbard model.  
In itself, the SCRPA is an approximation to the general Dyson Equation
Approach (DEA) to correlation functions where the full (exact) mass
operator is replaced by its instantaneous contribution. For the
single particle Green's function this strategy leads to the standard
Hartree--Fock theory. In analogy, it has been argued in the past that
the SCRPA corresponds to a HF theory for fermion pair clusters.
Recently, this theory has produced very interesting results in various
domains of many--body physics\cite{Krueger,Dukelsky}.

Unfortunately, being a (non--linear) mean--field theory for non--local
correlation functions, SCRPA is numerically very demanding. 
As a first step, we therefore had to proceed to further rather drastic
simplifications. The latter consist in retaining correlations 
{\em only} in the single--particle occupation numbers. 
This approximation to SCRPA is known in the literature as 
{\em renormalized} RPA\cite{Rowe,Toivanen}.  
In spite of this, the essentials of the self--consistency and closure
aspects remain intact.  
As a further virtue, the $f$--sum rule is shown to be fulfilled in
Self--Consistent as well as in renormalized RPA.  
This also implies that the Goldstone theorem is fulfilled and
symmetries are treated correctly.

We solved numerically the renormalized RPA equations for the
one--dimensional single--band Hubbard model in the paramagnetic phase,
for different fillings and interactions. 
\revised{Although we were aware of the difficulty of describing one
dimensional models because of the extreme importance of quantum 
correlations\cite{Krueger}, there were multiple reasons for this
choice. 
In a first place, the exact solution of Lieb and Wu\cite{LiebWu}
provides a benchmark for our results, which, in higher dimensions,
does not exist.
As was argued in section~\ref{phGFcharge}, we expect the renormalized
RPA to perform better with increasing dimensionality. Therefore, $1d$
can be considered as a ``worst case check'' for our approximation.
The second reason is mainly technical: The self-consistency equation
for the occupation numbers (\ref{OccSpectralTheorem}) illustrates that
the numerical effort increases with the square of the spatial
dimension. Therefore, the experience in $1d$ is desirable before
attacking higher dimensions.
In the last place, even in one dimension, we are able to test
explicitly the essentials of our method and the convergence of the
iteration cycle.  

As expected, most of the particularities of the one-dimensional model,
as e.g. Luttinger liquid behaviour away from half filling,
could not be reproduced. 
The scenario predicted by our theory rather confirms the results which
were obtained from methods designed for higher dimensional models,
like e.g. the Hubbard-III approximation\cite{HubbardIII,Cyrot}:
For small $U$, we find a Fermi-liquid like metallic ground state. 
The system undergoes a Mott-Hubbard transition for an interaction
slightly smaller than the bandwidth. The exact value of the critical
interaction depends on the filling.
For stronger interactions, our theory predicts an insulator for all
fillings. 

Despite these deficiencies, we obtain several interesting results.}
In the strong coupling regime of the half filled model, for example,
we are able to express the central quantities of our theory
analytically. The renormalized RPA momentum distribution function,
given by eq.~(\ref{nk_UinfRPA}), agrees, apart from a prefactor, with
the $n_{k}\propto\cos k$ behaviour known from the large $U$ expansion
of the Bethe ansatz solution.  

Moreover, the mean--field spin instability around $2 k_{F}$, which
causes the breakdown of the pure RPA for any finite interaction, turns
out to be cured in renormalized RPA in the sense that no more purely
imaginary eigenfrequencies occur in the spin channel.  
On the other hand, the renormalization is still rather weak such that
a strongly overdamped spin pole remains at low energy.
These low--lying spin excitations give rise to a slow (logarithmic)
divergence of the two--body correlation functions, which is
carried into the number of double occupancies and thus also into the 
ground--state energy. The stronger renormalization contained in the
SCRPA would certainly cure this pathology, since it also renormalizes
(screens) the interaction self--consistently. In this sense, the SCRPA
bears some similarities with the approach of Tremblay et al.\cite{Vilk}.

Several of the renormalized RPA results may nonetheless be improved by
applying the Hellmann-Feynman theorem. 
Using this approach, which may simulate a step towards the full
solution of the SCRPA, the ground--state energy compares for all
interaction strengths reasonably well with the exact results, known
from Bethe ansatz.  
Moreover, the number of doubly occupied sites shows a qualitatively
correct behaviour over the whole $U$ range. 
Especially for $U\to 0$, the mean--field value is recovered, and, for
$U\to\infty$, the double occupancies vanish like $1/U^{2}$, as
predicted by the exact solution.   

In the strong coupling regime of the half filled model, this approach 
reproduces the exact results for the momentum distribution function,
the ground--state energy, the kinetic and potential energy, and the
number of doubly occupied sites, apart from a general prefactor.
We would obtain the right prefactor by substituting the bare Hubbard
$U$ with a screened interaction, or, in other words, by multiplying
$U$ in our theory by a factor $1/(2\ln 2)\approx 0.72$.

Since in renormalized RPA the Hubbard interaction is approximated by a
simple spin--flip interaction, it is not astonishing that the best
results are produced at half filling\cite{BaeriswylLinden}. 
Away from half filling, other scattering processes, still taken into
account in SCRPA but neglected in renormalized RPA, become important.
Therefore, the renormalized RPA is less effective. 
The Luttinger liquid behaviour, exhibited by the exact momentum
distribution even for $U\to\infty$\cite{OgataShiba89}, is not
obtained. In the strong coupling limit, the renormalized RPA produces 
smooth occupation numbers, although still having a steeper slope at
$k_{F}$ as at half filling.   
For small interactions, the renormalized RPA still predicts a
discontinuity in the momentum distribution at $k_{F}$. 
This implies that the Luttinger theorem is satisfied.

\revised{A further point to be discussed is the fact that we
completely neglected the dynamical interaction part, 
$\Heff_{A B}^{\rm res}(\omega)$, defined in eq.~(\ref{Heff}).
It is certainly true, as recently pointed out by Logan for the
infinite dimensional Hubbard model\cite{Logan}, that this dynamical
interaction should be taken into account, in order to describe the low
energy scale in the spin channel correctly.  
Nevertheless, in the present work, some dynamical effects in the spin
channel are considered via the coupling of the occupation
numbers to the RPA ground state correlations. A full inclusion of the
dynamical effects goes, however, beyond the scope of this paper.}

At this point, it may be appropriate to shortly come back to some 
technical aspects of the SCRPA which we did not develop in the main
text in order not to overduely extend the size of the paper.
In the introduction, we mentioned that the Equation of Motion Method
(EOM), on which our formalism is based, goes back rather far in
time. The first major theoretical input was developed by D.J.~Rowe in
his review article\cite{Rowe}, where the calculation of
density--density correlation functions is described in the context of
nuclear physics. The method was later applied to strongly correlated
electrons by Roth\cite{Roth}. She evaluated the single--particle
Green's function by coupling it in an approximate way to the three
particle propagator (see also Beenen and Edwards\cite{Beenen} for a
more recent application to the Hubbard model).
Since then, in solid state physics, the EOM has, to the best of our
knowledge, exclusively been used for the calculation of single
particle properties\cite{Kuzemsky}.  

As we point out in ref.\cite{DRS98}, the optimal procedure will be to
combine single--particle and fermion--pair channels. Indeed, as is well
known, the single particle mass operator can be expressed exactly by
the two--particle $T$-matrix\cite{Migdal}. Replacing it by its SCRPA
counterpart then naturally leads to a self--consistent coupling of both
channels.
In this scheme, as we will describe in more detail in a future
publication, the single--particle occupation numbers will not be
evaluated from the particle--hole propagator 
(see eqs.~(\ref{OccSpectralTheoremGen}) and
(\ref{OccSpectralTheorem}), respectively) but directly from the
single--particle Green's function. A further advantage of this
strategy is that single--particle and fermion--pair properties are
obtained simultaneously and on equal footing. We did not follow this
route in this paper, since it again would have strongly increased the
numerical difficulties. In spite of these possible 
improvements, the present investigation shows that the essentials of
the SCRPA theory (i.e. the self--consistency procedure) work correctly
in a numerical application to a homogeneous system of strongly
interacting fermions. Once the method will be solvable in its full
complexity, the possible applications are very numerous. Indeed, the
SCRPA is a very flexible formalism applicable to strongly correlated
Fermi systems, but also to Bose or spin systems. The study of such
systems are planned in the future.

A very appealing application of the SCRPA may be the Hubbard model in
infinite dimensions, since, on one hand, the physics in $d=\infty$ is
expected to be somewhat similar to $d=3$. On the other hand, spatial
fluctuations are suppressed in $d=\infty$ which allows to reduce the
many--body problem to either a dynamical single--site problem, or, to
an effective one dimensional problem. We thus may expect, that the
numerical solution of the SCRPA will be feasible in infinite
dimensions, in contrast to finite dimensions, where the effort will
grow with an exponent $3d$, since the effective Hamiltonian contains
correlation function depending on three momenta.

% %%%%%%%%%%%%%%%%%%%%%%%%%%%%%%%%%%%%%%%%%%%%%%%%%%%%%%%%%%%%
% Acknowledgments
% %%%%%%%%%%%%%%%%%%%%%%%%%%%%%%%%%%%%%%%%%%%%%%%%%%%%%%%%%%%%
\acknowledgments

We are specially grateful to Florian Gebhard for many interesting
discussions and stimulating comments, and the careful reading of the
manuscript. 
\revised{We also thank Jorge Dukelsky, Mireille Lavagna and Wilhelm
Brenig} for criticism and many helpful comments.

% %%%%%%%%%%%%%%%%%%%%%%%%%%%%%%%%%%%%%%%%%%%%%%%%%%%%%%%%%%%%
% APPENDIX
% %%%%%%%%%%%%%%%%%%%%%%%%%%%%%%%%%%%%%%%%%%%%%%%%%%%%%%%%%%%%
\begin{appendix}

% %%%%%%%%%%%%%%%%%%%%%%%%%%%%%%%%%%%%%%%%%%%%%%%%%%%%%%%%%%%%
% free susceptibility
% %%%%%%%%%%%%%%%%%%%%%%%%%%%%%%%%%%%%%%%%%%%%%%%%%%%%%%%%%%%%
\section{Derivation of the SCRPA from a variational principle}
\label{Variational}

\revised{
In this section, we will briefly outline the derivation of the
particle-hole SCRPA equations from a variational principle. 
An analogous method was derived by Baranger\cite{Baranger} for the
single particle case.

Let us therefore consider the spectral representation of the retarded
Green's function
\begin{equation}
  \label{GSpectral}
  \GFome{X}{X^{+}}^{\rm ret}
  \; = \;
  \sum\limits_{n}
  \left[ \,
    \frac{ \expect{0}{X}{n} \expect{n}{X^{+}}{0} }{\omega-\omega_{n 0}+i0^{+}}
    \, -\epsilon \,
    \frac{ \expect{0}{X^{+}}{n} \expect{n}{X}{0} }{\omega+\omega_{n 0}+i0^{+}}
    \,
  \right]
  \; \mbox{,}
\end{equation}
where $\ket{0}$ is the exact ground state of $H$, and $\omega_{n 0}$
denotes the excitation energy for the exact eigenstates $\ket{n}$.
For the particle-hole problem, we set $\epsilon=+1$. The excitation
operators $X^{+}$ are given by
\begin{equation}
  \label{phExcitOp}
  X^{+} \; = \; \sum\limits_{k p} \, x_{k p} \, a^{+}_{k} a_{p}
  \; \mbox{.}
\end{equation}

The normalized mean excitation energy is given by:
\begin{eqnarray}
  \label{VarSumRule}
  S_{1} & = &  
  \frac{ \sum\limits_{n} \, \omega_{n 0} \,
    \left( 
      \abs{\expect{n}{X^{+}}{0}}^{2} \,-\, \abs{\expect{n}{X}{0}}^{2} 
    \right) }
  { \sum\limits_{n} 
    \left( 
      \abs{\expect{n}{X^{+}}{0}}^{2} \,-\, \abs{\expect{n}{X}{0}}^{2} 
    \right) }
  \nonumber \\
  & = &
  \frac{ \int \limits_{-\infty}^{\infty} {\rm d}\omega \; 
    \omega \, \imag \GFome{X}{X^{+}}^{\rm ret} }
  {      \int \limits_{-\infty}^{\infty} {\rm d}\omega \; 
    \imag \GFome{X}{X^{+}}^{\rm ret} }
\end{eqnarray}
The equivalence between the first and the second line can be seen by
substituting the spectral representation (\ref{GSpectral}) in the
second line.
The denominator can easily be evaluated with the spectral
theorem~(\ref{RetSpectralTheorem}). This yields the norm  
$\expect{0}{\comm{X}{X^{+}}}{0}$, which would be equal to unity if
$X^{+}$ were ideal bose operators.

We now minimize eq.~(\ref{VarSumRule}) with respect to the excitation
operators. The $X^{+}$ with the lowest mean excitation energy obey the
condition: 
\begin{equation}
  \label{VarXmin}
  \frac{\partial S_{1}}{\partial x_{k p}} \; = \; 0
\end{equation}
It is straightforward to verify that eq.~(\ref{VarXmin}) leads to the
SCRPA for the particle-hole propagator derived in
section~\ref{sec:phGFGen}. 
}
                                
% %%%%%%%%%%%%%%%%%%%%%%%%%%%%%%%%%%%%%%%%%%%%%%%%%%%%%%%%%%%%
% free susceptibility
% %%%%%%%%%%%%%%%%%%%%%%%%%%%%%%%%%%%%%%%%%%%%%%%%%%%%%%%%%%%%
\section{Free susceptibility in one dimension} 
\label{ChiFree}

The free susceptibility is obtained by performing the integral in
eq.~(\ref{phChiZero}) for step--like occupation
numbers~(\ref{OccHF}). 
Because of the symmetry of the dispersion relation
$\varepsilon_{k}$, the real part of $\ChiZe(q,\omega)$ is
symmetric in $q$ and $\omega$, whereas the imaginary part is
symmetric in $q$ and antisymmetric in $\omega$.

In the paramagnetic phase, $n_{k\uparrow}=n_{k\downarrow}$, and in
one dimension, the explicit expression for the real part is: 
\begin{eqnarray}
  \label{ReChiFree}
  \real \ChiZe(q,\omega) 
  \, = \,
  \frac{1}{\pi\sqrt{\abs{z}}} \,
  & \Bigg[ & 
    {\mathrm atan}\left(\frac{\sqrt{\abs{z}}\sin k_{F}}
      {\omega\cos k_{F}\,-\,4\sin^{2}(q/2)}\right) 
    \nonumber \\
    & - & 
    {\mathrm atan}\left(\frac{\sqrt{\abs{z}}\sin k_{F}}
      {\omega\cos k_{F}\,+\,4\sin^{2}(q/2)}\right) \,
  \Bigg]
\end{eqnarray}
with $z=\left(4\sin\sfrac{q}{2}\right)^{2}-\omega^{2}$ and
\begin{equation}
  \label{atan}
  {\mathrm atan\,}x \, = \, 
  \cases{  \arctan x          & \mbox{for $z<0$} \cr
          {\mathrm artanh\,}x & \mbox{for $z>0$ and $\abs{x}<1$} \cr 
          {\mathrm arcoth\,}x & \mbox{for $z>0$ and $\abs{x}>1$} }  
\end{equation}
For the imaginary part, we find in agreement with B\'enard et
al.\cite{Benard} 
\begin{eqnarray}
  \label{ImChiFree}
  \imag \ChiZe(q,\omega) 
  \, = \, -\frac{\Theta\left(z\right)}{2\sqrt{\abs{z}}}
  & \Bigg[ &
  \Theta\left(\varepsilon_{F}\,+\,\frac{\omega}{2}
    \,+\,\frac{1}{2}\sqrt{\abs{z}}\cot\frac{q}{2} \right)
  \nonumber \\
  & + &
  \Theta\left(\varepsilon_{F}\,+\,\frac{\omega}{2}
    \,-\,\frac{1}{2}\sqrt{\abs{z}}\cot\frac{q}{2} \right)
  \nonumber \\
  & - &
  \Theta\left(\varepsilon_{F}\,-\,\frac{\omega}{2}
    \,+\,\frac{1}{2}\sqrt{\abs{z}}\cot\frac{q}{2} \right)
  \nonumber \\
  & - &
  \Theta\left(\varepsilon_{F}\,-\,\frac{\omega}{2}
    \,-\,\frac{1}{2}\sqrt{\abs{z}}\cot\frac{q}{2} \right)
  \,\Bigg] 
  \; \mbox{.}
\end{eqnarray}

% %%%%%%%%%%%%%%%%%%%%%%%%%%%%%%%%%%%%%%%%%%%%%%%%%%%%%%%%%%%%
% Large U limit 
% %%%%%%%%%%%%%%%%%%%%%%%%%%%%%%%%%%%%%%%%%%%%%%%%%%%%%%%%%%%%
\section{Self--consistency of the large $U$ limit}
\label{LargeUOcc}

In this section, we will briefly outline that the occupation
numbers which are given by eq.~(\ref{nk_UinfRPA}) for the half filled
Hubbard chain in the large $U$ limit are indeed a fully
self--consistent solution of the renormalized RPA equations.

If we assume the occupation numbers from eq.~(\ref{nk_UinfRPA}),
we find the renormalized free susceptibility~(\ref{ChiZeroInf}) by
calculating the $k$--sum in eq.~(\ref{phChiZero}). 
The large $U$ limit of the renormalized RPA Green's functions is then
obtained by substituting $\ChiZe(q,\omega)$ in eq.~(\ref{phGFrpa}).

We are now able to calculate a new set of occupation numbers by
inserting this renormalized RPA Green's function in
eq.~(\ref{OccSpectralTheorem}). 
For convenience, however, we will rather use the commutator spectral
theorem, eq.~(\ref{RetSpectralTheorem}), itself:
\begin{eqnarray}
  \label{SpectralSelfConsist}
  \left( n_{k\sigma} - n_{k+q\sigma} \right)^{\mathrm new}
  & = &  - \frac{1}{\pi} \sum \limits_{p}
  \int \limits_{-\infty}^{\infty} {\rm d} \omega \;
  \imag \phGF_{k\sigma\,p\sigma}(q,\omega)
  \nonumber \\
  & = &  - \frac{1}{\pi} 
  \int \limits_{-\infty}^{\infty} {\rm d} \omega \;
  \imag 
  \left[
    \frac{\left( n_{k\sigma} - n_{k+q\sigma} \right)^{\mathrm old}}
    {\omega\,-\,\left[\varepsilon_{k+q}-\varepsilon_{k}\right]\,+\,i0^{+}} \;
    \frac{1}{1-U\ChiZe_{\uparrow}(q,\omega)\,U\ChiZe_{\downarrow}(q,\omega)}
  \right]
\end{eqnarray}

For the self--consistency to be fulfilled, we now have to show that the
``new'' occupation numbers are equal to the ``old'' ones.
As we are in the paramagnetic phase, we may drop the spin indices 
and convert the integrand in a partial fraction. 
After substituting $x=\omega/\abs{4\sin\frac{q}{2}}$ and 
$y_{k q}
=\left[\varepsilon_{k+q}-\varepsilon_{k}\right]/\abs{4\sin\frac{q}{2}}$,
we find for the ratio between new and old occupation numbers:
\begin{equation}
  \label{I_ykq}
  I(y_{k q}) \, = \, -\frac{1}{2\pi}
  \int \limits_{-\infty}^{\infty} {\rm d} x \;
  \imag 
  \left[
    \frac{1}{x\,-\,y_{k q}+\,i0^{+}} \;
    \left( 
      \frac{1}{1-U\ChiZe} \, + \,  \frac{1}{1+U\ChiZe}
    \right)
  \right]
\end{equation}

In the following, we will treat the two fractions in the integrand
separately, considering in analogy to eq.~(\ref{cdwRPASusc}) 
the first term as the charge term, and the second term as the spin
term. 
With the renormalized free susceptibility from eq.~(\ref{ChiZeroInf}),
the denominator of the charge term may be written as 
\begin{equation}
  \label{ChargeDenominator}
  1-U\ChiZe(x) \, = \, 
  \cases{ 
    2 + i\frac{x     }{\sqrt{1-x^2}}  & \mbox{for $\abs{x}<1$} \cr
    2 - \frac{\abs{x}}{\sqrt{x^2-1}} 
    + i0^{+} \, \sign x               & \mbox{for $\abs{x}>1$} }
  \mbox{.}
\end{equation}
In the same way, the denominator of the spin term is
\begin{equation}
  \label{SpinDenominator}
  1+U\ChiZe(x) \, = \, 
  \cases{ 
    -i\frac{x     }{\sqrt{1-x^2}}     & \mbox{for $\abs{x}<1$} \cr
     \frac{\abs{x}}{\sqrt{x^2-1}} 
     -i0^{+} \, \sign x               & \mbox{for $\abs{x}>1$} }
  \mbox{.}
\end{equation}

To solve the integral, we have to account for three different
contributions: 
The first one comes from the pole of the free ph Green's function, 
$1/(x\,-\,y_{k q}+\,i0^{+})$, which lies always in the ph continuum,
i.e. $\abs{y_{k q}}\le 1$.
With Dirac's identity,
\begin{equation}
  \label{DiracIdentity}
  \frac{1}{f(x)\pm i0^{+}} 
  \,=\, {\cal P}\frac{1}{f(x)} \, \mp i \pi \delta \left[ f(x) \right]
  \; \mbox{,}
\end{equation}
we find for the charge contribution
\begin{equation}
  \label{I_ykq_phPole}
  I^{\rm ch}_{1}(y_{k q}) \, = \,
  \frac{1\,-\,y_{k q}^2}{4\,-\,3\,y_{k q}^2}
  \; \mbox{.}
\end{equation}
The corresponding spin contribution is zero, since $1+U\ChiZe(x)$ is
purely imaginary within the ph continuum.

Secondly, we have to integrate over the charge and spin continuum,
respectively. Therefore, we combine the imaginary part of 
$1/(1\pm U\ChiZe)$ with the real part of the free ph Green's
function. This yields
\begin{eqnarray}
  \label{I_ykq_Cont}
  I^{\rm ch}_{2}(y_{k q}) & = & 
  \frac{2\,-\,3\,y_{k q}^2}{6\,\left(4\,-\,3\,y_{k q}^2\right)}
  \nonumber \\
  I^{\rm sp}_{2}(y_{k q}) & = & 
  \frac{1}{2} 
\end{eqnarray}

The last contribution comes from the collective poles. Again, the spin
contribution is zero, as the renormalized spin response does not show
any magnon peak in the strong coupling limit. 
As pointed out in section~\ref{LargeU}, the plasmon peak occurs at
$x=\pm 2/\sqrt 3$. Using the Dirac identity~(\ref{DiracIdentity})
once more, we get
\begin{equation}
  \label{I_ykq_collPole}
  I^{\rm ch}_{3}(y_{k q}) \, = \,
  \frac{2}{3\,\left(4\,-\,3\,y_{k q}^2\right)}
  \; \mbox{.}  
\end{equation}

Summing the three contributions, 
(\ref{I_ykq_phPole}), (\ref{I_ykq_Cont}) and (\ref{I_ykq_collPole}),  
yields unity for the ratio of the new and the old occupation numbers,
$I(y_{k q})$. By this means, we have found a fully self--consistent
solution of the renormalized RPA equations.

\end{appendix}

% %%%%%%%%%%%%%%%%%%%%%%%%%%%%%%%%%%%%%%%%%%%%%%%%%%%%%%%%%%%%
% REFERENCES
% %%%%%%%%%%%%%%%%%%%%%%%%%%%%%%%%%%%%%%%%%%%%%%%%%%%%%%%%%%%%

% %%%%%%%%%%%%%%%%%%%%%%%%%%%%%%%%%%%%%%%%%%%%%%%%%%%%%%%%%%%%
% FIGURES
% %%%%%%%%%%%%%%%%%%%%%%%%%%%%%%%%%%%%%%%%%%%%%%%%%%%%%%%%%%%%

\newpage

\begin{figure}[htbp]
  \begin{center}
    \leavevmode
    \epsfig{file=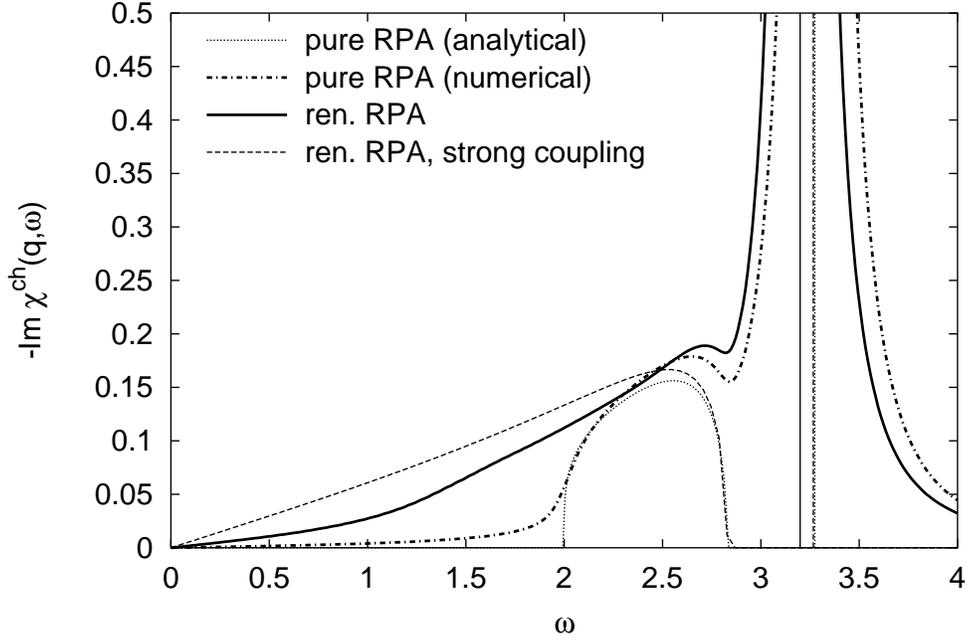,width=13cm}
    \caption{
      Charge response $\ChiCh(q=\pi/2,\omega)$ in pure and
      renormalized RPA for the half filled Hubbard chain at $U=3$. 
      The vertical lines indicate the plasmon peaks.
      Accidentally, the pure RPA plasmon and the strong coupling
      plasmon occur at almost the same energy, such that the dotted
      and the dashed vertical line cannot be resolved from another.
      The thin continuous vertical line illustrates the precise
      position of the renormalized RPA plasmon, which itself is
      represented by the singularity in the renormalized RPA charge
      response (continuous line). } 
    \label{fig:ImChiC_U3}
  \end{center}
\end{figure}

\begin{figure}[htbp]
  \begin{center}
    \leavevmode
    \epsfig{file=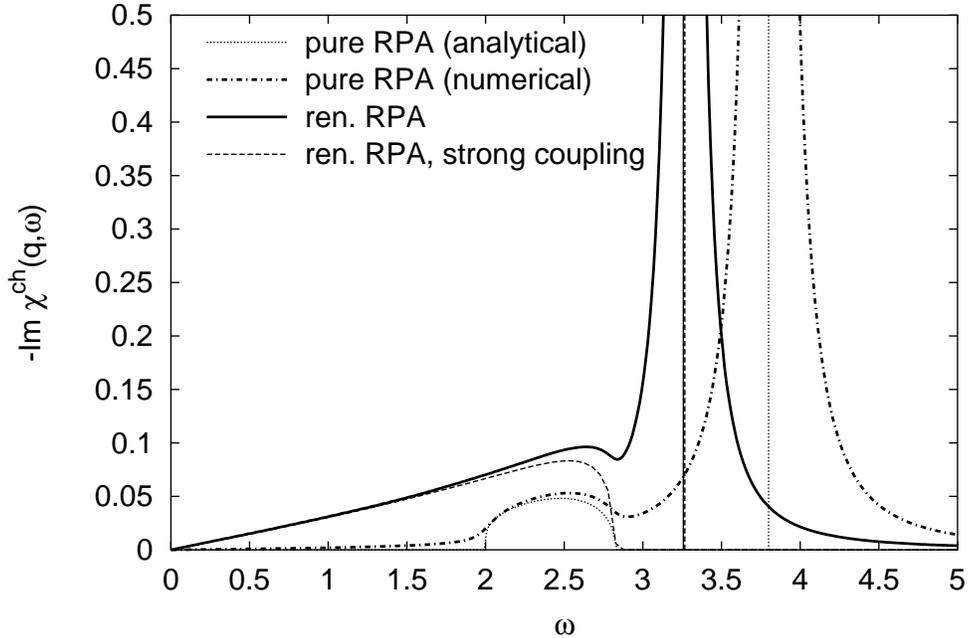,width=13cm}
    \caption{
      Charge response $\ChiCh(q=\pi/2,\omega)$ in pure and
      renormalized RPA for the half filled Hubbard chain at $U=6$. 
      The vertical lines indicate the plasmon peaks (see
      Fig.~\ref{fig:ImChiC_U3}). 
      The strong coupling limit of our theory is reached, and the
      renormalized RPA plasmon (continuous vertical line) cannot be
      resolved from the strong coupling plasmon (dashed vertical line). }  
    \label{fig:ImChiC_U6}
  \end{center}
\end{figure}

% \end{multicols}

\end{document}